\documentclass[12pt]{article}
\usepackage{graphicx} 
\usepackage{feynmf}
\usepackage{amssymb}
\usepackage{amsmath}
\usepackage{xcolor}
\usepackage{dsfont}
\usepackage{cite}
\newcommand{\cals}{\text{$\cal S$}}

\newcommand{\calf}{\mbox{${\cal F}$}}

\newcommand{\dilog}{\mbox{Li}_2}

\newcommand{\tX}{\tilde{X}}

\newcommand{\tn}{\bar{n}}

\newcommand{\bbar}{\overline{b}}
\newcommand{\detg}{\det{(G)}}
\newcommand{\dets}{\det{(\cals)}}
\newcommand{\sign}{\mbox{sign}}

\newcommand{\baru}{\bar{u}}

\newcommand{\tD}{\widetilde{D}}
\newcommand{\bbj}[2]{\overline{b}_{#1}^{\{#2\}}}
\newcommand{\bbjsq}[2]{\overline{b}_{#1}^{\{#2\} 2}}
\newcommand{\detgj}[1]{\det{(G^{\{#1\}})}}
\newcommand{\detsj}[1]{\det{(\cals^{\{#1\}})}}
\newcommand{\myref}[1]{(\ref{#1})}

\renewcommand\Re{\operatorname{Re}}
\renewcommand\Im{\operatorname{Im}}
\newcommand{\tuz}{\widetilde{u}_0}
\newcommand{\bR}{\bar{R}}
\newsavebox{\Gammap}
\newsavebox{\Gammam}
\numberwithin{equation}{section}

\begin{document}

\setlength{\unitlength}{1mm}
\begin{fmffile}{samplepics}

\begin{titlepage}

\vspace{1.cm}

\long\def\symbolfootnote[#1]#2{\begingroup%
\def\thefootnote{\fnsymbol{footnote}}\footnote[#1]{#2}\endgroup} 

\begin{center}

{\large \bf A novel approach to \\
the computation of one-loop three- and four-point functions. \\
\vspace{0.1cm}
I - The real mass case}\\[2cm]

{\large  J.~Ph.~Guillet$^{a}$, E.~Pilon$^{a}$, 
Y.~Shimizu$^{b}$ and M. S. Zidi$^{c}$ } \\[.5cm]

\normalsize
{$^{a}$ Univ. Grenoble Alpes, Univ. Savoie Mont Blanc, CNRS, LAPTH, F-74000 Annecy, France}\\
{$^{b}$ KEK, Oho 1-1, Tsukuba, Ibaraki 305-0801, Japan\symbolfootnote[2]{Y. Shimizu passed away during the completion of this series of articles.}}\\
{$^{c}$ LPTh, Universit\'e de Jijel, B.P. 98 Ouled-Aissa, 18000 Jijel, Alg\'erie}\\
      
\today
\end{center}

\vspace{2cm}

\begin{abstract} 
\noindent
This article is the first of a series of three presenting an alternative 
method to compute the one-loop scalar integrals.
This novel method enjoys a couple of
interesting features as compared with
the method closely following 't Hooft and Veltman adopted previously. 
It directly proceeds in terms of the quantities driving algebraic 
reduction methods.
It applies to the three-point functions and, in a similar way, to the four-point functions.
It also extends to complex masses without much 
complication. Lastly, it extends to kinematics more general than the one of
physical e.g.\ collider processes relevant at one loop. This last feature may be
useful when considering the application of this method beyond one loop using 
generalised one-loop integrals as building blocks. 
\end{abstract}

\vspace{1cm}

\begin{flushright}
%arXiv: yymm.xxxx\\
LAPTH-038/18\\
\end{flushright}

\vspace{2cm}

\end{titlepage}
\savebox{\Gammap}[7mm][r]{%
  \begin{fmfgraph*}(7,7)
  \fmfset{arrow_len}{3mm}
  \fmfset{arrow_ang}{10}
  \fmfipair{o,xm,xp,ym,yp}
  \fmfipath{c[]}
  \fmfipair{r[]}
  \fmfiequ{o}{(.1w,-.2h)}
  \fmfiequ{xm}{(0,-.2h)}
  \fmfiequ{ym}{(.1w,0)}
  \fmfiequ{r3}{(.35w,-.2h)}
  \fmfiset{c1}{fullcircle scaled 1.5w shifted o}
  \fmfi{fermion}{subpath (length(c1)/4,0) of c1}
  \fmfiequ{r1}{point 0 of c1}
  \fmfiequ{r2}{point length(c1)/4 of c1}
  \fmfi{fermion}{r1--r3}
  \fmfi{fermion}{o--r2}
  \fmfiv{l={\tiny 0},l.a=-120,l.d=0.05w}{o}
  \fmfiv{l={\tiny 1},l.a=-70,l.d=0.06w}{r3}
\end{fmfgraph*}}

\savebox{\Gammam}[7mm][r]{%
  \begin{fmfgraph*}(7,7)
  \fmfset{arrow_len}{3mm}
  \fmfset{arrow_ang}{10}
  \fmfipair{o,xm,xp,ym,yp}
  \fmfipath{c[]}
  \fmfipair{r[]}
  \fmfiequ{o}{(.1w,.5h)}
  \fmfiequ{xm}{(0,.5h)}
  \fmfiequ{ym}{(.1w,0)}
  \fmfiequ{r3}{(.35w,.5h)}
  \fmfiset{c1}{fullcircle scaled 1.5w shifted o}
  %\fmfpen{thick}
  \fmfiequ{r1}{point 0 of c1}
  \fmfiequ{r2}{point 3length(c1)/4 of c1}
  \fmfi{fermion}{subpath (3length(c1)/4,length(c1)) of c1}
  \fmfi{fermion}{r1--r3}
  \fmfi{fermion}{o--r2}
  \fmfiv{l={\tiny 0},l.a=90,l.d=0.05w}{o}
  \fmfiv{l={\tiny 1},l.a=70,l.d=0.06w}{r3}
\end{fmfgraph*}}

\newpage

\section{Introduction}\label{intro}

Automated evaluations of loop multileg processes demand a
fast and numerically stable evaluation of
Feynman integrals. In particular, the calculation of 
two-loop three- and four-point functions in the general complex mass case 
remains challenging. Getting a reliable result using a multidimensional 
numerical integration and sector decomposition 
\cite{Borowka:2015mxa,Borowka:2012yc,Soper:1999xk,Bogner:2007cr,Smirnov:2008py}
has a high computing cost. The derivation of a fully analytic result remains 
beyond reach so far in the general mass case. In between, approaches based on
Mellin-Barnes techniques 
\cite{Czakon:2005rk,Gluza:2007rt,Smirnov:2009up,Freitas:2010nx,Gluza:2016fwh} 
allow to perform part of the integrals analytically, yet, as far as we 
understand, the number of integrals left over 
for numerical quadratures depends on the topologies considered and can remain 
rather costly. 
An alternative approach performing some/many of the Feynman
parameter integrations analytically in a systematic way to reduce the
number of integrations to be performed numerically would therefore be useful.
 
\vspace{0.3cm}

\noindent
Such a working program was initiated in \cite{letter} for 
the calculation of massive two-loop $N$-point functions 
using analytically computed one-loop building blocks. This
approach is based on the implementation of two-loop scalar $N$-point 
functions in $n$ dimensions $^{(2)}I_{N}^{n}$ as double integrals of the form:
\[
^{(2)}I_{N}^{n}
\sim 
\sum \int_{0}^{1} d \rho \int_{0}^{1} d \xi \, W(\rho,\xi) \;
^{(1)}\widetilde{I}_{N^\prime}^{n^\prime}(\rho,\xi)
\]
where $W(\rho,\xi)$ are some weighting functions whereas the 
$^{(1)}\widetilde{I}_{N^{\prime}}^{n^\prime}(\rho,\xi)$ are``generalised
one-loop type" $N^{\prime}$-point Feynman-type integrals\footnote{The 
``effective number of external legs" $N^{\prime}$ and ``effective dimension 
$n^{\prime}$ depend on the two-loop topology in particular the number of 
internal lines $I$, and on the dimension $n$, see ref. \cite{letter}.}. 
The latter
are ``generalised" in the sense that the integration domain spanned by 
the Feynman parameters
is no longer the usual simplex 
$\{ 0 \leq z_{j} \leq 1, j = 1,\cdots,N^\prime; 
\sum_{j=1}^{N^\prime} z_{j}=1\}$ 
at work for the one-loop $N^\prime$-point function
but another domain, e.g.\ a hypercube or a cylinder with triangular basis,
which depends on the topology of the two-loop $N$-point function considered. 
The generalisation also concerns the underlying kinematics, 
which, besides external momenta, depends on two extra Feynman parameters 
$\rho$ and $\xi$. The parameter space spanned by this kinematics is 
larger than the one spanned in one-loop $N^{\prime}$-particle processes at 
colliders - for example Gram determinants may be all positive whereas this
never happens in the physical region for one-loop 
$N^\prime$-particle collider processes. 
Both these generalisations may be addressed by tuning case-by-case adaptations
of the methods well-established in the standard one-loop calculations 
\cite{tHooft:1978jhc} and using careful analytic continuations \cite{Denner:2010tr} to all 
cases considered, whose implementation can nevertheless be tricky. 
Alternatively, we hereby and in two related companion papers propose to 
develop a novel approach to address both these generalisations in a systematic 
way. 
More generally, for both the three- and four-point functions
the approach presented will consider integrals of the 
form\footnote{The contour prescription $D:= D-i\lambda$ is implicit in the 
sloppy integrals (\ref{sloppy1}) and (\ref{sloppy2}). Accordingly, 
the prescription $\Delta_{\tn} :=  \Delta_{\tn} + i \, \lambda$ is induced 
in identity (\ref{eqDEFREL1}) when the latter is substituted in integrals 
(\ref{sloppy1}) and (\ref{sloppy2}), see appendix \ref{ap1}.}
\begin{equation}\label{sloppy1}
I \sim \int_{\Sigma} \frac{d^{\tn}x}{D^{\, \alpha+1}}
\end{equation}
where $\tn=1,2$ or $3$, 
$D =X^{\;T} \cdot G \cdot X - 2 \, V^{T} \cdot X - C$, with an $\tn \times \tn$
Gram matrix $G$, a column $\tn$-vector  $V$, and the 
$x_{j}, \, j=1,\cdots,\tn$ are the components of a column $\tn$-vector $X$
spanning the simplex 
$\Sigma=\{0 \leq x_{j} \leq 1, \; j=1,\cdots,\tn \, | \; 
\sum_{j=1}^{\tn}x_{j}\leq 1\}$.
The method will make extensive use of the following Stokes-type identity
\footnote{The integration over the simplex defining the phase 
space of the Feynman parameters of this identity gives the relation between 
the $N$-point one-loop scalar functions in $n$ and $n+2$ dimensions 
\cite{Kotikov:1991hm,Kotikov:1991pm,Tarasov:1996br,Bern:1993kr,Binoth:1999sp,Duplancic:2003tv,Giele:2004iy,Denner:2005nn}.}
proven in appendix A:
\begin{equation}
  \frac{1}{D^{\alpha+1}} 
  = 
  \frac{1}{2 \, \alpha \, \Delta_{\tn}} 
  \left[ 
   \frac{\tn-2 \, \alpha}{D^{\alpha}} 
   -  
   {\nabla}^{\,T} . 
   \left( \frac{X - G^{-1} \cdot V}{D^{\alpha}} \right) 
  \right]
  \label{eqDEFREL1}
\end{equation}
where $\Delta_{\tn} = V^{T} \cdot G^{-1} \cdot V + C$ and ${\nabla}$ stands for 
the gradient w.r.t. $X$.

\vspace{0.3cm}

\noindent
The present article and two companion papers  
\cite{paper2,paper3} aim at presenting the method advocated to compute the 
building blocks $^{(1)}\widetilde{I}_{N^{\prime}}^{n^{\prime}}$ 
by applying it to the calculation of the 
standard one-loop three- and four-point functions as a ``proof of concept".
It has been designed to be straightforwardly applied so as 
to, on one hand, trivialise the Feynman parameter integrations as boundary 
terms in the integrals defining the above building 
blocks $^{(1)}\widetilde{I}_{N^{\prime}}^{n^{\prime}}(\rho,\xi)$ in a 
systematic way i.e. 
regardless of the shape of the integration domain of the Feynman parameters; 
and on the other hand to obtain all necessary analytic
continuations in a systematic way as well. 
The primary aim of its comparison with the well-established methods on this 
well studied case is not so much to readily
provide an alternative competing first-line method to compute standard one-loop 
$N$-point functions, but rather use this comparison as a test-bench seeing 
the well-established approach as a benchmark in the matter of efficiency,
which provides guidance to improve and optimise the novel method before its 
application to compute the two-loop ingredients which it has been designed for.
We aim at showing its ability to circumvent the subtleties of the various 
analytic continuations in the kinematical variables in a systematic way. 
We also aim at controlling the proliferation of dilogarithms in the closed form
expressions, which otherwise
would hamper its use in further two-loop calculations. The computation of the 
building 
blocks $^{(1)}\widetilde{I}_{N^{\prime}}^{n^{\prime}}(\rho,\xi)$ by itself 
using this method will be presented in a future article.

\vspace{0.3cm}

\noindent
Let us remind that the motivation behind this work is to study two-loop 
massive three- and four-point functions in a scalar theory. The case where 
some internal masses vanish may lead to soft/collinear divergent functions 
for which $n$ and $n^{\prime}$ have to be taken away from $4$. The computation 
of the ``generalised one-loop function'' in this case (restricted to 
the case where the phase space volume of the Feynman parameters is a simplex) 
is presented in a companion paper \cite{paper3}. 
Nevertheless, even if no internal mass vanishes, in a general scalar theory 
with three- and four-leg vertices, some two-loop three- and four-point functions 
diverge in the UV region. In this case, the space-time dimension has to be 
taken slightly under $4$ to regularise the Feynman integrals. It can be shown 
by power counting that, in this theory, the two-loop UV-divergent three- and 
four-point diagrams have four propagators which implies that the associated 
``generalised one-loop functions'' are two-point functions \cite{letter}. 
To compute analytically the latter, after doing the $n^{\prime}-4$ expansion around $0$, 
an integration over one Feynman parameter of a logarithm, at most to the power 2, 
whose argument is a second order polynomial in this Feynman parameter has to be 
performed. This integration can be carried out without any difficulty. In more 
complicated field theories, for example gauge theories, the UV divergences can 
come from tensorial integrals. Although, in principle, the method presented can 
be used, this case is postponed to a future work. Since in this article, we focus 
on fully massive three- and four-point one-loop functions which are related to UV 
finite scalar two-loop functions, we set $n$ and $n^{\prime}$ to $4$ in the rest 
of this article.

\vspace{0.3cm}

\noindent
As already said, the work presented in the present article does not quite provide 
an alternative competing 
first-line method to compute standard one-loop $N$-point functions, nevertheless it also 
provides a few interesting byproducts, we think, which are not manifest 
on the existing results.
Most of the general results for three- and
four-point scalar functions, valid for complex mass case, expressed in terms of
dilogarithms \cite{tHooft:1978jhc,Nhung:2009pm,Denner:1991qq}
%,Denner:2010tr}. Most of these results 
are valid in the physical domain, except those given in
ref. \cite{Denner:2010tr} where the authors presented very compact results for
the massive four-point scalar function whose validity has been extended by
analytical continuation for kinematical ranges accessible beyond one-loop. 
Let us mention another general result for scalar box in arbitrary dimensions 
\cite{Fleischer:2003rm}
expressed in term of generalised hypergeometric functions but this result is
hardly usable for practical computations. 
In a previous article
\cite{Guillet:2013mta} closely following  't Hooft and Veltman
\cite{tHooft:1978jhc}, one-dimensional integral representations free of
numerical instabilities caused by negative powers of Gram determinants were
obtained for three-point functions in $4,6$ and $8$ dimensions in the general
complex mass case and they have been used in the {\tt Golem95} library 
\cite{Cullen:2011kv,Binoth:2008uq}. The various analytic handlings used in 
\cite{Guillet:2013mta} following \cite{tHooft:1978jhc}: the
successive cuttings and pastings of integrals and the corresponding changes of 
variables involved, were rather intricate, they are even more so for the 
- yet unpublished - case of the four-point functions in 4 dimensions or more. 
Furthermore, in \cite{tHooft:1978jhc,Nhung:2009pm} and \cite{Denner:1991qq}, 
the connection between the analytic results and the algebraic quantities 
$\dets$ (the
determinant of the kinematic matrix associated with a given Feynman diagram), 
$\detg$ (the associated Gram determinant), and the algebraic reduction 
coefficients $b_i$ involved in algebraic reduction methods such as in 
{\tt Golem} \cite{Binoth:2005ff} (as well as similar ones corresponding to 
associated 
pinched diagrams), is badly blurred. Tracing back these algebraic ingredients
in the coefficients and in the arguments of logarithms and dilogarithms in the
final expressions obtained then requires a cumbersome extra work.
Our alternative approach
circumvents at least some of the above difficulties. 
In this respect the method presented enjoys a couple 
of interesting features as compared with the method primitively adopted in 
\cite{Guillet:2013mta}.
It directly proceeds in terms of the algebraic quantities $\dets$, 
$\detg$, $b_i$ etc. The derivation of the one-loop integral representations
happens to be more systematic and applies to the three- and four-point 
function cases in a similar way. 
It also extends to complex masses without much complication. It also applies to
kinematical configurations beyond those relevant for collider processes at the
one-loop order.
The method exploits an integral identity that, as a byproduct, also allows one
to compute angular integrals of eikonal terms describing real (massless) 
emission off massive emitters in closed form in a (relatively) simple way, 
while keeping the origin of the kinematical arguments involved in the final 
result more transparent.

\vspace{0.3cm}

\noindent
At first glance a price to pay with the novel method seems to be an inherent 
increase of the number of dilogarithms involved compared to the 't 
Hooft-Veltman results - a doubling 
in the three-point case and worse in the four-point case - when the 
one-dimensional integral representations are further computed in closed 
form. However this increase can be counteracted in a simple way, at least in 
the three-point case. The four-point case still needs more 
work to resolve this issue.

\vspace{0.3cm}

\noindent
The account of this approach amounts to an extensive technical matter.
In particular, the treatment of the four-point function involves a multistep 
process, whereas 
the upgrade from the real mass case to the general complex mass case
implies a patchwork of cases. Both deserve some elaboration. Gathering 
everything at once in a single article might seem indigestible,
we therefore chose to split the presentation into a triptych for 
ease of reading. In part I forming the present article we revisit the 
one-loop scalar three-point function $I_{3}^{4}$ with real internal masses 
to present the method, and apply it to the four-point function 
$I_{4}^{4}$ with internal masses all real. 
The extension to general complex masses, especially for $I_{4}^{4}$  will 
constitute part II, presented in a companion article.
The presence of some zero internal masses causes the appearance of infrared
soft and/or collinear singularities therefore requires some specific 
treatments although the general line of thought sticks to the one of
fully massive case. These specific cases with zero masses will thus be 
addressed separately in part III forming yet another companion article. 

\vspace{0.3cm}

\noindent
In this article we start by considering the three-point function $I_{3}^{4}$ 
with real internal masses considered as a warm-up in sec. \ref{sectthreepoint}.
We successively present two variants of the method. The simplest variant, 
which we call the ``direct way", is presented in subsubsec. 2.2.1. It is 
well suited for the three-point function, unfortunately it is not suitable 
for a handy extension to the case of the four-point function. 
In subsubsec 2.2.2. we therefore tailor a more sophisticated alternative 
called the ``indirect way". The latter comes across as rather artificial in the 
three-point case but it has the virtue of extending in a straightforward - 
albeit somewhat foaming - way to the four-point case. 
We conclude this section by commenting on the apparent doubling of 
dilogarithms and the way to counteract this doubling. 
In sec.\ \ref{sectfourpoint}, we consider the four-point function with internal masses all real, we follow the 
``indirect way" which comprises one stage more than in the three-point case because of the larger number of Feynman parameters. 
Its implementation in four steps is presented in subsec.~\ref{fourpointstep1} to \ref{fourpointstep4}.
We then discuss our results with respect to those of ref. \cite{tHooft:1978jhc}
and about the proliferation of dilogarithms in subsec.~\ref{discfourpdilog}. Lastly, we conclude.
Various appendices gather a number of utilities: tools, proofs of steps, etc. 
We removed them from the main text to facilitate its reading but we 
consider them useful to the reader. Accordingly, 
in appendix \ref{ap1}, we present a Stokes-type identity which is the
master identity for the derivation of our result. Appendix \ref{ap2} explains how
to tune the power of the denominators to apply the master identity of appendix \ref{ap1}.
Appendix \ref{detsdetg} provides a detailed discussion on how to solve the equation $\sum_{j=1}^{N} \, \cals_{ij} \, b_j = 1$, $\forall i = 1 \cdots N$
even in the case of peculiar kinematical configurations
for which $\detg$ or $\dets$ and $\detg$ vanish. 
Appendix \ref{appendJ} gives the derivation of two integrals useful for the computation
of three-point and four-point integrals. And appendix \ref{appF} shows how to compute
the last integration in a closed form in terms of dilogarithms. Lastly, appendix \ref{ImofdetS}
contains a discussion about the prescription of the imaginary part of $\dets$.

%\section{Warm-up: $I_{3}^{4}$}\label{sectthreepoint}
\section{Warm-up: the scalar three-point function $I_{3}^{4}$} \label{sectthreepoint}

\begin{figure}[h]
\centering
\parbox[c][43mm][t]{80mm}{\begin{fmfgraph*}(60,80)
  \fmfleftn{i}{1} \fmfrightn{o}{1} \fmftopn{t}{1}
  \fmf{fermion,label=$p_1$}{t1,v1}
  \fmf{fermion,label=$p_2$}{i1,v2}
  \fmf{fermion,label=$p_3$}{o1,v3}
  \fmf{fermion,tension=0.5,label=$q_1$}{v1,v2}
  \fmf{fermion,tension=0.5,label=$q_2$}{v2,v3}
  \fmf{fermion,tension=0.5,label=$q_3$,label.side=right}{v3,v1}
\end{fmfgraph*}}
\caption{\footnotesize 
A triangle showing the one-loop three-point function.}
\label{fig1} 
\end{figure}
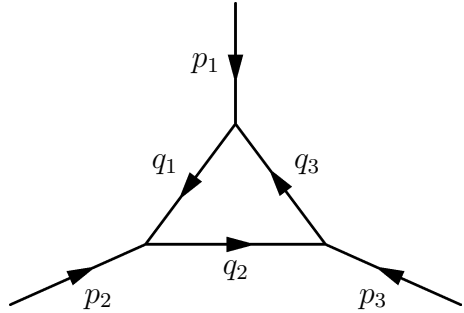

\noindent
Each internal line with momentum $q_i$ stands for the propagator of a particle
of mass $m_i$. We define the kinematic matrix $\cals$, which 
encodes all the information on the kinematics associated with this diagram by:
\begin{equation}
\cals_{i \, j} = (q_i-q_j)^2 - m_i^2 - m_j^2
\label{eqDEFCALS}
\end{equation}
The squares of differences of two internal momenta can be written in terms of
the internal masses $m_i$ and the external invariants $p_i^2$ so that 
$\cals$ reads:
\begin{equation}\label{eqcals}
\cals 
= 
\left( 
 \begin{array}{ccc}
 -2 \, m_1^2 & p_2^2 - m_1^2 - m_2^2 & p_1^2 - m_1^2 - m_3^2 \\
 p_2^2 - m_1^2 - m_2^2 & - 2 \, m_2^2 & p_3^2 - m_2^2 - m_3^2 \\
 p_1^2 - m_1^2 - m_3^2 & p_3^2 - m_2^2 - m_3^2 & - 2 \, m_3^2
 \end{array}
\right)
\end{equation}

\noindent
The usual Feynman integral\footnote{In this article and companions, we use 
the notation $- i \, \lambda$ for the Feynman prescription coming from 
propagators.} representation of $I_3^4$ is:
\begin{eqnarray}
I^4_3 
& = & 
-\int_0^1 \, \prod_{i=1}^3 \, d z_i \, 
\delta(1- \sum_{i=1}^3 \, z_i) 
\left( 
 - \, \frac{1}{2} \, Z^{\;T} \cdot \cals \cdot Z - i \, \lambda 
\right)^{-1}
\label{eqSTARTINGPOINT3}
\end{eqnarray}
Here $Z$ stands for a column 3-vector whose components are the $z_{i}$, 
$\cals$ is the $3 \times 3$ kinematic matrix associated with the diagram of fig.
\ref{fig1}, and the superscript ``$^{T}$" stands for the matrix transpose.
Let us single out the subscript value 
$a$ ($ a \in S_3 =\{1,2,3\}$) and write $z_a$
as $1 - \sum_{i \ne a} z_i$. We find:
\begin{eqnarray}
- \, Z^{\;T} \cdot \cals \cdot Z 
& = & 
\sum_{i,j \in S_3 \setminus \{a\}} G_{i\,j}^{(a)} \, z_i \, z_j  - 
2 \sum_{j \in S_3 \setminus \{a\}} V_j^{(a)} \, z_j  -  C^{(a)}
\label{eqVECZT3}
\end{eqnarray}
where the $2 \times 2$ Gram matrix $G^{(a)}$ and the column 2-vector $V^{(a)}$
are defined by
\begin{eqnarray}
G_{i\,j}^{(a)} 
& = & 
-  \, (\cals_{i\,j}-\cals_{a\,j}-\cals_{i\,a}+\cals_{a\,a}), \;\; i,j \neq a
\nonumber
\\
V_j^{(a)} 
& = & 
\;\;\;\;\;  \cals_{a\,j} - \cals_{a \, a}, \;\;  j \neq a
\label{eqVJA3}\\
C^{(a)}& = & 
\;\;\;\;\;  \cals_{a \, a} 
\nonumber
\end{eqnarray}
We label $b$ and $c$ the two elements of $S_3 \setminus \{a\}$, with $b < c$. 
We write the polynomial (\ref{eqVECZT3}) with the 
$2 \times 2$ matrix $G^{(a)}$, the column 2-vector $V^{(a)}$, the scalar 
$C^{(a)}$ as:
\begin{equation}
D^{(a)}(z_b,z_c) 
=  
X^{(a)\;T} \cdot G^{(a)} \cdot X^{(a)} - 
2 \; V^{(a) \, T} \cdot X^{(a)} - C^{(a)}
\;\; , \;\; 
X^{(a)} =
 \left[
 \begin{array}{c}
   z_b \\
   z_c 
 \end{array}
 \right]
\label{eqDsupa}
\end{equation}
We note the integration simplex 
$\Sigma_{bc}=\{ 0 \leq z_{b}, z_{c}, z_{b}+z_{c} \leq 1\}$. 
$I^4_3$ can be written:
\begin{eqnarray}
I_3^4 
& = & 
- \, 2 
\int_{\Sigma_{bc}} \frac{d z_b \, d z_c}{D^{(a)}(z_b,z_c) - i \, \lambda}
\label{eqI341}
\end{eqnarray}

\subsection{Step 1}\label{sect3pstep1}

Let us substitute identity (\ref{eqDEFREL1}) into eqs. (\ref{eqI341}) 
or (\ref{sloppy1}). Were the power $\alpha+1$ in the l.h.s of 
eq. (\ref{eqDEFREL1}) such that $\tn - 2 \alpha = 0$, only the 
boundary term - i.e. the second term in eq. (\ref{eqDEFREL1}) - would remain, 
thus making one integration in $I$ trivial. In the case of the $N=3$-point 
function, cf. eq. (\ref{eqI341}), $\tn=N-1$ is equal to 2. Imposing 
$\tn \, - \, 2 \alpha = 0$ thus forces $\alpha+1 = 2$ in  eq. (\ref{eqDEFREL1}). 
However in eq. (\ref{eqI341}), $1/D$ appears raised to the power 1, not 2. 
The idea is thus to adjust the power of the denominator by introducing an 
appropriate integral representation of the form
\begin{equation}\label{sloppy2}
\frac{1}{D^{\alpha+1}} 
\sim
\int_{0}^{+\infty} \frac{d \xi}{\left( D + \xi^{\nu} \right)^{\alpha^{\prime}+1}}
\end{equation}
($\xi^{\nu}$ being some suitable power of $\xi$) so that the power 
$\alpha^{\prime}$  of the effective denominator in this representation 
matches the request $\tn - 2 \alpha^{\prime} = 0$. The auxiliary identity 
(\ref{sloppy2}) is properly made explicit and derived in its general form in 
appendix \ref{ap2}.
In the case of the three-point function in four dimensions $I_{3}^{4}$
the relevant form for identity (\ref{sloppy2}) is simply the familiar integral 
representation:
\begin{equation}
\frac{1}{D - i \, \lambda}
= 
\int_0^{+\infty} \, \frac{d \xi}{(D + \xi - i \, \lambda)^{2}} 
\label{eqFOND1-simple}
\end{equation}
The substitution of identity (\ref{eqFOND1-simple}) into eq. (\ref{eqI341}) 
provides a representation of $I_{3}^{4}$ where $(D^{(a)} + \xi)^{2}$ now
replaces $D^{(a)}$ in the integrand:
\begin{eqnarray}
I_3^4 
& = & 
- \, 2 \int_0^{+\infty} d \xi \, \int_{\Sigma_{bc}} d x_b \, d x_c \, 
\frac{1}{(D^{(a)}(x_b,x_c) + \xi - i \, \lambda)^2}
\label{eqI3420}
\end{eqnarray}
Identity (\ref{eqDEFREL1}) is then applied to the integrand of (\ref{eqI3420}) 
considered as a function of the two variables $x_{b},x_{c}$ to be integrated 
first keeping $\xi$ fixed, yielding:
\begin{eqnarray}
I_3^4 
& = & 
\int_0^{+\infty} d \xi \, \int_{\Sigma_{bc}} d x_b \, d x_c \, 
\nonumber \\
& & 
\mbox{} \times 
\left[ \frac{1}{\Delta_2 - \, \xi + i \, \lambda} 
 \sum_{j \in S_3 \setminus\{a\}} \frac{\partial}{\partial x_{j}}  
 \left( 
  \frac{(X \, - \, (G^{(a)})^{-1} \cdot V^{(a)})_{j}}
  {D^{(a)}(x_b,x_c) + \xi - i \, \lambda} 
 \right) 
\right]
\label{eqI342}
\end{eqnarray}
The use of the Stokes identity (\ref{eqDEFREL1}) in the  integral representation
(\ref{eqFOND1-simple}) induces an  apparent pole at $\xi = \Delta_{2}$ in eq.
(\ref{eqI342}). 
Insofar as $\Delta_{2} \neq 0$ this apparent pole is no issue,
as can be  justified directly as follows.   
It is manifestly no issue in the real mass case for which the 
$- \, i \, \lambda$ contour prescription avoids this apparent pole anyway.
In the general complex mass case the $- \, i \, \lambda$ contour prescription is 
overruled by the finite imaginary part of $\Delta_2$, whose sign may however 
change in a continuous way with the kinematics: one might then worry about what
happens whenever $\Im(\Delta_2)$ vanishes with a change of sign while
$\Re(\Delta_2) > 0$. In this respect we shall however notice that
\[
D^{(a)}(x_b,x_c) + \Delta_2 
= 
\left( X^{(a)} - (G^{(a)})^{-1} \cdot V^{(a)} \right)^{T} 
\cdot G^{(a)} \cdot
\left( X^{(a)} - (G^{(a)})^{-1} \cdot V^{(a)} \right)
\]
so that
\begin{eqnarray}
\lefteqn{
\sum_{j \in S_3 \setminus\{a\}} \frac{\partial}{\partial x_{j}}  
 \left( 
  \frac{(X \, - \, (G^{(a)})^{-1} \cdot V^{(a)})_{j}}
  {D^{(a)}(x_b,x_c) + \Delta_{2} - i \, \lambda} 
 \right) 
}
\nonumber\\
& = &
\frac{1}{2} \,
\sum_{j,k \in S_3 \setminus\{a\}} 
\frac{\partial}{\partial x_{j}} 
\left(
 (G^{(a)})^{-1}_{jk} \,
 \frac{\partial}{\partial x_{k}} \,
  \ln \left( D^{(a)}(x_b,x_c) + \Delta_{2} - i \, \lambda \right) 
\right) 
\label{laplgen}
\end{eqnarray}
This distribution (\ref{laplgen}), whether by direct calculation of the l.h.s. 
or by recognition on the r.h.s., proves to vanish 
identically outside its singular support contained in the region where 
$D^{(a)}(x_b,x_c) + \Delta_{2} = 0$. 
On the other hand $\Im(D^{(a)}(x_b,x_c)) < 0$ on the integration simplex 
$\Sigma_{bc}$ where it is a convex linear combination of the imaginary parts of 
the masses squared. Thus, when $\Im(\Delta_{2})$ vanishes, 
$\Im(D^{(a)}(x_b,x_c) + \Delta_{2} - i \, \lambda) <0$ on $\Sigma_{bc}$ hence 
the residue of the apparent pole $\xi = \Delta_2$ in eq. (\ref{eqI342}) 
vanishes identically in $x_b,x_c$ on $\Sigma_{bc}$.

\vspace{0.3cm}

\noindent
The cases where $\Delta_2$ vanishes require a separate examination.
This implies that $\dets$ vanishes. However the condition $\dets = 0$ does not 
correspond to a kinematical singularity of the initial Feynman integral if the
corres\-ponding eigenvector of $\cals$ points outside the quadrant 
$\{z_j \geq 0\}$. In this case $D^{(a)}(x_b,x_c)$ never vanishes on the simplex
$\Sigma_{bc}$ and the above argument based on eq. (\ref{laplgen}) still 
applies. The case corresponding to the presence of infrared singularities in
conjunction with some internal masses is addressed separately in the third article of the series. 
Lastly, if the condition $\dets = 0$ corresponds to a
threshold singularity, the original Feynman integral is indeed singular and 
$\xi=0$ is an end-point singularity of the integral representation 
(\ref{eqI342}).
 
\vspace{0.3cm}

\noindent
For each term of the sum in eq. (\ref{eqI342}) the integration performed first 
is on the variable $x_{j}$ on which the derivative acts, and we get:
\begin{eqnarray}
I_3^4 
& = & 
\int_0^{+\infty} \frac{d \xi}{\Delta_2 - \, \xi + i \, \lambda} 
%\, 
\nonumber\\
&&
\;\;\;\;
\Biggl[ 
 \int^1_0 d x_c \, 
 \left( 
  \frac{(1-x_c) - ((G^{(a)})^{-1} \cdot V^{(a)})_b}
  {D^{(a)}(1-x_c,x_c) + \xi - i \, \lambda} 
  - 
  \frac{- \, ((G^{(a)})^{-1} \cdot V^{(a)})_b}
  {D^{(a)}(0,x_c) + \xi - i \, \lambda} 
\right) 
\nonumber \\
& & 
\mbox{} 
\;\;
+ 
 \int^1_0 d x_b \, 
 \left( 
  \frac{(1-x_b) - ((G^{(a)})^{-1}\cdot V^{(a)})_c}
  {D(x_b,1-x_b) + \xi - i \, \lambda} 
  - 
  \frac{- \, ((G^{(a)})^{-1}\cdot V^{(a)})_c}
  {D(x_b,0) + \xi - i \, \lambda} 
 \right) \, 
\Biggr]
\label{eqI343}
\end{eqnarray}
After the change of variables $x_b=x$ and $x_c=1-x$, 
the first and third terms in the integrand of the r.h.s. of eq. (\ref{eqI343}) 
recombine and we get:
\begin{eqnarray}
I_3^4 
& = & 
\int_0^{+\infty}
\frac{d \xi}{\Delta_2 -  \, \xi + i \, \lambda} \,  \int^1_0 d x \, 
\Biggl( 
 \frac{1 - ((G^{(a)})^{-1}\cdot V^{(a)})_b - ((G^{(a)})^{-1}\cdot V^{(a)})_c}
 {D^{(a)}(x,1-x) + \xi - i \, \lambda} 
\nonumber \\
& & 
\mbox{} 
\;\;\;\;\;\;\;\;\;\;\;\;\;\;\;\;\;\;\;\;\;\;\;\;\;\;\;\;\;\;\;\;
+ \,
 \frac{((G^{(a)})^{-1}\cdot V^{(a)})_b}{D^{(a)}(0,x) + \xi - i \, \lambda} 
+ 
 \frac{((G^{(a)})^{-1}\cdot V^{(a)})_c}{D^{(a)}(x,0) + \xi - i \, \lambda}  
\Biggr)
\label{eqI344}
\end{eqnarray}
The quantities $\Delta_{2}$ and $(G^{(a)})^{-1}\cdot V^{(a)}$ in eq. 
\myref{eqI344} are expressed simply\footnote{We remind that 
$\detg \equiv \det \, G^{(a)}$ is independent of $a$.} in terms of $\dets$, 
$\detg$ and the coefficients $\overline{b}_{j}$ such that 
$\sum_{j=1}^{3} \cals_{i \, j} \, \overline{b}_{j} = \dets \, e_i$ with $e_i = 1$ for $i = 1, \, 2, \, 3$ 
(cf.\ eqs.~(\ref{eqe10-02}) and (\ref{eqsolbbar}) of appendix~\ref{detsdetg}):
\begin{eqnarray}
\Delta_2 
& = & 
\frac{\dets}{\detg} 
\nonumber\\
((G^{(a)})^{-1}\cdot V^{(a)})_{b} 
& = & 
\frac{\bbar_b}{\detg} 
\nonumber\\
((G^{(a)})^{-1}\cdot V^{(a)})_{c}
& = & 
\frac{\bbar_c}{\detg} 
\nonumber\\
1 - ((G^{(a)})^{-1}\cdot V^{(a)})_b - ((G^{(a)})^{-1}\cdot V^{(a)})_c
& = & 
\frac{\bbar_a}{\detg} 
\label{eqdefdelta2} 
\end{eqnarray}
The denominators $D^{(a)}(0,x)$ and $D^{(a)}(0,x)$ straightforwardly read:
\begin{eqnarray}
D^{(a)}(0,x)
& = & 
- \left( \cals_{cc} - 2 \, \cals_{ca} + \cals_{aa} \right) \, x^{2}
- 2 \left( \cals_{ca} - \cals_{aa} \right) \, x - \cals_{aa}
\nonumber\\
& = & 
G^{\{b\}(a)} \, x^{2} - 2 \, V^{\{b\}(a)} \, x - C^{\{b\}(a)} 
\; \equiv \; D^{\{b\} (a)}(x)
\label{Db}\\
D^{(a)}(x,0)
& = & 
- \left( \cals_{bb} - 2 \, \cals_{ba} + \cals_{aa} \right) \, x^{2}
- 2 \left( \cals_{ba} - \cals_{aa} \right) \, x - \cals_{aa}
\nonumber\\
& = & 
G^{\{c\}(a)} \, x^{2} - 2 \, V^{\{c\}(a)} \, x - C^{\{c\}(a)} 
\; \equiv \; D^{\{c\} (a)}(x)
\label{Dc}
\end{eqnarray}
and for $D^{(a)}(x,1-x)$ a simple calculation yields:
\begin{eqnarray}
D^{(a)}(x,1-x)
& = &
- \left( \cals_{bb} - 2 \, \cals_{bc} + \cals_{cc} \right) \, x^{2}
- 2 \left( \cals_{bc} - \cals_{cc} \right) \, x - \cals_{cc}
\nonumber\\
& = & 
G^{\{a\}(c)} \, x^{2} - 2 \, V^{\{a\}(c)} \, x - C^{\{a\}(c)} 
\; \equiv \; D^{\{a\} (c)}(x)
\label{Da}
\end{eqnarray}
The expressions in eqs. (\ref{Db})-(\ref{Da}) are one-variable counterparts 
of eq. (\ref{eqDsupa}). For example the  equation making explicit $D^{(a)}(x,1-x)$
involves the pinched matrix $\cals^{\{a\}}$ in which line and column $c$ are
singled out so as to define the associated pinched Gram ``matrix"
$G^{\{a\}(c)}$, the ``vector" $V^{\{a\}(c)}$ and the scalar $C^{\{a\}(c)}$ as
in eqs.  (\ref{eqVECZT3})-(\ref{eqVJA3}). The pinched matrix $\cals^{\{a\}}$ is itself 
built from the $\cals$ matrix by removing the line and column $a$ cf.\ ref.\ \cite{Binoth:2005ff}.
Similar explications hold for
$D^{(a)}(0,x)$  and $D^{(a)}(x,0)$.
The quantities in eqs. (\ref{Db})-(\ref{Da}) are precisely those involved in 
the integral representations of the one-loop two-point integrals corres\-ponding
to all possible pinchings of the triangle diagram in fig. 1.
In the three-point case at hand, the above pinched Gram ``matrices" and 
``vectors" have actually one 
component only, thus the multiple superscript notation ``$^{\{a\}(c)}$", etc. 
may seem a clumsy sophistication.
Yet it keeps track of the line and column that have been singled out and 
it encodes how quantities with the same pinching but 
distinct singled out lines and columns are related. For example,
\begin{equation}\label{samepinch}
D^{\{a\} (c)}(x)  =  D^{\{a\} (b)}(1-x)
\end{equation}
i.e.
\begin{eqnarray}
G^{\{a\}(b)} & = & \;\;\;  G^{\{a\}(c)}
\nonumber\\
V^{\{a\}(b)} & = & - \, V^{\{a\}(c)} +  G^{\{a\}(c)}
\label{samepinch1}\\
C^{\{a\}(b)} & = & \;\;\; C^{\{a\}(c)} + 2 \,V^{\{a\}(c)} -  G^{\{a\}(c)}
\nonumber
\end{eqnarray}
Likewise we have: 
\begin{eqnarray}
D^{\{b\} (a)}(x) \, = \, D^{\{b\} (c)}(1-x)
& , & D^{\{c\} (a)}(x) \, = \, D^{\{c\} (b)}(1-x)
\label{samepinch2}
\end{eqnarray}
with corresponding relations among the 
$G^{\{b\}(j)},V^{\{b\}(j)},C^{\{b\}(j)}$ for $j \in S_3 \setminus \{b\}$, 
and among the 
$G^{\{c\}(k)},V^{\{c\}(k)},C^{\{c\}(k)}$ for $k \in S_3 \setminus \{c\}$ 
respectively. The three-point integral (\ref{eqI344}) 
may then be written as a weighted sum over the $\bbar_i$:
\begin{eqnarray}
I_3^4 
& = & 
\sum_{i \in S_3} \, \frac{\bbar_i}{\detg} \, 
\int_0^{+\infty} \frac{d \xi}{\Delta_2 - \, \xi + i \, \lambda} \, 
\int_0^1 \frac{dx}{D^{\{i\} (i^{\prime})}(x) + \xi - i \, \lambda} 
\label{eqI345}
\end{eqnarray}
For the sake of definiteness, eqs. (\ref{Db})-(\ref{Da}) provide
a value of $i^{\prime}$ for each $i$ in eq. (\ref{eqI345}). 
Let us however note that the assignment for $i^{\prime}$ 
in eq. (\ref{eqI345}) is not unique. It can be 
changed relying on properties (\ref{samepinch}) and (\ref{samepinch2}) 
in combination with changes of variables $y=1-x$ in the terms in eq. 
(\ref{eqI345}). For further convenience, we trade the assignment provided by 
eqs. (\ref{Db})-(\ref{Da}) for the alternative choice 
$i^{\, \prime} \equiv 1 + (i$ modulo $3)$ in what follows. At this 
stage, one may argue that we have progressed by next to nothing as we performed 
one integration trivially yet at the price of introducing an extra integral
so that two integrals still remain to be performed. 
However representation (\ref{eqI345}) amounts to a handier form as we will 
see next.

\subsection{Step 2}\label{sect3pstep2}

We can proceed further in two different ways.
The first way henceforth called ``direct'' has the virtue to lead in a very few
simple steps to the result which we obtained using the method {\em \`a la} 
't Hooft and Veltman cf. ref. \cite{Guillet:2013mta}. Unfortunately, in the case of
the four-point function we did not succeed in proceeding as simply. 
We have therefore formulated an alternative to the direct way, henceforth 
called ``indirect". 
It is somewhat academic to follow the latter for the three-point function, all 
the more so as it is uselessly more cumbersome than the direct way in this 
case. Yet we do the exercise as a warm-up before tackling the more involved case 
of the four-point function. 
As will be seen, the result that we obtain at first for the three-point 
function in the ``indirect way'' is {\em not quite} the same formula as the one 
obtained via the ``direct way''. 
In particular the ``indirect way'' would lead to a doubling of dilogarithmic 
terms compared to result via the ``direct way'' if the last integral were
explicitly performed in closed form. A comparison 
between the two may help understand how to recombine terms from the ``indirect 
way'' so as to counteract this doubling. The 
experience gained on this example may guide us to reverse a similar unwanted 
proliferation of dilogarithms in the case of the four-point function where only 
the indirect way is available. Let us now present the two ways in this
subsection. 

\subsubsection{Direct way}\label{subsubdirectway}

We integrate directly the l.h.s. of eq. (\ref{eqI345}) over the variable $\xi$ 
first, keeping $x$ fixed. This requires us to perform the following type of 
integral:
\begin{equation}
K = - \,  \int^{+\infty}_0 d \xi \, 
\frac{1}
{\left(\xi - \Delta_2 - i \, \lambda \right) \, 
 \left(\xi + D^{\{i\}(i^{\prime})}(x) - i \, \lambda \right)
}
\label{eqintafaire0}
\end{equation}
which, using eq. (\ref{eqdefk4}), gives\footnote{We assume that the elements of
the kinematic matrix $\cals$ have been made dimensionless by an appropriate
rescaling so that the arguments of the logarithms in eq. (\ref{eqintafaire3})
are dimensionless as well.}:
\begin{equation}
K = - \,  \frac{1}{D^{\{i\}(i^{\prime})}(x)+\Delta_2} \, 
\left[ 
 \ln \left(D^{\{i\}(i^{\prime})}(x)- i \, \lambda \right)
 - 
 \ln \left(- \Delta_2 - i \, \lambda \right) 
\right]
\label{eqintafaire3}
\end{equation}
Last we substitute eq. (\ref{eqintafaire3}) into eq. (\ref{eqI345}).
To make the connection with the notations used in \cite{Guillet:2013mta}, we 
identify $\Delta_{2} = 1/B$, $\bbar_i/\detg = b_{i}/B$ (cf.\ also appendix~\ref{detsdetg}) and write
\begin{equation}
I_3^4 
= 
- \, \sum_{i \in S_3} \, 
b_i \int_0^1
\frac{d x}{B \, D^{\{i\}(i^{\prime})}(x) + 1} \, 
\left[ 
 \ln \left( D^{\{i\}(i^{\prime})}(x)- i \, \lambda \right)
 - 
 \ln \left( - \, \frac{1}{B} - i \, \lambda \right)
\right]
\label{eqI346}
\end{equation}
$D^{\{i\} (i^{\prime})}(x)$ in eq. (\ref{eqI346}) shall be 
identified with $2 \times g_{(i)}(x)$ in 
\cite{Guillet:2013mta}\footnote{In ref. \cite{Guillet:2013mta} we defined 
$D(x_{b},x_{c}) = \frac{1}{2} \, X^{T} \cdot G^{(a)} \cdot X  - 
V^{(a) \, T} \cdot X - \frac{1}{2} \, \cals_{aa}$, 
whereas in the present article the more convenient normalisation
$D^{(a)}(x_{b},x_{c}) = X^{T} \cdot G^{(a)} \cdot X - 
2 \, V^{(a) \, T} \cdot X - \cals_{aa}$ is used, hence this factor 2 mismatch.}. 
The above result thus coincides with the result obtained in ref. 
\cite{Guillet:2013mta} after symmetrisation over the parameter $\alpha$ using 
the method {\em \`a la} 't Hooft and Veltman. The present derivation is somewhat 
faster, though. The derivation of eq. (\ref{eqI346}) holds true also in the 
complex mass case. In the real mass case, the difference of logarithms in eq. 
(\ref{eqI346}) can be replaced by the logarithm of the ratio.

\subsubsection{Indirect way}\label{subsubsectindway}

We perform the integration over $x$ in eq. (\ref{eqI345}) by writing the 
integrand as a derivative using identity \myref{eqDEFREL1} anew - this time for
``$\tn=1$". Therefore identity (\ref{eqFOND1-simple}) is not the relevant one to 
provide an integral representation of $1/(D^{\{i\} (i^{\prime})} + \xi)$ with the 
appropriate power in the denominator so as to apply identity \myref{eqDEFREL1}. 
Indeed, in order to keep only the boundary term
in identity (\ref{eqDEFREL1}) we shall choose $\alpha = 1/2$ whereas the power
of the denominator in eq. (\ref{eqI345}) is 1 not 3/2. We thus have to 
customise an alternative representation of the type (\ref{sloppy2})
providing shifts in powers of denominators by 1/2 instead of 1. 
A generalised representation including this type
is derived in appendix \ref{ap2}. It happens to be also the 
one suited to the four-point function case as will be seen below.
It takes the following form (see eq.~(\ref{eqFOND1})):
\begin{equation}
\frac{1}{(D^{\{i\} (i^{\prime})})^{\mu-1/\nu}}
= 
\frac{\nu}{ 
B \left( \mu - \frac{1}{\nu},\frac{1}{\nu} \right)}
\int^{+\infty}_0 \, \frac{d \rho}{(D^{\{i\} (i^{\prime})} + \rho^{\nu})^{\mu}} 
\label{eqFOND1biss}
\end{equation}
where $B(y,z) = \Gamma(y) \, \Gamma(z)/\Gamma(y+z)$ and $\Gamma(y)$ are the 
Euler Beta and Gamma functions \cite{gradshteyn2007}.
The parameters $\mu$ and $\nu$ are chosen such that  $\mu - 1/\nu =$ 
the power of $(D^{\{i\} (i^{\prime})} + \xi)$ in the integrand of eq. 
(\ref{eqI345}) 
i.e. 1. The representation of $1/(D^{\{i\} (i^{\prime})}+\xi)$ thus obtained 
is substituted into eq. (\ref{eqI345}). This provides a representation of 
$I_{3}^{4}$ with factors 
$1/(D^{\{i\} (i^{\prime})} + \xi + \rho^{\nu})^{\mu}$. 
Identity (\ref{eqDEFREL1}) 
is then applied to this new integrand considered as a function of 
$x$ seen as single integration variable (i.e. $n=1$), with $\rho$ (and $\xi$) 
seen as fixed. 
In order that the first term of identity (\ref{eqDEFREL1})  
vanish, $\mu-1 = \alpha$ shall be chosen equal to $1/2$ thus
$\mu = 3/2$ and $\nu= 2$. We substitute this integral representation 
into eq. (\ref{eqI345}) and perform the integration over $x$
explicitly. We get:
\begin{eqnarray}
I_3^4 
& = & 
- \sum_{i \in S_3} \, 
\frac{\bbar_i}{\detg} \, \frac{2}{B(1,1/2)} \, 
\int_0^{+\infty} 
\frac{d \xi}{\Delta_2 - \, \xi + i \, \lambda} \, 
\int_0^{+\infty} 
\frac{d \rho}{\Delta_1^{\{i\}} -  \, \xi -  \, \rho^2 + i \, \lambda} 
\nonumber \\
& & 
\;\;\;\;\;\;
\times 
\left[ 
 \frac{\left( 1 - (G^{\{i\}(i^{\prime})})^{-1} \, V^{\{i\}(i^{\prime})} \right)}
 {\left( 
  D^{\{i\} (i^{\prime})}(1) + \xi + \rho^2 - i \, \lambda 
  \right) ^{1/2}
 } 
 + 
 \frac{(G^{\{i\}(i^{\prime})})^{-1} \, V^{\{i\}(i^{\prime})}}
 {\left( 
   D^{\{i\} (i^{\prime})}(0) + \xi + \rho^2 - i \, \lambda 
  \right) ^{1/2}
 } 
\right]
\label{eqI347}
\end{eqnarray}
with $ \Delta_{1}^{\{i\}} = 
(G^{\{i\}(i^{\prime})})^{-1} \, (V^{\{i\}(i^{\prime})})^{2} + 
C^{\{i\}(i^{\prime})}$. In eq. (\ref{eqI347}), one recognises familiar 
algebraic quantities associated with the three possible pinchings of the 
triangle diagram (cf.\ eqs.\ (\ref{eqe10-02}) and (\ref{eqsolbbar})):
\begin{equation}
  \Delta_1^{\{i\}} = - \, \frac{\detsj{i}}{\detgj{i}}
  \label{eqdefdelta1}
\end{equation}
and
\begin{eqnarray}
1 - ( G^{\{1\}(2)} )^{-1} \,  V^{\{1\}(2)}
= - \, \frac{\bbj{2}{1}}{\detgj{1}} 
&,& 
(G^{\{1\}(2)})^{-1} \, V^{\{1\}(2)}
= - \, \frac{\bbj{3}{1}}{\detgj{1}} 
\nonumber\\
1 - (G^{\{2\}(3)})^{-1} \, V^{\{2\}(3)}
= - \, \frac{\bbj{3}{2}}{\detgj{2}} 
&,& 
(G^{\{2\}(3)})^{-1} \, V^{\{2\}(3)}
= - \, \frac{\bbj{1}{2}}{\detgj{2}} 
\label{Rij}\\
1 - (G^{\{3\}(1)})^{-1} \, V^{\{3\}(1)}
= - \, \frac{\bbj{1}{3}}{\detgj{3}} 
&,& 
(G^{\{3\}(1)})^{-1} \, V^{\{3\}(1)}
= - \, \frac{\bbj{2}{3}}{\detgj{3}} 
\nonumber
\end{eqnarray}
Let us introduce\footnote{This definition is chosen so that
$\tD_{ij} = 2 \, m_{k}^{2}$ where $k \in S_{3} \setminus \{i,j\}$.} 
\begin{eqnarray}
\tD_{12} \; \equiv \; D^{\{1\}(2)}(1) & = & D^{\{2\}(1)}(1) \; \equiv \; \tD_{21}
\nonumber\\
\tD_{23} \; \equiv \; D^{\{2\}(3)}(1) & = & D^{\{3\}(2)}(1) \; \equiv \; \tD_{32}
\label{Dtilde}\\
\tD_{31} \; \equiv \; D^{\{3\}(1)}(1) & = & D^{\{1\}(3)}(1) \; \equiv \; \tD_{13}
\nonumber
\end{eqnarray}
From eq. (\ref{samepinch}) we also have:
\begin{eqnarray}
D^{\{1\}(2)}(0) \; = \; D^{\{1\}(3)}(1) & = & \tD_{13}
\nonumber\\
D^{\{2\}(3)}(0) \; = \; D^{\{2\}(1)}(1) & = & \tD_{21}
\label{Dtilde2}\\
D^{\{3\}(1)}(0) \; = \; D^{\{3\}(2)}(1) & = & \tD_{32}
\nonumber
\end{eqnarray}
We can write eq. \myref{eqI347} as the following weighted sum over the 
coefficients $\bbar_{i}$ and $\bbj{j}{i}$:
\begin{align}
&I_3^4  = \, \sum_{i \in S_3} \, 
\sum_{j \in S_3 \setminus \{i\}} \, \frac{\bbar_i}{\detg} \, 
\frac{\bbj{j}{i}}{\detgj{i}} \; 
L_3^4 \left( \Delta_2, \Delta_{1}^{\{i\}}, \tD_{ij} \right)
\label{eqI349}
\end{align}
with
\begin{eqnarray}
L_3^4 \left( \Delta_2, \Delta_{1}^{\{i\}}, \tD_{ij} \right)
& = &
\int_0^{+\infty} \frac{d \xi}{(-\Delta_2 +  \, \xi - i \, \lambda)}
\nonumber\\
&&  \times
 \int_0^{+\infty} \frac{d \rho}{ 
 (- \Delta_1^{\{i\}} +  \, \xi +  \, \rho^2 - i \, \lambda) \,
 (\tD_{ij} + \xi + \rho^2 - i \, \lambda)^{1/2}
}
\label{eqdeflij1}
\end{eqnarray}
We first perform the $\rho$ integration in eq. (\ref{eqdeflij1})
using appendix~\ref{appendJ}. In the real mass case at hand
$\Im(- \Delta_1^{\{i\}} - i \, \lambda)$ and 
$\Im(\tD_{ij} - i \, \lambda)$ have the same (negative) sign, 
we can then use relation (\ref{eqdeffuncjp2}) to rewrite 
$L_3^4 \left( \Delta_2, \Delta_{1}^{\{i\}}, \tD_{ij} \right)$ as:
\begin{equation}
L_3^4 \left( \Delta_2, \Delta_{1}^{\{i\}}, \tD_{ij} \right)
= 
\int_0^{+\infty} d \xi \, \frac{d \xi}{-\Delta_2 +  \, \xi - i \, \lambda}  
\int_0^{1} 
\frac{d z}
{\xi - (1-z^2) \, \Delta_1^{\{i\}} + z^2 \, \tD_{ij} - i \, \lambda}
\label{eqdeflij2}
\end{equation}
We perform the $\xi$ integration first, using eq. (\ref{eqdefk4}) to get:
\begin{align}
L_3^4 \left( \Delta_2, \Delta_{1}^{\{i\}}, \tD_{ij} \right)
&= 
\int_0^{1} 
\frac{d z}{(\tD_{ij} + \Delta_1^{\{i\}}) \, z^2 + \Delta_2 - \Delta_1^{\{i\}}}
\notag \\
& \quad {}\quad {}
\left[  
 \ln 
  \left( 
   (\tD_{ij} + \Delta_1^{\{i\}}) \, z^2  - \Delta_1^{\{i\}} - i \, \lambda 
  \right) 
  - 
  \ln \left( -\Delta_2 - i \, \lambda \right) 
\right]
\label{eqdeflij3}
\end{align}
In the real mass case, the difference of logarithms in eq. 
(\ref{eqdeflij3}) can be rewritten as the logarithm of a ratio.

\vspace{0.3cm}

\noindent
The quantities $\Delta_2$ and $\Delta_1^{\{i\}}$ and  $\widetilde{D}_{ij}$ are
expressed in terms of the various determinants and $\bar{b}$ coefficients:
\begin{eqnarray}
\widetilde{D}_{ij} + \Delta_1^{\{i\}} 
& = & 
\frac{\bbjsq{j}{i}}{\detgj{i}} 
\label{eqtruc1}\\
\Delta_2 - \Delta_1^{\{i\}} 
& = & 
\frac{\bbar_i^2}{\detg \, \detgj{i}} 
\label{eqtruc2}\\
- \, \Delta_1^{\{i\}} 
& = & 
\frac{\detsj{i}}{\detgj{i}}
\label{eqtruc3} \\
\Delta_2 & = & \frac{\dets}{\detg}
\label{eqtruc4}
\end{eqnarray}
To derive eqs.\ (\ref{eqtruc1}) and (\ref{eqtruc2}), the following identities are used:
\begin{align}
  \left( \bbj{k}{i} \right)^{2} &= - \detsj{i}  + \tD_{ik} \, \detgj{i} \label{magicid1} \\
  \bbar_i^2 &= \dets \, \detgj{i} + \detg \, \detsj{i} \label{magicid2}
\end{align}
They are particular cases of the 
so-called Jacobi identity for determinants \cite{yan}, 
of which various cases of interest  
for the issues discussed here were specified in appendix A.2 of 
\cite{Guillet:2013mta}.\\

\noindent
Using eqs. (\ref{eqtruc1}) - (\ref{eqtruc4}) we obtain:
\begin{align}
I_3^4 
& =
\sum_{i \in S_3} \, \sum_{j \in S_3 \setminus \{i\}} \, 
\bbar_i \, \bbj{j}{i} \, 
\int_0^{1} \frac{dz}{\detg \, \bbjsq{j}{i} \, z^2 + \bbar_i^2} \,
\notag\\
& \;\;\;\;\;\;\;\;\;\;\;\;\;\;\;\;\;\;
\times
\left[
 \ln 
 \left( 
   \frac{\bbjsq{j}{i} \, z^2  + \detsj{i}}{\detgj{i}}  - i \, \lambda
 \right)
 - 
 \ln \left( - \, \frac{\dets}{\detg} - i \, \lambda \right)
\right]
\label{eqI3414}
\end{align}
Eq. (\ref{eqI3414}) can be recast in the form (\ref{eqI346}) obtained 
according to the direct way. 
To achieve this goal, we first make the change of variable 
$s = z \, \bbj{j}{i}$, so that $I_3^4$ reads:
\begin{eqnarray}
I_3^4 
& = & 
\sum_{i \in S_3} \, \sum_{j \in S_3 \setminus \{i\}} \, \bbar_i  \, 
\int_0^{\bbj{j}{i}} \frac{d s}{s^2  \detg + \bbar_i^2} \,
\nonumber \\ 
& & 
\;\;\;\;\;\;\;\;\;\;\;\;\;\;\;\;
\times
\left[
 \ln 
 \left( 
   \frac{s^2  + \detsj{i}}{\detgj{i}}  - i \, \lambda
 \right)
 - 
 \ln \left( - \, \frac{\dets}{\detg} - i \, \lambda \right)
\right]
\label{eqI3415}
\end{eqnarray}

\noindent
In the real mass case, the $\bbj{j}{i}$ are real: the integration contours of 
the two integrals corresponding to the two values of 
$j \in S_3 \setminus \{i\}$ in eq. (\ref{eqI3415}) both run along the real 
axis. The poles in the integrand of eq. \myref{eqI3415}
are fake as their residues vanish by construction, and
the straight contours of integration in eq. \myref{eqI3415} 
do not cross the cuts\footnote{In anticipation, let us stress that the latter 
still holds in the complex mass case, although the justification of the second 
point requires a little more care.} of the logarithms. The two real-contour 
integrals corresponding to the two values of $j \in S_3 \setminus \{i\}$ can 
thus be joined end-to-end into a single one in a straightforward way. 
To refer to the two
elements of $S_3 \setminus \{i\}$ in a definite way, let us introduce 
$k \equiv 1 + ((i+1)$ modulo $3)$ and $l \equiv 1 + (i$ modulo $3)$.
Accordingly $I_3^4$ can be recast into:
\begin{eqnarray}
I_3^4 
& = & 
\sum_{i \in S_3} \, \bbar_i  \, \int_{-\bbj{k}{i}}^{\bbj{l}{i}} 
\frac{d s}{s^2  \detg + \bbar_i^2} \,
\nonumber\\
&&
\;\;\;\;\;\;\;\;\;\;\;\;
\times
\left[
 \ln 
 \left( 
   \frac{s^2  + \detsj{i}}{\detgj{i}}  - i \, \lambda
 \right)
 - 
 \ln \left( - \, \frac{\dets}{\detg} - i \, \lambda \right)
\right]
\label{eqI3416}
\end{eqnarray}
We then make the following change of variable: 
\begin{equation}\label{cvar1}
s= - \, \bbj{k}{i} + \left( \bbj{k}{i} + \bbj{l}{i}\right) \, u
= - \, \bbj{k}{i} - \detgj{i} \, u
\end{equation} 
such that $s=- \, \bbj{k}{i} \leftrightarrow u = 0$ and 
 $s=\bbj{l}{i} \leftrightarrow u = 1$. The numerator 
of the argument of the logarithm in eq. (\ref{eqI3416})
becomes:
\begin{equation}\label{newdef}
\frac{s^2 + \detsj{i}}{\detgj{i}}
=
\detgj{i} \, u^2 +  2 \, \bbj{k}{i} \, u + 
\frac{\bbjsq{k}{i} + \detsj{i}}{\detgj{i}} 
\end{equation}
Thanks to the identity (\ref{magicid1}),
we recognise
\begin{eqnarray*}
\frac{\bbjsq{k}{i}+\detsj{i}}{\detgj{i}} 
& = & 
\tD_{ik} \; = \; - \; C^{{\{i\}}(l)}
\end{eqnarray*}
We can further identify
\begin{eqnarray*}
\detgj{i}  
& = & 
\;\;\;\, G^{{\{i\}}(l)}\\
\bbj{k}{i} 
& = & 
- \;   V^{{\{i\}}(l)}\\
\end{eqnarray*}
so that the complete identification reads:
\begin{eqnarray}
\frac{s^2 + \detsj{i}}{\detgj{i}}
& = & 
D^{{\{i\}}(l)}(u)
\label{eqdefgi}
\end{eqnarray}
Furthermore, thanks to the identity (\ref{magicid2}),
we recast the denominator $[s^2  \detg + \bbar_i^2]$ in the integrand of eq. 
(\ref{eqI3416}) as:
\[
\detg \, \left( - \, \bbj{k}{i} - \detgj{i} \, u \right) ^2 + \bbar_i^2 
= 
\detgj{i} \, \left[  \detg \, D^{\{i\}(l)}(u) + \dets \right]
\]
Reminding that $B = \detg/\dets$ and $\bbar_i /\dets = b_i$, 
we finally get:
\begin{eqnarray}
I_3^4 
& = & 
- \sum_{i \in S_3} \, b_i  \, \int_{0}^{1} 
\frac{du}{B \, D^{\{i\}(l)} (u) + 1} 
\left[
 \ln \left(  D^{\{i\}(l)}(u) - i \, \lambda \right)
 -
 \ln \left( - \, \frac{1}{B} - i \, \lambda \right)
\right]
\nonumber
\end{eqnarray}
which is nothing but eq. \myref{eqI346}.

\subsection{Taming the proliferation of dilogarithms}\label{sss223} 

A comment is in order here regarding the dilogarithms generated by the
computation of eqs.\ (\ref{eqI346}) and (\ref{eqI3414}) in closed form,
performing the leftover integration after partial fraction decomposition.
In the present case the number of dilogarithms generated by 
eq. (\ref{eqI346}) seems quite non minimal, and by eq. (\ref{eqI3414}) 
even less so. The latter contains 6 terms (cf. summation over 
$i$ and $j$) versus 3 (summation over $i$ only) for both eq. (\ref{eqI346}) 
and the original calculation of 't Hooft and Veltman. Furthermore each term 
of eqs. (\ref{eqI346}) or (\ref{eqI3414}) yields 8 dilogarithms: 4 generated 
by each of the {\em two} poles i.e. the zeroes of the second degree polynomial 
$B \,  D^{\{i\}(l)}(x) + 1$, whereas the denominators in original 
calculation 
of 't Hooft and Veltman were first degree in $x$ only. This seemingly amounts 
to a twofold proliferation of dilogarithms (24 instead of 12) using eq. 
(\ref{eqI346}) and, even a fourfold proliferation (48 instead of 12) using eq. 
(\ref{eqI3414}), w.r.t. the original calculation of 't Hooft and Veltman. 
The situation is actually not so bad.

\vspace{0.3cm}

\noindent
We previously explained how pairs of terms in eq. (\ref{eqI3414}) can be
recombined so as to recover eq. (\ref{eqI346}).
Thus we are left only with the
relative twofold proliferation seemingly occurring  in eq. (\ref{eqI346}) w.r.t
to \cite{tHooft:1978jhc}. Let us show that this apparent doubling is actually 
fake. Let us rewrite eq. (\ref{eqI346}) 
as\footnote{To be definite in case $\detg > 0$, $\sqrt{- \, \detg}$ is 
understood to be $+i \, \sqrt{\detg}$. This choice actually does not matter 
as the two roots are exchanged under $x \leftrightarrow 1-x$.}
\begin{align}
I_3^4 
&= 
-\, \frac{1}{2 \, \sqrt{- \, \detg}} 
\notag\\
&
\;\;\;\;\;\;
\sum_{i \in S_3}
\int_{0}^{1} 
dx \, \left[ \frac{1}{x - x_{+}} - \frac{1}{x - x_{-}} \right] 
\left[ 
 \ln \left( D^{\{i\}(l)}(x) - i \, \lambda \right)
 -
 \ln \left( - \, \frac{1}{B} - i \, \lambda \right)
\right]
\label{eqapp4i340}
\end{align}
where
\[
  x_{\pm} = - \, \frac{1}{\detgj{i}} \, 
\left( \bbj{k}{i} \pm \frac{\bbar_i}{\sqrt{- \, \detg}} \right)
\]
are the roots of the denominator in eq. (\ref{eqI3416}) and 
the index $k$ is the element of $S_3 \setminus \{i,l\}$. 
Eq. (\ref{eqapp4i340}) involves 24 dilogarithms i.e. twice as many as the 
result of ref. \cite{tHooft:1978jhc}. To reduce their number let us first compute the 
following quantity:
\begin{align}
K 
& = 
\sum_{i \in S_3} \, 
\int_{0}^{1} 
dx \, \left[ \frac{1}{x - x_{+}} + \frac{1}{x - x_{-}} \right] 
\left[
 \ln \left( D^{\{i\}(l)}(x) - i \, \lambda \right)
 - 
\ln \left( - \, \frac{1}{B} - i \, \lambda \right)
\right] 
\label{eqapp4k0}
\end{align}
Combining the two terms in the
square bracket into a single denominator and using eq. (\ref{eqdefgi}) yields:
\begin{align}
K 
& = 
\sum_{i \in S_3} \, 
\int_{0}^{1} 
dx \, \frac{d \, D^{\{i\}(l)}(x)}{d x} \, 
\frac{1}{D^{\{i\}(l)}(x) + \frac{1}{B}} 
\notag\\
&\;\;\;\;\;\;\;\;\;\;\;\;\;\;\;\;\;\;\;\;\;\;\;\;\;\;\;\;\;\;\;\;\;\;\;\;
\;\;\;\;
\left[
 \ln \left( D^{\{i\}(l)}(x) - i \, \lambda \right)
 - 
 \ln \left( - \, \frac{1}{B} - i \, \lambda \right)
\right]  
\label{eqapp4k1}
\end{align}
We now make the change of variable $t = D^{\{i\}(l)}(x)$. 
The integrals over $[0,1]$ in eq. (\ref{eqapp4k1}) are traded for contour
integrals in the complex plane\footnote{In the real mass case, the integration
contours of the integrals in eq. (\ref{eqapp4k3}) still run along the real 
$t$-axis. In the complex mass case the contours more generally draw 
parabolic arcs in the complex $t$-plane.}. 
$K$ takes the form:
\begin{eqnarray}
K 
& = & 
\left\{ 
\int_{D^{\{1\}(2)}(0)}^{D^{\{1\}(2)}(1)}  dt 
 + 
\int_{D^{\{2\}(3)}(0)}^{D^{\{2\}(3)}(1)}  dt 
 + 
\int_{D^{\{3\}(1)}(0)}^{D^{\{3\}(1)}(1)}  dt 
\right\}
\, F(t) 
\label{eqapp4k3}
\end{eqnarray}
where
\begin{eqnarray}
F(t)
& = &
\frac{1}{t + \frac{1}{B}} 
\left[
 \ln \left(  t - i \, \lambda \right)
 - 
 \ln \left(- \, \frac{1}{B} - i \, \lambda \right)
\right]
\nonumber
\end{eqnarray}
Note that $D^{\{i\}(l)}(0)$ and $D^{\{i\}(l)}(1)$ are 
internal masses, and from the correspondence (\ref{Dtilde2}) in
subsubsec. \ref{subsubsectindway} we see that the three integrals in 
eq. (\ref{eqapp4k3}) combine into an integral along a closed 
contour (triangle).
Note also that the pole in $F(t)$ is fake the residue is
zero by construction thus we do not have to worry about the positions of the 
pole with respect to the triangle. In addition, 
in the case of complex masses, this triangle lies entirely below the real axis 
in the complex $t$ plane: in no case it crosses the real axis, consequently
there is no worry with the discontinuity cuts of the various logarithms.
Thus $K=0$. To summarise, defining
\begin{align}
I_i^{\pm} 
& = 
- \frac{1}{2 \, \sqrt{- \, \detg}}  \, \int_{0}^{1} \frac{dx}{x - x_{\pm}} 
\left[ 
 \ln \left( D^{\{i\}(l)}(x) - i \, \lambda \right) 
 - 
 \ln \left( - \, \frac{1}{B} - i \, \lambda \right)
\right]
\label{eqdefipm0}
\end{align}
``$K=0$" reads:
\begin{equation}
I_1^{+} + I_2^{+} + I_3^{+} = - \, \left( I_1^{-} + I_2^{-} + I_3^{-} \right)
\label{eqsumipm1}
\end{equation}
and these two mutually opposite sums of three terms happen to be
subtracted one from another to yield $I_3^4$ in eq. (\ref{eqI346}), so that
the three-point function is given by:
\begin{eqnarray}
I_3^4 
& = & 
\;\; (I_1^{+} + I_2^{+} + I_3^{+}) - (I_1^{-} + I_2^{-} + I_3^{-}) 
\nonumber \\
& = & 
2 \, (I_1^{+} + I_2^{+} + I_3^{+})
\label{eqapp4i34fin}
\end{eqnarray}
which involves 4 dilogarithms in each of the three terms $I_1^{+}$, $I_2^{+}$ 
and $I_3^{+}$, hence 12 distinct dilogarithms only, not 24 as hastily thought.
In some intermediate step, the authors of \cite{tHooft:1978jhc} made use of some change of 
variable involving either of the two roots $\alpha_{\pm}$ of some second degree 
polynomial. Both changes of variables were usable and lead to namely 
one combination of three $I_i^{\pm}$ and its complementary with respect to eq. 
(\ref{eqsumipm1}), respectively\footnote{We do not specify which combination namely 
leads to the result of \cite{tHooft:1978jhc} here because it depends on the assignment adopted 
for the superscript $l$. The exercise is left to the masochistic reader!}.
The two combinations, albeit
seemingly distinct analytically, were guaranteed to take equal values in the 
approach of \cite{tHooft:1978jhc} since the three-point function does 
not depend on the choice of the change of variable made.  
Instead, our approach does not rely on any such choice triggering a breakdown 
of symmetry between two options. It thus yields a
result which preserves this symmetry, being half the sum of the two 
expressions corresponding to $\alpha_{+}$ and to $\alpha_{-}$ in \cite{tHooft:1978jhc},
yet at the expense of doubling the number of terms. 
The use of identity (\ref{eqsumipm1}) to counteract the doubling 
is somehow a counterpart of the choice of $\alpha_{+}$ vs. 
$\alpha_{-}$ in \cite{tHooft:1978jhc}. 

\vspace{0.3cm}

\noindent
Anticipating the absence of a direct way in the four-point function case, 
one may wonder whether the fourfold proliferation in eq. (\ref{eqI3414}) 
compared to ref. \cite{tHooft:1978jhc} can be counteracted directly instead of 
recovering eq. (\ref{eqI346}) first and then acting as explained just above.
In this respect it shall be stressed that the actual number of dilogarithms
generated by eq. (\ref{eqI3414}) is twice smaller than naively 
counted because the integrand is even in $x$. Indeed, each of 
the 6 terms in eq. (\ref{eqI3414}), of the form 
\[
T = \int_{0}^{1} dx \, \frac{\ln \left(P_{1}(x) \right) }{P_{2}(x)}
\]
where both $P_{j}(x)$ are even second degree trinomials in $x$, can be
rewritten
\[
T = \frac{1}{2} 
\int_{-1}^{1} dx \, \frac{\ln \left(P_{1}(x) \right) }{P_{2}(x)}
\]
Let us note $\pm x_{2}$ the two mutually opposite roots of $P_{2}(x)$. 
Partial fraction decomposition of $1/P_{2}(x)$ leads to an integral of the form 
\begin{eqnarray}
T 
& \propto & 
\int_{-1}^{1} dx \, 
\frac{\ln \left(P_{1}(x) \right) }{x-x_{2}} - \int_{-1}^{1} dx \, 
\frac{\ln \left(P_{1}(x) \right) }{x+x_{2}}
\nonumber\\
& = &
2 \, \int_{-1}^{1} dx \, 
\frac{\ln \left(P_{1}(x) \right) }{x-x_{2}}
\nonumber
\end{eqnarray}
which yields only 4 dilogarithms instead of 8: eq. (\ref{eqI3414}) thus
yields $6 \times 4 = 24$ dilogarithms ``only". 
This fortunate feature of parity in the integration variable and its consequence will also be 
encountered in the case of the four-point function. 
One may wonder whether there is a straight way to reduce the number of 
dilogarithms, bypassing the step that reshuffles eq. (\ref{eqI3414}) into eq. 
(\ref{eqI346}), which would instead exploit
the parity in $z$ of the integrand in eq. (\ref{eqI3414}) combined with some 
other trick. For the three-point function, an independent derivation of 
identity (\ref{eqsumipm1}) alternative to the one above might be worked out 
directly from eq. (\ref{eqI3414}) 
computed in closed form, using a version of the five-dilogarithm Hill identity 
\cite{lewin} supplemented by a couple of cancellations thereby made manifest
\cite{future}.

%\section{Leg up: $I_4^4$}\label{sectfourpoint}
\section{Leg up: the scalar four-point function $I_4^4$}\label{sectfourpoint}

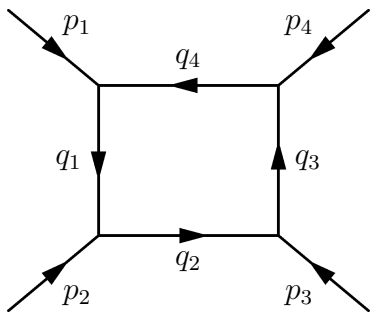
\begin{figure}[h]
\centering
\parbox[c][43mm][t]{40mm}{\begin{fmfgraph*}(60,40)
  \fmfleftn{i}{2} \fmfrightn{o}{2}
  \fmf{fermion,label=$p_1$}{i2,v1}
  \fmf{fermion,label=$p_2$}{i1,v2}
  \fmf{fermion,label=$p_3$}{o1,v3}
  \fmf{fermion,label=$p_4$}{o2,v4}
  \fmf{fermion,tension=0.5,label=$q_1$}{v1,v2}
  \fmf{fermion,tension=0.5,label=$q_2$}{v2,v3}
  \fmf{fermion,tension=0.5,label=$q_3$}{v3,v4}
  \fmf{fermion,tension=0.5,label=$q_4$}{v4,v1}
\end{fmfgraph*}}
\caption{The box picturing the one-loop four-point function.}
\label{fig2} 
\end{figure}

\noindent
The usual integral representation of $I_4^4$ in terms of Feynman parameters is 
given by:
\begin{eqnarray}
I_4^4 & = & \int_0^1 \, \prod_{i=1}^4 \, dz_{i} \, \delta(1- \sum_{i=1}^4 z_i) 
\left( 
 - \, \frac{1}{2} \, Z^{\;T} \cdot \cals \cdot Z - i \, \lambda 
\right)^{-2}
\label{eqSTARTINGPOINT}
\end{eqnarray}
where $Z$ is now a column 4-vector whose components are the $z_{i}$.
In a way similar to sec. \ref{sectthreepoint} we arbitrarily single out the
subscript value $a$ ($a \in S_{4}=\{1,2,3,4\}$), and write $z_a$ as 
$1 - \sum_{j \neq a} z_j$. We find:
\begin{eqnarray}
- \, Z^{\;T} \cdot \cals \cdot Z 
& = & 
\sum_{i,j \in S_4 \setminus \{a\}} G_{i\,j}^{(a)} \, z_i \, z_j  - 
2 \sum_{j \in S_4 \setminus \{a\}} V_j^{(a)} \, z_j  -  C^{(a)}
\label{eqVECZT4}
\end{eqnarray}
where the $3 \times 3$ Gram matrix $G^{(a)}$ and the column 3-vector $V^{(a)}$
are defined by
\begin{eqnarray}
G_{i\,j}^{(a)} 
& = & 
-  \, (\cals_{i\,j}-\cals_{a\,j}-\cals_{i\,a}+\cals_{a\,a}), \;\; i,j \neq a
\nonumber
\\
V_j^{(a)} 
& = & 
\;\;\;\;\;  \cals_{a\,j} - \cals_{a \, a}, \;\;  j \neq a
\label{eqVJA4}\\
C^{(a)}& = & 
\;\;\;\;\;  \cals_{a \, a} 
\nonumber
\end{eqnarray}
We label $b$, $c$ and $d$ the three elements of $S_4 \setminus \{a\}$ with 
$b < c < d$. The polynomial (\ref{eqVECZT4}) reads:
\begin{eqnarray}
  D^{(a)}(X) 
= 
X^{\;T} \cdot G^{(a)} \cdot X - 2 \, V^{(a) \, T} \cdot X - C^{(a)}
& , &  X =
 \left[
 \begin{array}{c}
   z_b \\
   z_c \\
   z_d
 \end{array}
 \right]
  \label{eqDEFD}
\end{eqnarray}
We note the integration simplex 
$\Sigma_{bcd} = \{0 \leq z_{b},z_c,z_d,z_{b}+z_{c}+z_{d} \leq 1\}$. 
$I_4^4$ can be written
%as
\begin{eqnarray}
I_4^4 
& = & 
4 \int_{\Sigma_{bcd}} 
\frac{dz_b \, d z_c \, d z_d}{(D^{(a)}(X)-i \, \lambda)^2}
\label{eqI441}
\end{eqnarray}
Again the dependence on $G^{(a)}$, $V^{(a)}$ and $C^{(a)}$ will arise through 
quantities independent of the actual choice of $a$. 
We will follow a similar strategy as for the three-point function in section 
\ref{sectthreepoint} {\em mutatis mutandis}.

\subsection{Step 1}\label{fourpointstep1}

The idea is again to adjust the power of the denominator in the l.h.s of eq. 
(\ref{eqDEFREL1}) in such way that only the boundary term in eq. 
(\ref{eqDEFREL1}) remains. In the four-point function case at hand, cf. eq. 
(\ref{eqI441}) $\tn = N-1$ is equal to 3. Imposing $\tn \, - \, 2 \alpha = 0$ 
implies $\alpha= 3/2$, hence a power $\alpha+1=5/2$ in the l.h.s. 
of eq. (\ref{eqDEFREL1}). In eq. (\ref{eqI441}) however  $D$ is raised to the 
power 2, not 5/2.
Identity (\ref{eqFOND1biss}) is used with $\mu=5/2$ and $\nu=2$ 
chosen such that $\mu - 1/\nu = 2$ and $2(\mu-1) = 3$ to customise a 
denominator raised to a power shifted from 2 to 5/2.
The representation of $1/D^{\,2}$ obtained is substituted into eq. 
(\ref{eqI441}). 
Identity (\ref{eqDEFREL1}) is then applied to the new integrand seen as a 
function of the three variables $z_{b}$, $z_{c}$, $z_{d}$. Integration will be
performed on the latter first, keeping $\xi$ fixed. 
This yields:
\begin{equation}
I_4^4 
=
\frac{8}{B(2,1/2)} \, \int_0^{+\infty} d \xi 
\int_{\Sigma_{bcd}}  
\frac{dz_b \, d z_c \, d z_d}{(D^{(a)}(X)+\xi^2-i \, \lambda)^{5/2}}
\label{eqI442}
\end{equation}
Identity (\ref{eqDEFREL1}) recasts eq. (\ref{eqI442}) into:
\begin{align}
I_4^4 
& =
- \, \frac{8}{3 \, B(2,1/2)} \, 
\int_0^{+\infty} \frac{d \xi}{\Delta_3 - \, \xi^2 + i \, \lambda} 
\notag\\
&\;\;\;\;\;\;\;\;\;\;\;\;\;\;\;\;\;\;\;\;\;\;
\times
\int_{\Sigma_{bcd}} dz_b \, d z_c \, d z_d
\left[
 \sum_{j \in S_{4} \setminus \{a\}} 
 \frac{\partial}{\partial z_{j}} \, 
  \left( 
   \frac{z_j - ((G^{(a)})^{-1} \cdot V^{(a)})_{j}}
   {(D^{(a)}(X) + \xi^2 - i \, \lambda)^{3/2}} 
   \right) 
\right]
 \label{eqI444}
\end{align}
 with $\Delta_3 = V^{(a) \, T} \cdot (G^{(a)})^{-1} \cdot V^{(a)} + C^{(a)}$.
For each of the three terms $j = b,c,d$ in eq. (\ref{eqI444}) this makes 
the first integration over the corresponding $z_{j}$ trivial, and we get:
\begin{eqnarray}
I_4^4 
& = & - \, \frac{8}{3 \, B(2,1/2)} \, 
\int^{+\infty}_0 \frac{d \xi}{\Delta_3 - \, \xi^2 + i \, \lambda} 
\sum_{j \in S_{4} \setminus \{a\}} {\cal T}_{j}
\label{eqI445-a}
\end{eqnarray}
with 
\begin{eqnarray}
{\cal T}_{b}
& = &
\int_{\Sigma_{cd}} dz_c \, dz_d \,
\left[ 
 \frac{(1-z_c-z_d) - ((G^{(a)})^{-1} \cdot V^{(a)})_{b}}
 {( D^{(a)}(1-z_c-z_d, \, z_c, \, z_d) + \xi^2 - i \, \lambda)^{3/2}} 
\right.
\nonumber \\
& & 
\;\;\;\;\;\;\;\;\;\;\;\;\;\;\;\;\;\;\;\;\;\;\;\;\;\;\;\;\;\;\;
\, + \, 
\left. 
 \frac{((G^{(a)})^{-1} \cdot V^{(a)})_{b}}
{( D^{(a)}(0, \, z_c, \, z_d) + \xi^2 - i \, \lambda)^{3/2}} 
\right] 
\nonumber\\
{\cal T}_{c}
& = &
\int_{\Sigma_{bd}} dz_b \, dz_d \,
\left[ 
 \frac{(1-z_b-z_d) - ((G^{(a)})^{-1} \cdot V^{(a)})_{c}}
 {( D^{(a)}(z_b, \, 1-z_b-z_d, \, z_d) + \xi^2 - i \, \lambda)^{3/2}} 
\right. 
\nonumber \\
& & 
\;\;\;\;\;\;\;\;\;\;\;\;\;\;\;\;\;\;\;\;\;\;\;\;\;\;\;\;\;\;\;
\, + \, 
\left. 
 \frac{((G^{(a)})^{-1} \cdot V^{(a)})_{c}}
{(D^{(a)}(z_b, \, 0, \, z_d) + \xi^2 - i \, \lambda)^{3/2}} 
\right] 
\label{eqI445-c}  \\
{\cal T}_{d}
& = & 
\int_{\Sigma_{bc}} dz_b \, dz_c \,
\left[ 
 \frac{(1-z_b-z_c) - ((G^{(a)})^{-1} \cdot V^{(a)})_{d}}
 {(D^{(a)}(z_b, \, z_c, \, 1-z_b-z_c) + \xi^2 - i \, \lambda)^{3/2}} 
\right. 
\nonumber \\
& & 
\;\;\;\;\;\;\;\;\;\;\;\;\;\;\;\;\;\;\;\;\;\;\;\;\;\;\;\;\;\;\;
\, + \, 
\left. 
 \frac{((G^{(a)})^{-1} \cdot V^{(a)})_{d}}
 {( D^{(a)}(z_b, \, z_c, \, 0) + \xi^2 - i \, \lambda)^{3/2}} 
\right]
\nonumber
\end{eqnarray}
By an appropriate relabelling of the variables, all three terms ${\cal T}_{j}$
in eq. (\ref{eqI445-a}) can be gathered in one integral:
\begin{align}
I_4^4 
& =  
- \, \frac{8}{3 \, B(2,1/2)} \, \int^{+\infty}_0 
\frac{d \xi}{\Delta_3 - \, \xi^2 + i \, \lambda} \,
 \int_{\Sigma_{12}} dx_1 \, d x_2 
\notag \\
&  
\times 
\Biggl\{ 
  \frac{(1-x_1-x_2) - ((G^{(a)})^{-1} \cdot V^{(a)})_{b}}
  {(D^{(a)}(1-x_1-x_2, \, x_1, \, x_2) + \xi^2 - i \lambda)^{3/2}} 
  + 
  \frac{((G^{(a)})^{-1} \cdot V^{(a)})_{b}}
  {(D^{(a)}(0, \, x_1, \, x_2) + \xi^2 - i \lambda)^{3/2}} 
\notag \\
&  
\;\;\;\; +  
  \frac{(1-x_1-x_2) - ((G^{(a)})^{-1} \cdot V^{(a)})_{c}}
  {(D^{(a)}(x_1, \, 1-x_1-x_2, \, x_2) + \xi^2 - i \lambda)^{3/2}} 
  + 
  \frac{((G^{(a)})^{-1} \cdot V^{(a)})_{c}}
  {(D^{(a)}(x_1, \, 0, \, x_2) + \xi^2 - i \lambda)^{3/2}} 
\notag\\
& 
\;\;\;\; +  
  \frac{(1-x_1-x_2) - ((G^{(a)})^{-1} \cdot V^{(a)})_{d}}
  {(D^{(a)}(x_1, \, x_2, \, 1-x_1-x_2) + \xi^2 - i \lambda)^{3/2}} 
  + 
  \frac{((G^{(a)})^{-1} \cdot V^{(a)})_{d}}
  {(D^{(a)}(x_1, \, x_2, \, 0) + \xi^2 - i \lambda)^{3/2}} 
\Biggr\} 
\label{eqI446} 
\end{align}
The number of terms in eq. (\ref{eqI446}) can be reduced. 
In the r.h.s. of eq. (\ref{eqI446}) let us consider the first term 
of the second line of the curly bracket:
\begin{equation}
 T_{c} \equiv 
\int_{\Sigma_{12}} d x_1 \, d x_2 \, 
%\int_0^1 dx_1 \, \int^{1-x_1}_0 d x_2 \, 
\frac{(1-x_1-x_2) - ((G^{(a)})^{-1} \cdot V^{(a)})_{c}}
{(D^{(a)}(x_1, \, 1-x_1-x_2,\, x_2) + \xi^2 - i \, \lambda)^{3/2}}
 \label{eqFIRSTTERM}
\end{equation}
and perform the following change of variables:
\[
  \left\{
  \begin{array}{ccl}
    y_1 & = & 1 - x_1 - x_2 \\
    y_2 & = & x_2
  \end{array}
  \right.
  \Leftrightarrow
  \left\{
  \begin{array}{ccl}
    x_1 & = & 1-y_1-y_2 \\
    x_2 & = & y_2
  \end{array}
  \right.
\]
The new variables $y_1$ and $y_2$ still span the simplex $\Sigma_{12}$.
Eq. (\ref{eqFIRSTTERM}) becomes :
\begin{eqnarray}
T_{c} 
& = & 
\int_{\Sigma_{12}} d y_1 \, d y_2 \, 
\frac{y_1 - ((G^{(a)})^{-1} \cdot V^{(a)})_c}
{(D^{(a)}(1-y_1-y_2,y_1,y_2)+\xi^2 - i \, \lambda)^{3/2}} 
 \label{tc}
\end{eqnarray}
In a similar way, using $t_1 = x_2$ and $t_2 = 1- x_1-x_2$ we recast the 
first term of the third line of the curly bracket in the r.h.s. of eq. 
(\ref{eqI446}) into:
\begin{eqnarray}
T_{d}
& \equiv &
\int_{\Sigma_{12}} dx_1 \, d x_2 \,
\frac{(1-x_1-x_2) - ((G^{(a)})^{-1} \cdot V^{(a)})_d}
{(D^{(a)}(x_1, \, x_2, \, 1-x_1-x_2) + \xi^2 - i \, \lambda)^{3/2}}
\nonumber \\
& = & 
\int_{\Sigma_{12}} dt_1 \, d t_2 \,
\frac{t_2 - ((G^{(a)})^{-1} \cdot V^{(a)})_d}
{(D^{(a)}(1-t_1-t_2,t_1,t_2)+\xi^2 - i \, \lambda)^{3/2}}
\label{eqSECTERM}
\end{eqnarray}
We then recombine $T_c$ and $T_d$ with the first term of the first line of 
the curly bracket in the r.h.s. of eq. (\ref{eqI446}), and $I_4^4$ reads:
\begin{eqnarray}
I_4^4 
& = & 
- \, \frac{8}{3 \, B(2,1/2)} \, \int^{+\infty}_0 
\frac{d \xi}{\Delta_3- \, \xi^2 + i \, \lambda} \, 
\int_{\Sigma_{12}} dx_1 \, d x_2 
\nonumber \\
& & 
\mbox{} \times 
\Biggl\{  
 \frac{1 - \sum_{j \in S_{4} \setminus \{a\}}((G^{(a)})^{-1} \cdot V^{(a)})_j}
 {(D^{(a)}(1-x_1-x_1, \, x_1, \, x_2) + \xi^2 - i \, \lambda)^{3/2}} 
 +
 \frac{((G^{(a)})^{-1} \cdot V^{(a)})_b}
 {(D^{(a)}(0, \, x_1, \, x_2) + \xi^2 - i \, \lambda)^{3/2}} 
\nonumber \\
& & 
\mbox{}  
\;\;\;\;
 \, + \, 
 \frac{((G^{(a)})^{-1} \cdot V^{(a)})_c}
 {(D^{(a)}(x_1, \, 0, \, x_2) + \xi^2 - i \, \lambda)^{3/2}} 
 + 
 \frac{((G^{(a)})^{-1} \cdot V^{(a)})_d}
 {(D^{(a)}(x_1, \, x_2, \, 0) + \xi^2 - i \, \lambda)^{3/2}} 
\Biggr\}
\label{eqI447}
\end{eqnarray}
The quantities $\Delta_3$ and $((G^{(a)})^{-1} \cdot V^{(a)})_j$ involved in 
eq. (\ref{eqI447}) are expressed in terms of $\dets$, $\detg$ and the 
$\bbar_{j} = b_{j} \, \dets$ (see eqs.~(\ref{eqe10-02}) and (\ref{eqsolbbar}) of appendix~\ref{detsdetg}):
\begin{eqnarray}
\Delta_3 
& = & 
- \, \frac{\dets}{\detg} \notag \\
((G^{(a)})^{-1} \cdot V^{(a)})_{j} 
& = & 
- \, \frac{\bbar_j}{\detg}, \;\; j \in S_{4} \setminus \{a\} \label{eqdefdelta3} \\
 1 - \sum_{j \in S_{4} \setminus \{a\}}((G^{(a)})^{-1} \cdot V^{(a)})_{j}
& = & 
- \, \frac{\bbar_a}{\detg} \notag
\end{eqnarray}
The four new polynomials of $x_{1},x_{2}$ appearing in the denominators
coincide with those involved in the integral representations of 
the triangle diagrams obtained by all possible pinchings of the box 
diagram of fig.~\ref{fig2}. Namely,
$D^{(a)}(0,x_1,x_2) = D^{\{b\}(a)}(x_1,x_2)$ comes out when pinching propagator 
$b$ in the box, which corresponds to suppressing line and column $b$ in the 
kinematic matrix $\cals$ in which line and column $a$ are singled out:
\begin{eqnarray}
D^{(a)}(0,x_1,x_2) 
& = &
\tX^{\,T} \cdot G^{\{b\}(a)} \cdot \tX \, - \,
2 \, V^{\{b\}(a) \; T} \cdot \tX \, - \, C^{\{b\}(a)} 
\nonumber\\
& = &
D^{\{b\}(a)}(x_1,x_2)
\;\; , \;\;
  \tX =
 \left[
 \begin{array}{c}
   x_1 \\
   x_2 \\
 \end{array}
 \right]
\label{pinched-a}
\end{eqnarray}
Likewise for $c,d,a$: 
\begin{eqnarray}
D^{(a)}(x_1,0,x_2)
& = &
\tX^{\,T} \cdot G^{\{c\}(a)} \cdot \tX \, - \,
2 \, V^{\{c\}(a) \; T} \cdot \tX \, - \, C^{\{c\}(a)}
\nonumber\\
& = &
D^{\{c\}(a)}(x_1,x_2)
\nonumber\\
D^{(a)}(x_1,x_2,0)
& = &
\tX^{\,T} \cdot G^{\{d\}(a)} \cdot \tX \, - \,
2 \, V^{\{d\}(a) \; T} \cdot \tX \, - \, C^{\{d\}(a)}
\nonumber\\
& = &
D^{\{d\}(a)}(x_1,x_2)
\nonumber
\end{eqnarray}
whereas a simple calculation yields:
\begin{eqnarray}
D^{(a)}(1-x_1-x_2,x_1,x_2)
& = &
\tX^{\,T} \cdot G^{\{a\}(b)} \cdot \tX \; - \;
2 \, V^{\{a\}(b) \; T} \cdot \tX \; - \; C^{\{a\}(b)}
\nonumber\\
& = &
D^{\{a\}(b)}(x_1,x_2)
\nonumber
\end{eqnarray}
$I_4^4$  thus reads:
\begin{align}
I_4^4 
& = 
\frac{8}{3 \, B(2,1/2)} \, \sum_{i=1}^{4} \, \frac{\bbar_i}{\detg} \, 
\int^{+\infty}_0 
\frac{d \xi}{\Delta_3-\, \xi^2 + i \, \lambda}
\int_{\Sigma_{kl}}
\frac{dx_k \,d x_l}
{(D^{\{i\}(i^{\prime})}(x_k,x_l) + \xi^2 - i \, \lambda)^{3/2}}
  \label{eqI448}
\end{align}
where, for definiteness, the $i^{\, \prime}\,$ assignment for each $i$ is the one
read on the four eqs.\ (\ref{pinched-a}) above: 
$i^{\, \prime} = a$ for $i = b, \, c, \, d$ and $i^{\, \prime} = b$ for $i = a$.
In a similar way as with 
the three-point function case, this 
$i^{\, \prime}\,$ assignment can actually be changed relying on 
identities such as $D^{\{d\}(a)}(x_1,x_2)= D^{\{d\}(b)}(1-x_1-x_2,x_2)= 
D^{\{d\}(c)}(x_1,1-x_1-x_2)$ etc. combined with corresponding appropriate 
changes of variables that leave the integration simplex unchanged.

\subsection{Step 2}\label{fourpointstep2}

We proceed further as we did for the three-point function following the 
``indirect way", iterating the procedure to integrate over one more $x_i$
explicitly. Let us define the quantities $J_i$ by:
\begin{equation}
J_i 
= 
\int_{\Sigma_{kl}}
\frac{dx_k \,d x_l}
{(D^{\{i\}(i^{\prime})}(x_k,x_l) + \xi^2 - i \, \lambda)^{3/2}}
\label{eqDEFJ1}
\end{equation}
Here again we do not apply identity (\ref{eqDEFREL1}) directly because the 
power of the numerator is not the appropriate one.
We now deal with $\tn=2$ variables of integration,
$\alpha$ shall thus be equal to 1 in order to keep the boundary term only, 
whereas the exponent of the denominator in eq. (\ref{eqDEFJ1}) happens to be 
3/2, not $\alpha+1=2$.
To get the appropriate power, we again make use of eq. (\ref{eqFOND1biss}) 
with $\mu = 2$, $\nu = 2$ and $J_i$ is represented by:
\begin{equation}
  J_i = \frac{2}{B(3/2,1/2)} \, \int^{+\infty}_0 d \rho \, 
  \int_{\Sigma_{kl}} 
  \frac{dx_k \, dx_l}
  {(D^{\{i\}(i^{\prime})}(x_k,x_l) + \xi^2 + \rho^2 - i \, \lambda)^{2}}
  \label{eqDEFJ2}
\end{equation}
We use identity (\ref{eqDEFREL1}) in which ``$D$" is interpreted as 
$(D^{\{i\}(i^{\prime})} + \xi^2 + \rho^2)$. We note 
$k,l \in S_{4} \setminus \{i,i^{\prime}\}$ with $k < l$. We get:
\begin{eqnarray}
J_i 
& = & 
- \, \frac{1}{B(3/2,1/2)} \,  \int_0^{+\infty} \, 
\frac{d \rho}{\Delta_{2}^{\{i\}}- \, \xi^2 - \, \rho^2 + i \, \lambda} \, 
\int_{\Sigma_{kl}} dx_k \, d x_l
\nonumber \\
& & 
\mbox{} 
\;\;\;\;\;\;\;\;\;\;\;\;\;\;\;\;\;\;\;
\times 
\Biggl[ \,
 \sum_{j \in S_{4} \setminus \{i,i^{\prime}\}} 
  \frac{\partial}{\partial \, x_j}  
  \left( 
  \frac{x_j - ((G^{\{i\}(i^{\prime})})^{-1} \cdot V^{\{i\}(i^{\prime})})_j}
  {(D^{\{i\}(i^{\prime})}(x_k,x_l) + \xi^2 + \rho^2 - i \, \lambda)} 
 \right) 
\Biggr]
\label{eqDEFJ3}
\end{eqnarray}
with 
$\Delta_{2}^{\{i\}} = 
V^{\{i\}(i^{\prime}) \; T} \cdot (G^{\{i\}(i^{\prime})})^{-1} \cdot 
V^{\{i\}(i^{\prime})} + \, C^{\{i\}(i^{\prime})}$. In each term ``$j$" 
the integration performed over $x_j$ first is trivialised as a boundary term 
and we get for $J_i$:
\begin{eqnarray}
J_i 
& = & 
- \, \frac{1}{B(3/2,1/2)} \,  \int_0^{+\infty}  \, 
\frac{d \rho}{\Delta_{2}^{\{i\}}- \, \xi^2 - \, \rho^2 + i \, \lambda} 
\nonumber \\
& & 
\mbox{} \times 
\left[ 
 \int^1_0 d x_l 
 \left( 
  \frac{(1-x_l) - 
  ( (G^{\{i\}(i^{\prime})})^{-1} \cdot V^{\{i\}(i^{\prime})} )_k}
  {(D^{\{i\}(i^{\prime})}(1-x_l, \, x_l) + \xi^2 + \rho^2 - i \, \lambda)} 
  +
\frac{( (G^{\{i\}(i^{\prime})})^{-1} \cdot V^{\{i\}(i^{\prime})})_k}
  {(D^{\{i\}(i^{\prime})}(0, \,x_l) + \xi^2 + \rho^2 - i \, \lambda)} 
 \right) 
\right.
\nonumber \\
& & 
\mbox{} + 
 \int^1_0 d x_k 
\left. 
 \left( 
  \frac{(1-x_k) - 
  ( (G^{\{i\}(i^{\prime})})^{-1} \cdot V^{\{i\}(i^{\prime})} )_l}
  {(D^{\{i\}(i^{\prime})}(x_k, \, 1-x_k) + \xi^2 + \rho^2 - i \, \lambda)} 
  + 
\frac{( (G^{\{i\}(i^{\prime})})^{-1} \cdot V^{\{i\}(i^{\prime})})_l}
  {(D^{\{i\}(i^{\prime})}(x_k, \, 0) + \xi^2 + \rho^2 - i \, \lambda)} 
 \right) 
\right]
\nonumber\\
\label{eqDEFJ4}
\end{eqnarray}
Making the change of variable $x_k=x$ and $x_l = 1-x$ in the first and third terms inside the square bracket 
of eq. (\ref{eqDEFJ4}) and gathering the terms in the same integral, we get:
\begin{align}
J_i 
& =  
- \, \frac{1}{B(3/2,1/2)} \,  \int_0^{+\infty}  
\frac{d \rho}{\Delta_{2}^{\{i\}}-  \, \xi^2 - \, \rho^2 + i \, \lambda} 
\notag \\
& 
\mbox{}\;\;\;\;\;\;\;\;\;\;\;\;\;\;\;\;\;\;\;\;\;\;\;\;
\times 
\int^1_0 d x 
\left[ 
 \frac{( (G^{\{i\}(i^{\prime})})^{-1} \cdot V^{\{i\}(i^{\prime})} )_k}
{(D^{\{i\}(i^{\prime})}(0,x) + \xi^2 + \rho^2)} 
 + 
 \frac{( (G^{\{i\}(i^{\prime})})^{-1} \cdot V^{\{i\}(i^{\prime})} )_l}
 {(D^{\{i\}(i^{\prime})}(x,0) + \xi^2 + \rho^2)} 
\right.
\notag \\
& 
\mbox{} 
\;\;\;\;\;\;\;\;\;\;\;\;\;\;\;\;\;\;\;\;\;\;\;\;\;\;\;\;\;\;\;\; 
\;\;\;\;\;\;\;\;
\left.
+ \frac{1 - 
  \sum_{j \in S_{4} \setminus \{i,i^{\prime}\}} 
  ( (G^{\{i\}(i^{\prime})})^{-1} \cdot V^{\{i\}(i^{\prime})})_j}
 {(D^{\{i\}(i^{\prime})}(x,1-x) + \xi^2 + \rho^2 - i \, \lambda)} 
\right]
\label{eqDEFJ5}
\end{align}
We recognise 
\begin{eqnarray}
\Delta_{2}^{\{i\}} 
& = & 
\frac{\det \, ({\cal S}^{\{i\}})}{\det \,(G^{\{i\}})} 
\nonumber\\
((G^{\{i\}(i^{\prime})})^{-1} \cdot V^{\{i\}(i^{\prime})})_{j} 
& = & 
\frac{\bbar_j^{\{i\}}}{\det \,(G^{\{i\}})}, \;\; 
j \in S_{4} \setminus \{i,i^{\prime}\} 
\label{eqdefdelta2j} \\
1 - 
\sum_{j \in S_{4} \setminus \{i,i^{\prime}\}}
((G^{\{i\}(i^{\prime})})^{-1} \cdot V^{\{i\}(i^{\prime})})_{j}
& = & 
\frac{\bbar_{i^{\prime}}^{\{i\}}}{\det \,(G^{\{i\}})} 
\nonumber
\end{eqnarray}
We recall that the coefficients $\bar{b}_{j}^{\{i\}}$ are equal to 
$b_{j}^{\{i\}}/\detsj{i}$ with the $b_{j}^{\{i\}}$ such that\\* 
$\sum_{l \in S_{4} \setminus \{i\}} \cals^{\{i\}}_{k \, l} b^{\{i\}}_{l} = 1$. 
In denominators, the polynomials $D^{\{i\}(i^{\prime})}(0,1-x)$, 
$D^{\{i\}(i^{\prime})}(x,0)$ and $D^{\{i\}(i^{\prime})}(x,1-x)$ also have
a simple interpretation. We collect the dictionary:
\begin{equation}
\begin{array}{llll}
D^{(a)}(0, \, 0, \, x) 
& = \,
D^{\{b\}(a)}(0, \, x) 
& = \, 
D^{\{c\}(a)}(0, \, x)
& = \,
D^{\{b,c\}(a)}(x)
\\
D^{(a)}(0, \, x, \, 0  ) 
& = \,
D^{\{b\}(a)}(x, \, 0)   
& = \,
D^{\{d\}(a)}(0, \, x) 
& =
D^{\{b,d\}(a)}(x) 
\\
D^{(a)}(x, \, 0, \, 0  )  
& = \, 
D^{\{c\}(a)}(x, \, 0)
& = \,
D^{\{d\}(a)}(x, \, 0)   
& = \,
D^{\{c,d\}(a)}(x)
\\
D^{(a)}(0, \, x, \, 1-x) 
& = \,
D^{\{b\}(a)}(x, \, 1-x) 
& = \, 
D^{\{a\}(b)}(x, \, 1-x) 
& = \,
D^{\{a,b\}(d)}(x)
\\
D^{(a)}(x, \, 0, \, 1-x) 
& = \, 
D^{\{c\}(a)}(x, \, 1-x) 
& = \,  
D^{\{a\}(b)}(0, \, 1-x) 
& = \,
D^{\{a,c\}(d)}(x) 
\\
D^{(a)}(x, \,1-x, \, 0) 
& = \, 
D^{\{d\}(a)}(x, \, 1-x) 
& = \,  
D^{\{a\}(b)}(1-x, \, 0)   
& = \,
D^{\{a,d\}(c)}(x)   
\end{array}
\label{dico}
\end{equation}
$D^{\{i,j\}(k)}(x)$ is involved in the integral representation of the two-point 
diagram obtained by pinching the two propagators $i$ and $j \neq i$ in 
the box diagram. The $k$ assigned to each pair $\{i,j\}$ in the fourth 
column of table (\ref{dico}) corresponds to singling out line and column 
$k \in S_{4} \setminus \{i,j\}$ in the twice pinched matrix 
$\cals^{\{i,j\}}$. Alternatively, $D^{\{i,j\}(l)}(x)$ with 
$l \in S_{4} \setminus \{i,j,k\}$ corresponds to singling out 
line and column $l$ in $\cals^{\{i,j\}}$ and is related to 
$D^{\{i,j\}(k)}(x)$ by:
\begin{eqnarray}
D^{\{i,j\}(k)}(x)
& = &
G^{\{i,j\}(k)} \, x^{2} - 2 \, V^{\{i,j\}(k)} \, x - 
C^{\{i,j\}(k)}
\nonumber\\
& = &
D^{\{i,j\}(l)}(1-x)
\label{Ddoublepinched}
\end{eqnarray}
For any pair $\{i,j\}$, $k$ can thus be traded for $l$ in table 
(\ref{dico}) modulo the change of variable $x^{\prime}= 1-x$ which leaves 
unchanged the corresponding term in eq. (\ref{eqDEFJ5}).
$I_4^4$ is rewritten as a sum of  twelve terms:
\begin{align}
I_4^4 
& =   
- \, \frac{8}{3 \, B(2,1/2) \, B(3/2,1/2)} 
\sum_{i \in S} \, \sum_{j \in S \setminus \{i\}}
\frac{\bbar_i}{\detg} \, \frac{\bbj{j}{i}}{\detgj{i}} \, 
\notag \\
&  
\times 
\int^{+\infty}_0 \frac{d \xi}{\Delta_3 - \xi^2 + i  \lambda}
\int^{+\infty}_0  
\frac{d \rho}{\Delta_{2}^{\{i\}}- \xi^2 - \rho^2 + i  \lambda}  
\int^1_0 
\frac{dx}{D^{\{i,j\}(k)}(x) + \xi^2 + \rho^2 - i  \lambda}
\label{eqI449}
\end{align}

\subsection{Step 3}\label{fourpointstep3}

We iterate once more the procedure, adjusting the power of the denominator of 
eq. (\ref{eqI449}) using identity (\ref{eqFOND1biss}) with $\mu=3/2$, $\nu=2$, 
so as to transform the integrand into an $x$ derivative using identity 
(\ref{eqDEFREL1}) for $\tn=1$, $\alpha=1/2$.
This trivialises the integration over $x$ - albeit at the price of yet an 
extra integration. Let us introduce
\begin{equation}
K_{ij} 
= 
\int^1_0 
\frac{dx}{D^{\{i,j\}(k)}(x) + \xi^2 + \rho^2 - i  \lambda}
\label{eqDEFK1}
\end{equation}
With the help of eq. (\ref{eqFOND1biss}), $K_{ij}$ reads:
\begin{equation}
K_{ij} = \frac{2}{B(1,1/2)} \, \int^{+\infty}_0 d \sigma \, 
\int_0^1 d x \, 
\frac{1}
{(D^{\{i,j\}(k)}(x) + \xi^2 + \rho^2 + \sigma^2 - i \, \lambda)^{3/2}}
\label{eqDEFK20}
\end{equation}
Using identity (\ref{eqDEFREL1}), we get:
\begin{align}
K_{ij} 
& = 
- \, \frac{2}{B(1,1/2)}  
\int^{+\infty}_0 
\frac{d \sigma}
{\Delta_1^{\{i,j\}} - \xi^2 - \rho^2- \sigma^2 + i \, \lambda} 
\notag \\
&  
\;\;\;\;\;\;\;\;\;\;\;\;\;\;\;\;\;\;\;\;
\mbox{} \times 
 \int_0^1 d x \, \frac{\partial}{\partial \, x} \, 
\left[ 
 \frac{ x \, - \, (G^{\{i,j\}(k)})^{-1} \, V^{\{i,j\}(k)}}
 {(D^{\{i,j\}(k)}(x) + \xi^2 + \rho^2 + \sigma^2 - i \, \lambda)^{1/2}} 
\right] 
\label{eqDEFK3}
\end{align}
with $\Delta_1^{\{i,j\}} = (G^{\{i,j\}(k)})^{-1} \,
(V^{\{i,j\}(k)})^{2} + C^{\{i,j\}(k)}$.
The trivial integration over $x$ reads:
\begin{align}
K_{ij}
& =  
-\, \frac{2}{B(1,1/2)} 
\int^{+\infty}_0 
\frac{d \sigma}
{\Delta_1^{\{i,j\}}- \xi^2 - \rho^2 - \sigma^2 + i \lambda} 
\notag \\
&
\;\;\;\;\;\;\;\;\;\;\;\;\;\;\;\;\;\;\;\;  
\mbox{} \times 
\left[ 
 \frac{1 \, - \, (G^{\{i,j\}(k)})^{-1} \, V^{\{i,j\}(k)}} 
 {(D^{\{i,j\}(k)}(1) + \xi^2 + \rho^2 + \sigma^2 - i \lambda)^{1/2}} 
\right.
\notag \\
&  
\;\;\;\;\;\;\;\;\;\;\;\;\;\;\;\;\;\;\;\;\;\;\;\;
+ 
\left.
 \frac{(G^{\{i,j\}(k)})^{-1} \, V^{\{i,j\}(k)}}
 {(D^{\{i,j\}(k)}(0) + \xi^2 + \rho^2 + \sigma^2 - i \lambda)^{1/2}} 
\right]
\label{eqDEFK30}
\end{align}
We recognise\footnote{``$\det \, (G^{\{i,j\}})$" is merely a fancy notation to
keep some unity in formulas, as $G^{\{i,j\}}$ reduces to one single scalar.}:
\begin{eqnarray}
\Delta_{1}^{\{i,j\}} 
& = & 
- \; \frac{\det \, ({\cal S}^{\{i,j\}})}{\det \,(G^{\{i,j\}})} 
\nonumber \\
(G^{\{i,j\}(k)})^{-1} \cdot V^{\{i,j\}(k)} 
& = & 
- \; \frac{\bbj{l}{i,j}}{\detgj{i,j}}, \;\;
l \in S_{4} \setminus \{i,j,k\} 
\label{eqdefdelta1ij} \\
1 \, - \, (G^{\{i,j\}(k)})^{-1} \cdot V^{\{i,j\}(k)}
& = & 
- \; \frac{\bbar_{k}^{\{i,j\}}}{\det \,(G^{\{i,j\}})} 
\nonumber
\end{eqnarray}
Furthermore, $D^{\{i,j\}(k)}(0)$ and $D^{\{i,j\}(k)}(1)$ are proportional to 
internal masses squared: 
\begin{eqnarray}
D^{\{i,j\}(k)}(1) & = & 2 \, m^{2}_{l} \; \equiv \; \widetilde{D}_{ijk} 
\nonumber\\
D^{\{i,j\}(k)}(0) & = & 2 \, m^{2}_{k} \; \equiv \; \widetilde{D}_{ijl} 
\label{dijk}
\end{eqnarray}
for $k \in S_{4} \setminus \{i,j\}$ and $l \in S_{4} \setminus \{i,j,k\}$.
Since $D^{\{i,j\}(k)}(0) = D^{\{i,j\}(l)}(1)$, $\widetilde{D}_{ijk}$
is completely symmetric in all three mutually distinct indices $i,j,k$. 
We thus rewrite $K_{ij}$ as a sum over the $\bbj{k}{i,j}$:
\begin{align}
K_{ij} 
& =  
\frac{2}{B(1,1/2)} 
\sum_{k \in S \setminus \{i,j\} } \frac{\bbj{k}{i,j}}{\detgj{i,j}}  
\notag\\
& \mbox{} \times 
\int^{+\infty}_0 
\frac{d \sigma}
{(\Delta_1^{\{i,j\}}- \xi^2 - \rho^2 - \sigma^2 + i \lambda) \,
 (\tD_{ijk} + \xi^2 + \rho^2 + \sigma^2 - i \lambda)^{1/2}} 
\label{eqDEFK5}
\end{align}
By substitution of eq. (\ref{eqDEFK5}) into eq. (\ref{eqI449}) $I_4^4$ reads:
\begin{align}
I_4^4 
& =  
\sum_{i \in S_{4}} \, \sum_{j \in S_{4} \setminus \{i\}} \, 
\sum_{k \in S_{4} \setminus \{i,j\}} 
\frac{\bbar_i}{\detg} \, \frac{\bbj{j}{i}}{\detgj{i}} \, 
\frac{\bbj{k}{i,j}}{\detgj{i,j}} \, L_4^4(\Delta_3,\Delta_2^{\{i\}},\Delta_1^{\{i,j\}},\tD_{ijk})
\label{eqI449new}
\end{align}
with
\begin{align}
L_4^4(\Delta_3,\Delta_2^{\{i\}},\Delta_1^{\{i,j\}},\tD_{ijk})
& = \kappa \, 
\int^{+\infty}_0 
\frac{d \xi }{(\xi^2 - \Delta_3  - i \lambda)} 
\int^{+\infty}_0 
\frac{d \rho}{(\rho^2 + \xi^2 - \Delta_{2}^{\{i\}} - i \lambda)} 
\label{eqDEFK2} \\
&
\;\;\:
\mbox{} \times 
\int^{+\infty}_0
\frac{d \sigma}
{(\sigma^2 + \rho^2 + \xi^2 - \Delta_1^{\{i,j\}} - i \lambda) \, 
(\sigma^2 + \rho^2 + \xi^2 + \tD_{ijk} - i \lambda)^{1/2}}
\notag\\
\kappa & = \frac{16}{3 \, B(2,1/2) \, B(3/2,1/2) \, B(1,1/2)}
\notag
\end{align}
In summary, steps 1 to 3 traded three nested integrals over the 
initial Feynman parameters spanning the three-dimensional simplex for three 
nested integrals over the real positive half line.
The procedure generated the natural appearance of familiar ingredients of the 
algebraic reduction: this has been the main benefit so far. We also note that
the r.h.s. of eq. (\ref{eqI449new}) is manifestly independent of the 
successive choices made to single out lines and columns in the kinematic 
matrix $\cals$ and its subsequent pinchings.
We now have to compute the triple integral on the first octant.

\subsection{Step 4 \label{fourpointstep4}}

Let us consider $L_4^4(\Delta_3,\Delta_2^{\{i\}},\Delta_1^{\{i,j\}},\tD_{ijk})$ and focus on the real mass case.
$\Delta_3$, $\Delta_{2}^{\{i\}}$, $\Delta_1^{\{i,j\}}$ and 
$\tD_{ijk}$ being all real, thus all imaginary parts come from the 
$- \, i \lambda$ prescription,
the situation is the simplest possible. In contrast the extension to the 
general complex mass case leads to a branching of cases, which, albeit 
systematic, is more profuse than what happened for the three-point 
function. We therefore 
postpone the extension to the general complex mass case to a separate article.
We split step 4 itself into a succession of substeps that will make the
further complex mass extension easier. 

\vspace{0.3cm}

\noindent
{\bf a.} Let us first consider the nested $\sigma$ integral in $L_4^4(\Delta_3,\Delta_2^{\{i\}},\Delta_1^{\{i,j\}},\tD_{ijk})$ given by:
\begin{equation}
M_1(\xi^2+\rho^2)
= 
\int^{+\infty}_0  
\frac{d \sigma}
{(\sigma^2+\xi^2+\rho^2-\Delta_1^{\{i,j\}} - i \lambda) \, 
(\sigma^2+\xi^2+\rho^2+\tD_{ijk} - i \lambda)^{1/2}}
\label{eqdefisigma}
\end{equation}
We use identity (\ref{eqdeffuncjp2}) with $A = \xi^2+\rho^2-\Delta_1^{\{i,j\}} - i \, \lambda$ 
and $B=\xi^2+\rho^2+\tD_{ijk} - i \, \lambda$ to transform the r.h.s. of (\ref{eqdefisigma}) 
into an integral of only one factor: 
\begin{equation}
M_1(\xi^2+\rho^2)
= 
\int^1_0 
\frac{d z}
{z^2 \, (\tD_{ijk}+\Delta_1^{\{i,j\}}) + 
\xi^2+\rho^2-\Delta_1^{\{i,j\}} - i \lambda}
\label{eqisig1}
\end{equation}
{\bf b.} Next we nest $M_1(\xi^2+\rho^2)$ in the $\rho$ integration which reads:
\begin{align}
M_2(\xi^2)
& = 
\int^{+\infty}_0
\frac{d \rho}{\rho^2 + \xi^2 - \Delta_2^{\{i\}} - i \lambda} \, 
M_1(\xi^2+\rho^2)
\label{eqdefirho}\\
& = 
\int^1_0 d z
\int^{+\infty}_0
\frac{d \rho}{(\rho^2 + \xi^2 - \Delta_2^{\{i\}} - i \lambda) \, 
\left( \rho^2 + z^2 \, (\tD_{ijk}+\Delta_1^{\{i,j\}}) + 
\xi^2-\Delta_1^{\{i,j\}} - i \lambda \right)}
\notag
\end{align}
The $\rho$ integration is performed first, using partial fraction decomposition and 
eq. (\ref{eqFOND1biss}). We get:
\begin{align}
M_2(\xi^2)
& = \frac{1}{2} \, B(1/2,1/2) \,
\notag\\
& \times 
\int^1_0 
\frac{d z}
{z^2 \, (\tD_{ijk}+\Delta_1^{\{i,j\}}) 
 + (\Delta_2^{\{i\}} - \Delta_1^{\{i,j\}})} 
\left[ 
 \frac{1}{( \xi^2 - \Delta_2^{\{i\}} - i \lambda )^{1/2}} 
\right.
\notag\\
& \;\;\;\;\;\;\;\;\;\;\;\;\;\;\;\;\;\;\;\;\;\;\;\;
\left.
 \; - \; 
 \frac{1}{( z^2 \, (\tD_{ijk} + \Delta_1^{\{i,j\}}) + 
          (\xi^2 - \Delta_1^{\{i,j\}} -i \lambda) )^{1/2}}
\right]
\label{eqisig3}
\end{align}
{\bf c.} We then substitute $M_2(\xi^2)$ given by eq. (\ref{eqisig3}) into the 
remaining $\xi$ integration 
providing $L_4^4(\Delta_3,\Delta_2^{\{i\}},\Delta_1^{\{i,j\}},\tD_{ijk})$:
\begin{equation}
L_4^4(\Delta_3,\Delta_2^{\{i\}},\Delta_1^{\{i,j\}},\tD_{ijk})
= \kappa
\int^{+\infty}_0 \frac{d \xi}{\xi^2 - \Delta_3 - i \lambda} \, M_2(\xi^2)
\label{eqdefixi}
\end{equation}
Normalisation factors greatly simplify in $L_4^4(\Delta_3,\Delta_2^{\{i\}},\Delta_1^{\{i,j\}},\tD_{ijk})$ as 
\[
\frac{1}{2} \, B(1/2,1/2) \, \kappa  = \; 2
\]
and we exchange the integrations over $\xi$ and $z$ in eq. (\ref{eqdefixi}):
\begin{align}
L_4^4(\Delta_3,\Delta_2^{\{i\}},\Delta_1^{\{i,j\}},\tD_{ijk})
& = 2
\int^1_0 
\frac{d z}
{z^2 \, (\tD_{ijk}+\Delta_1^{\{i,j\}}) 
 + (\Delta_2^{\{i\}} - \Delta_1^{\{i,j\}})} 
\label{eqdefixi2}\\
& 
\;\;
\times
\int^{+\infty}_0 \frac{d \xi}{\xi^2 - \Delta_3 - i \lambda}
\left[ 
 \frac{1}{( \xi^2 - \Delta_2^{\{i\}} - i \lambda )^{1/2}} 
\right.
\notag\\
& \;\;\;\;\;\;\;\;\;\;\;\;\;\;\;\;\;\;\;\;\;\;\;\;\;\;\;\;\;\;\;\;\;\;\;
\left.
 \; - \; 
 \frac{1}{( z^2 \, (\tD_{ijk} + \Delta_1^{\{i,j\}}) + 
          (\xi^2 - \Delta_1^{\{i,j\}} -i \lambda) )^{1/2}}
\right]
\notag
\end{align} 
{\bf d.} Keeping $z$ fixed, the nested $\xi$ integral in the last two lines of eq. 
(\ref{eqdefixi2}) can be split into two integrals, each of which being of the 
same type as the one in eq. (\ref{eqdefisigma}). Each integral can thus be
recast in a form similar to eq. (\ref{eqisig1}). For convenience we 
introduce:
\begin{eqnarray}
\left.
\begin{array}{lclcl}
 P_{ijk} & = & \;\;\; \tD_{ijk}            & + & \Delta_1^{\{i,j\}} \\
 R_{ij}  & = & \;\;\;\Delta_{2}^{\{i\}} & - & \Delta_1^{\{i,j\}}   \\
 Q_i     & = & \;\;\; \Delta_3       & - & \Delta_{2}^{\{i\}}       \\
 T       & = & - \, \Delta_3 &&
\end{array}
\right\}
& \Leftrightarrow &
\left\{
\begin{array}{lcl}
 P_{ijk} + R_{ij} + Q_i + T & = & \;\;\; \tD_{ijk} \\
 R_{ij} \; + \, Q_i\; + \, T  & = & - \, \Delta_1^{\{i,j\}}   \\
 Q_i \;\; + T & = & - \, \Delta_{2}^{\{i\}}      \\
 T       & = & - \, \Delta_3
\end{array}
\right.
\label{newparamdefPQRT}
\end{eqnarray}
The recasting reads:
\begin{align}
L_4^4(\Delta_3,\Delta_2^{\{i\}},\Delta_1^{\{i,j\}},\tD_{ijk})
& =
2 \int^1_0 \frac{d z}{z^2 \, P_{ijk} + R_{ij}} \label{eqlijk10}\\ 
&\quad {} \times \left[ 
\;
 \int^1_0 \frac{d y}{y^2 \, (Q_i+T) + (1-y^2) \, T - i \lambda} 
\right. \notag \\
& \qquad \quad
- 
\left. 
 \int^1_0 \frac{d y}
 {y^2 \, (z^2 \, P_{ijk} + R_{ij} + Q_i + T)+ (1-y^2) \, T - i \lambda} 
\right]
\notag
\end{align}
{\bf e.} We trade $z$ for $u = y \, z$ so that $L_4^4(\Delta_3,\Delta_2^{\{i\}},\Delta_1^{\{i,j\}},\tD_{ijk})$ takes the form:
\begin{align}
L_4^4(\Delta_3,\Delta_2^{\{i\}},\Delta_1^{\{i,j\}},\tD_{ijk})
&= - \, 2   
\int_0^1 d u 
\int_{u}^1 
\frac{dy \, y}{u^{2} \, P_{ijk} + y^{2} \, R_{ij}}
\notag \\
& \;\;\;\;\;\;\;\;\;\;\;\;\;\;\;\;\;\;\;\;\;\;\;\;
\;\; {} \times
\left[ 
 \frac{1}{u^2 \, P_{ijk} + y^{2} \, (R_{ij} + Q_i) + T - i \lambda} \right. \notag \\ 
 &\qquad \qquad \qquad \qquad \quad {} - \left. 
 \frac{1}{y^2 \, Q_i + T - i \lambda}  
\right]
\label{eqlijk11}
\end{align}
{\bf f.} We trade $y$ for $x=y^2$:
\begin{align}
L_4^4(\Delta_3,\Delta_2^{\{i\}},\Delta_1^{\{i,j\}},\tD_{ijk})
&= - 
\int^1_0 d u 
\int^1_{u^2} \frac{d x}{u^2 \, P_{ijk} + x \, R_{ij}} \, 
\label{eqlijk13} \\
& \;\;\;\;\;\;\;\;\;\;\;\;\;\;\;\;\;\;\;\;\;\;\; {} \times
\left[  
 \frac{1}{u^2 \, P_{ijk} + x \, (R_{ij} + Q_i) + T - i \lambda} 
 - \frac{1}{x \, Q_i + T - i \lambda}  
\right]
\notag
\end{align}
{\bf g.} We perform a partial fraction decomposition w.r.t. $x$ in the integrand of 
eq. (\ref{eqlijk13}):
\begin{align}
&\frac{1}{u^2 \, P_{ijk} + x \, R_{ij}} \, 
\left[  
 \frac{1}{u^2 \, P_{ijk} + x \, (R_{ij} + Q_i) + T - i \lambda} 
 - \frac{1}{x \, Q_i + T - i \lambda}  
\right]
\notag\\
&=
\frac{1}{u^2 \, P_{ijk} \, Q_i - R_{ij} \, (T - i \lambda)} \,
\left[
 \frac{R_{ij} + Q_i}{u^2 \, P_{ijk} + x \, (R_{ij} + Q_i) + T - i \lambda} 
 - 
 \frac{Q_i}{x \, Q_i + T - i \lambda}
\right]
\label{pfdx}
\end{align}
{\bf h.} Lastly the $x$ integration is readily performed and we finally get:
\begin{align}
L_4^4(\Delta_3,\Delta_2^{\{i\}},\Delta_1^{\{i,j\}},\tD_{ijk})
&= - \int^1_0 \frac{d u}{u^2 \, P_{ijk} \, Q_i - R_{ij} \, (T - i \lambda)} \,
\notag \\
& \;\;\;\;\;\;\;\;\;\;\;\;\; {} \times
\left[  
\ln 
\left(
 \frac{\; u^2 \, P_{ijk} + (R_{ij} + Q_i \; + T) - i \lambda}
{u^2 \, (P_{ijk} + R_{ij} + Q_i) + T - i \lambda} 
\right) \right. \notag \\ 
&\qquad \qquad \quad \; {} - \left.
\ln 
\left(\frac{\;\;\;\, Q_i + T - i \lambda}{u^2 \, Q_i + T - i \lambda} \right)  
\right]
\label{eqlijk14}
\end{align}
Using the following identities:
\begin{align}
\bar{b}_{i}^{2}
& = 
- \det (\cals^{\{i\}}) \, \detg - \dets \, \det (G^{\{i\}}) 
\notag\\
\bar{b}_{j}^{\{i\} \, 2}
& = 
\det (\cals^{\{i,j\}}) \, \det (G^{\{i\}}) + 
\det (\cals^{\{i\}}) \, \det (G^{\{i,j\}}) 
\label{magic_id-ji}\\
\bar{b}_{k}^{\{i,j\} \, 2}
& = \tD_{ijk} \, \det (G^{\{i,j\}}) - \det (\cals^{\{i,j\}}) 
\notag
\end{align}
(see appendix A.2 on Jacobi identities for determinants in ref. \cite{Guillet:2013mta}),
the coefficients $P_{ijk}, \, R_{ij}, \, Q_{i}$ and $T$ 
defined in eqs. (\ref{newparamdefPQRT}) can be recast in terms of
the $\bar{b}_{i},\bar{b}_{j}^{\{i\}}$ and $\bar{b}_{k}^{\{i,j\}}$ using the group of equations~(\ref{magic_id-ji}):
\begin{align}
P_{ijk} 
& = \frac{\bar{b}_{k}^{\{i,j\} \, 2}}{\det (G^{\{i,j\}})}
\notag\\
R_{ij} 
& = \frac{\bar{b}_{j}^{\{i\} \, 2}}{\det (G^{\{i,j\}}) \, \det (G^{\{i\}})}
\notag\\
Q_{i} 
& = \frac{\bar{b}_{i}^{2}}{\det (G^{\{i\}}) \, \detg}
\notag
\end{align}
With  
$\Delta_{3}, \, \Delta_{2}^{\{i\}}, \, \Delta_{1}^{\{i,j\}}$ and $\tD_{ijk}$,
given by the groups of eqs. (\ref{eqdefdelta3}), (\ref{eqdefdelta2j}), (\ref{eqdefdelta1ij}) 
and (\ref{dijk}) \\
respectively, eq. (\ref{eqlijk14}) then translates into:
\begin{align}
L_4^4(\Delta_3,\Delta_2^{\{i\}},\Delta_1^{\{i,j\}},\tD_{ijk})
&= - \, \det (G) \, \det (G^{\{i\}}) \, \det (G^{\{i,j\}})
\notag\\
&\;\;\;\;
\times 
\int_{0}^{1} 
\frac{du}
{u^2 \, \bar{b}_{i}^{2} \, \bar{b}_{k}^{\{i,j\} \, 2} - 
\bar{b}_{j}^{\{i\} \, 2}
\left( \dets - i \lambda \; \sign(\detg) \right)} \,
\notag \\
& \;\;\;\;
{} \times
\left[  
\ln 
\left(
 \frac{\;\;\; u^2 \, \tD_{ijk} -(1-u^2) \Delta_{1}^{\{i,j\}} - i \lambda}
{u^2 \, \tD_{ijk} - (1-u^2) \Delta_{3} - i \lambda} 
\right) \right. \notag \\ 
&\qquad \;\;\; {} - \left. 
\ln 
\left(\frac{ - \, \Delta_{2}^{\{i\}} - i \lambda}
{- \, u^2 \, \Delta_{2}^{\{i\}} -(1-u^2) \, \Delta_{3} - i \lambda} \right)  
\right]
\label{eqlijk14alternaivebbar}
\end{align}
which is useful in view of making practical numerical implementation. 

\vspace{0.3cm}

\noindent
Let us note that the apparent poles in the integrand of the r.h.s. of eq. 
(\ref{eqlijk14}) are fake as their residues vanish\footnote{This can be checked 
explicitly on eq.~(\ref{eqlijk14}) in which the residue contributions of the 
two logarithms cancel against each other. These fake poles are an artefact of the
partial fraction decomposition (\ref{pfdx}).}. 
Furthermore, in the real mass case at hand, 
$(u^2 \, P_{ijk} + R_{ij}+Q_i+T)$, $(u^2 \, (P_{ijk} + R_{ij} + Q_i) + T)$ and 
$(u^2 \, Q_i + T)$ are all real when $u$ spans $[0,1]$: each logarithm of 
ratios in eq. (\ref{eqlijk14}) can be safely split into a difference of two 
logarithms, the arguments of which keeping an (infinitesimal) imaginary part of
constant (negative) sign as $u$ spans $[0,1]$.
The remaining integration can be performed in closed form in terms of 
dilogarithms. This computation is provided in appendix \ref{appF}.

\subsection{Comparison with ref. \cite{tHooft:1978jhc} and proliferation of 
dilogarithms}\label{discfourpdilog}

We compared our result (\ref{eqlijk14}) with ref. \cite{tHooft:1978jhc}. Beyond mere
numerical comparisons, the respective analytical expressions do not compare
easily, and formula (\ref{eqlijk14}) involves a proliferation of
dilogarithms, namely 288 vs. 108 for ref. \cite{tHooft:1978jhc}.

\vspace{0.3cm}

\noindent
Let us first remind a few features of the latter. The method of ref. \cite{tHooft:1978jhc}
linearises the dependence of the integrand in some integration variables to
facilitate two integrations. For this purpose two parameters $\alpha$ and
$\beta$ are introduced and successively adjusted as roots of some second degree
equations, which generates a fourfold arbitrariness in the procedure. All
choices of $\alpha$ and $\beta$ are equivalent, they lead to results which look
formally the same, and provide the same numerical result. Yet the analytic 
forms when $\alpha$ and $\beta$ are made explicit in terms of the primary 
parameters depend on the choice made. 
In contrast our iterative decomposition closely related to algebraic reduction 
and pinching operations preserves a kind of symmetry - a ``democracy" among 
all possible single pinchings, and double pinchings. This 
symmetry is explicitly broken in ref. \cite{tHooft:1978jhc} whenever any choice for 
$\alpha$ and $\beta$ is made. Therefore, the comparison of our result shall be 
carried out with the double symmetrisation of ref. \cite{tHooft:1978jhc} w.r.t. both 
$\alpha$ and $\beta$, not with ref. \cite{tHooft:1978jhc} {\em per se}. 
This double symmetrisation increases the number of terms w.r.t. ref. 
\cite{tHooft:1978jhc}. The price to pay for our method compared to ref. \cite{tHooft:1978jhc} is thus 
that some extra work shall be carried out to counteract the corresponding 
proliferation. 
This already occurred for the three-point function, and we saw that the issue 
was easily overcome in that case. The situation for the four-point function 
is more complex and requires some elaboration and we need to carry out the
comparison in some more detail. 
 
\vspace{0.3cm}

\noindent
Consecutively to the choice of $\alpha$ the calculation of ref. \cite{tHooft:1978jhc} 
splits into three 
contributions depending on the choice of $\alpha$ - let us call them 
``$\alpha$-sectors". In each $\alpha$-sector, the subsequent choice of $\beta$
(which depends both on the sector considered and explicitly on 
$\alpha$) leads again to three contributions - or ``$\beta$-subsectors". 
This provides $I_4^4$ as a sum of one-dimensional integrals. In each of 
the $\beta$-subsectors the integrand is given by a logarithm of some rational 
fraction, divided by some second degree polynomial. The rational fractions in the 
logarithms are of the form  
$({\cal L} - i \lambda)/({\cal Q}- i \lambda)$ in which ${\cal L}$  and 
${\cal Q}$ are first and second degree polynomials respectively. 
After partial fraction decomposition of 
these denominators the values of the logarithms at each pole are subtracted
so that the poles are made manifestly fake in each term separately.
Whereas the ${\cal Q}$ are independent of the choices of $\alpha$ and
$\beta$, the ${\cal L}$ are independent on the choice of $\beta$ but do
depend explicitly on $\alpha$. The locations of the fake poles in each
$\alpha$-sector depend on $\beta$ explicitly but not on $\alpha$, yet 
there are relations between different poles from different $\alpha$-sectors.
Each one of the three $\alpha$-sectors in ref. \cite{tHooft:1978jhc} leads to three 
${\cal L}$ and three ${\cal Q}$, however this does not lead to nine distinct 
${\cal L}$ and ${\cal Q}$ overall. Among the nine ${\cal Q}$, three of them 
appear twice (among the last two lines in eq. (6.26) of ref. \cite{tHooft:1978jhc})
whereas the three others (in the second line in eq. (6.26)) appear only once. 
Each of the ${\cal Q}$ happens to correspond to a double-pinching 
pattern of the box diagram, and indeed there are only six such distinct double 
pinchings. There is no connection between splitting in sectors and double 
pinchings, though, and the correspondence between the polynomials ${\cal Q}$ 
and the double pinchings are identified a posteriori. The ${\cal L}$ happen to 
coincide whenever the ${\cal Q}$ correspondingly involved in the rational
fractions respectively coincide, yet the ${\cal L}$ look ``atypical" and have 
no simple interpretation e.g. in terms of pinching or otherwise, at least to us. 

\vspace{0.3cm}

\noindent
The terms $\ln(u^2 \, P_{ijk} + (R_{ij} + Q_i +T - i \lambda))$
in eq. (\ref{eqlijk14}) in our approach coincide with the $\ln({\cal Q})$
in \cite{tHooft:1978jhc}. This can be seen by first absorbing $\bar{b}_{k}^{\{i,j\}}$
by the rescaling of the integration variable $u$ as 
$t = \bar{b}_{k}^{\{i,j\}} \, u$, then performing the sum
$\sum_{k}$ explicitly introducing $y \in [0,1]$ such that 
$t = - \bar{b}_{l}^{\{i,j\}} - \det ( G^{\{i,j\}} ) \, y$. 
However, the pole terms weighting these identical logarithms do not
match in any clear way between  eq. (\ref{eqlijk14}) in our approach and 
\cite{tHooft:1978jhc}.
Furthermore, the two other logarithms $\ln(u^2 Q_i+T - i \lambda)$ and 
$\ln(u^2 \, (P_{ijk} + (R_{ij} + Q_i) +T - i \lambda))$ in our approach
are atypical and their contributions do not match with those of the atypical 
$\ln ({\cal L})$ of ref. \cite{tHooft:1978jhc} in any clear way either. 

\vspace{0.3cm}

\noindent
As for taming the proliferation of dilogarithms in eq. (\ref{eqlijk14}) in our 
approach, the use of the five-dilogarithm Hill identity combined with the
property of parity in $u$ of the involved integrals allows to reduce the number
of dilogarithms by a factor two, from 288 to 144, still somewhat more than in ref. 
\cite{tHooft:1978jhc}: some extra work would still be required to
further reduce this number.
Note also that the advantage of the method used here is that
the roots of the denominators and of the arguments of the logarithms are
expressed in terms of the determinants of the Gram matrix and the $\cals$ matrix as well as those of
the corresponding pinched matrices. From that, some relations between these different roots
can be deduced. These relations could be used to reduce in a systematic way
using modern tools like symbols the number of dilogarithms. This future work is postponed 
in a future publication \cite{future}.

\section{Summary and outlook}

In this article we have presented a novel approach to the computation of one-loop 
three- and four-point functions. The method naturally proceeds in terms of 
algebraic kinematical invariants involved in reduction algorithms and applies 
to general kinematics beyond the one relevant for one-loop collider processes, 
it thereby offers a potential application to the calculation of processes 
beyond one loop using one-loop (generalised) $N$-point functions as building 
blocks. 
In addition, limiting behaviours in a number of specific regimes with $\dets \to 0$ or 
$\detg \to 0$ or both simultaneously is presented. They are restricted to cases
that do not lead to Landau singularities and that are relevant for NLO kinematics.
The treatment of the leftover cases will be postponed to subsequent work if relevant.
For the sake of pedagogy, the method was hereby exposed on ``ordinary" 
three- and four-point functions in four dimensions in the real mass case.
It can be extended in several respects, 
and each of these respects will be tackled in two companion papers briefly 
advertised in what follows.

\vspace{0.3cm}

\noindent
In the first companion paper we will extend the presented framework to the 
general complex mass case. Whereas the extension will branch into a profusion of 
cases, the general philosophy of the approach extends straightforwardly, and the
profusion of cases can be tamed and reunified by means of one-dimensional 
contour integrals in the complex plane encoding analyticity properties. 
This companion paper will also provide a formula to express
$\Im(\dets)$ in terms of widths.

\vspace{0.3cm}

\noindent
In the second companion paper we will extend the presented framework to the case
where some vanishing internal masses cause infrared soft and/or collinear
divergences. The method extends in a straightforward way, once
a few intermediate steps and tools are appropriately adapted.  

\vspace{0.3cm}

\noindent
The proliferation of dilogarithms in the expression of the four-point function
computed in closed form with the present method requires some extra work to be
better apprehended, in order to counteract it. This issue will be addressed in a
future article.

\vspace{0.3cm}

\noindent
This method can of course also be applied to compute $N$-point one-loop 
functions with any $N \ge 5$ but this
is pointless because in these cases 
the first step 
boils down to recovering well-known formulae that reduce the $N$-point 
one-loop function to 
one-loop ones with fewer points.
\cite{vanNeerven:1983vr,Bern:1993kr,Tarasov:1996br,Binoth:1999sp}. 

\vspace{0.3cm}

\noindent
The last goal is to provide the generalised one-loop building blocks 
entering as integrands in the computation of two-loop three- and four-point 
functions by means of an extra numerical double integration.

\section*{In memoriam}

Various ideas and techniques used in this work were initiated by Prof. Shimizu
after a visit to LAPTh. He explained us his ideas about the numerical 
computation of scalar two-loop three- and four-point functions, he shared his 
notes partly in English, partly in Japanese with us and he encouraged us to 
push this project forward. J.Ph. G. would like to thank Shimizu-sensei for 
giving him a taste of the Japanese culture and for his kindness.

\section*{Acknowledgements}

We would like to thank P. Aurenche for his support during this project and for a careful reading of the manuscript.

\appendix

\section{A Stokes-type identity}\label{ap1}

In this appendix, we derive the master identity (\ref{eqDEFREL1}) that enables 
us to perform one integral trivially. Let us consider the column $\tn$-vector
\[
 X
=
\left[
 \begin{array}{c}
   x_1 \\
   x_2 \\
   \vdots \\
   x_{\tn}
 \end{array}
\right]
\]
and the polynomial $D$ of $\tn$ variables defined by:
\begin{equation}
  D = X^{\;T} \cdot G \cdot X - 2 \, V^{T} \cdot X - C
  \label{eqDEFD1}
\end{equation}
where $G$ is a symmetric $\tn \times \tn$ matrix and $V$ a column 
$\tn$-vector. The subscript $T$ stands for the transpose. Throughout this
appendix we assume the matrix $G$ to be invertible. 
Let $\nabla$ stand for:
\[
\nabla = 
\left[ 
 \begin{array}{c}
 
 \frac{\partial}{\partial \, x_1} \\
 \frac{\partial}{\partial \, x_2}, \\
 \vdots \\
 \frac{\partial}{\partial \, x_{\tn}} 
 \end{array}
\right]
\]
so that $\nabla \, f$ is the gradient of the scalar field $f$ whereas 
$\nabla^{T} \cdot F$ is the divergence of the $\tn$-vector field $F$. 
The gradient of $D$ is given by:
\[
\nabla \, D = 2 \, \left( G \cdot X - V \right) 
\]
so that
\begin{equation}
\nabla^{T} \cdot 
\left( \frac{X - G^{-1} \cdot V}{D^{\alpha}} \right)
=
\frac{\tn}{D^{\alpha}} - 
\frac{2 \, \alpha}{D^{\alpha+1}} \,
\left[
 \left( G \cdot X - V \right)^{T} \cdot \left( X - G^{-1} \cdot V \right)
\right]
\label{eqapp2}
\end{equation}
The square bracket in the r.h.s. of eq. (\ref{eqapp2}) may be written:
\begin{eqnarray}
 \left( G \cdot X - V \right)^{T} \cdot \left( X - G^{-1} \cdot V \right)
& = &
D + \Delta_{\tn}
\label{eqapp3}\\
\Delta_{\tn} 
& = &
V^{T} \cdot G^{-1} \cdot V + C
\label{eqapp4}
\end{eqnarray}
Substituting eqs. (\ref{eqapp3}), (\ref{eqapp4}) into eq. (\ref{eqapp2}) thus 
leads to
\begin{eqnarray}
\nabla^{T} \cdot 
\left( \frac{X - G^{-1} \cdot V}{D^{\alpha}} \right)
& = &
\frac{\tn - 2 \, \alpha}{D^{\alpha}} - 
\frac{2 \, \alpha \, \Delta_{\tn}}{D^{\alpha+1}}
\label{eqapp5}
\end{eqnarray}
from which identity (\ref{eqDEFREL1}) immediately follows whenever 
$\Delta_{\tn} \neq 0$. Let us remark in passing that 
\[
\Delta_{\tn} = (-1)^{\tn} \frac{\det \, (\cal S)}{\det \, (G)}
\]
as shown in appendix \ref{detsdetg}.

\section{How to tune powers of denominators}\label{ap2}

The following identity is used with real $\mu,\nu$ such that $\mu > 1/\nu > 0$ 
to tune powers in denominators:
\begin{equation}
 \int^{\infty}_0 \, \frac{d \xi}{(D + \xi^{\nu})^{\mu}} 
= 
\frac{1}{\nu} \, B\left( \mu-\frac{1}{\nu},\frac{1}{\nu} \right) \, 
\frac{1}{D^{\mu-1/\nu}}
\label{eqFOND1}
\end{equation}
where $B(x,y)$ is the Euler Beta function defined by 
\[
  B(x,y) = \frac{\Gamma(x) \, \Gamma(y)}{\Gamma(x+y)}
\]
The integral in the l.h.s. of eq. (\ref{eqFOND1}) is convergent for 
$\mu > 1/\nu > 0$.
Identity (\ref{eqFOND1}) is easily established for $D$ real writing
\begin{eqnarray}
\int^{\infty}_0 \, \frac{d \xi}{(D + \xi^{\nu})^{\mu}} 
& = & 
\frac{1}{\Gamma(\mu)} \int^{\infty}_0 \, d \xi 
\int_{0}^{+\infty} du \, u^{\mu-1} \, e^{- \, u(D+\xi^{\nu})}
\label{eqdreal}
\end{eqnarray}
then making the change of variable $\xi = v^{1/\nu}$ and performing the
integration over $v$ first.
For $D$ complex, identity (\ref{eqFOND1}) still holds provided
$\nu \, \mu > 1$. Indeed, we may equivalently 
\begin{itemize}
\item
either analytically continue the r.h.s. of identity (\ref{eqFOND1}) to 
$|\arg(D)| < \pi$ 
\item
or consider the integral of 
$F(\xi) = (D+\xi^{\nu})^{-\mu}$ along the closed contour of integration in the
complex $\xi$-plane defined by figure \ref{fig3}. 
According to the Cauchy theorem,
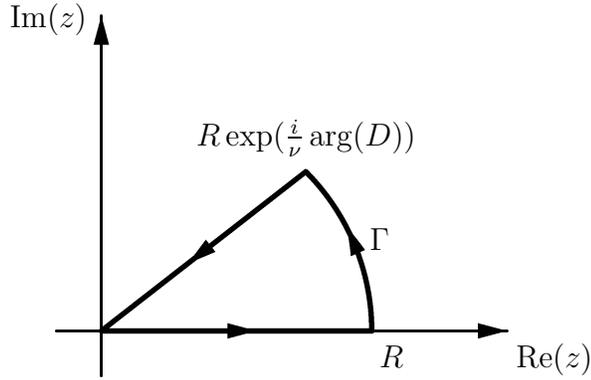
\begin{figure}[h]
\centering
\parbox[c][60mm][t]{60mm}
{\begin{fmfgraph*}(60,60)
  \fmfipair{o,xm,xp,ym,yp}
  \fmfipath{c[]}
  \fmfipair{r[]}
  \fmfiequ{o}{(.1w,.1h)}
  \fmfiequ{xm}{(0,.1h)}
  \fmfiequ{xp}{(w,.1h)}
  \fmfiequ{ym}{(.1w,0)}
  \fmfiequ{yp}{(.1w,.8h)}
  \fmfiequ{r3}{(.2w,.1h)}
  \fmfiv{l=$\Re(z)$,l.a=-55,l.d=2mm}{xp}
  \fmfiv{l=$\Im(z)$,l.a=-165,l.d=2mm}{yp}
  \fmfiset{c1}{fullcircle scaled w shifted r3}
  \fmfpen{thin}
  \fmfiset{c2}{xm--xp}
  \fmfiset{c3}{ym--yp}
  \fmfcmd{draw c2; cfill(harrow (c2,1.)); draw c3; cfill(harrow (c3,1.));}
  \fmfpen{thick}
  \fmfi{fermion,label=$\Gamma$}{ subpath (0,length(c1)/8) of c1}
  \fmfiequ{r1}{point 0 of c1}
  \fmfiequ{r2}{point length(c1)/8 of c1}
  \fmfi{fermion}{o--r1}
  \fmfi{fermion}{r2--o}
  \fmfiv{l=$R$}{r1}
  \fmfiv{l=$R \exp(\frac{i}{\nu} \arg(D))$,l.a=90,l.d=2mm}{r2}
\end{fmfgraph*}}
\vspace{0.2cm}
\caption{\footnotesize 
Integration contour $\mathcal{C}$ in the $z$ complex plane}
\label{fig3} 
\end{figure}

\begin{eqnarray}
0 
& = & 
\oint_{\mathcal{C}} d \xi \, F(\xi) 
\nonumber\\
& = & 
\int^{R}_0 d \xi \, F(\xi) + \int_{\Gamma} d \xi \, F(\xi) + 
\int_{R \, e^{i \, \arg(D)/\nu}}^{\,0} d \xi \, F(\xi) 
\label{eqcont1}
\end{eqnarray}
When $R \to + \, \infty$, the first integral of eq. (\ref{eqcont1}) leads to 
the sought-after one, the second integral is ${\cal O}(R^{1- \nu \, \mu}) \to 0$ 
provided $\nu > 1/\mu > 0$, and the path of the third integral may be 
parametrised by $\xi = \rho \, D^{1/\nu}$, 
$0 \leq \rho \leq R \, |D|^{-1/\nu}$ so as to get:
\begin{eqnarray}
\int_{R \, e^{i/\nu \, \arg(D)}}^{\,0} d \xi \, F(\xi) 
& = & 
- \, \int_{0}^{R \, |D|^{-1/\nu}} d \rho \, 
\frac{ d \rho \, D^{1/\nu}}{\left[ (\rho^{\nu} +1 ) D \right]^{\mu}} 
\nonumber \\ 
& \to & 
-
 \, \frac{1}{D^{\mu-1/\nu}} \, 
\int_0^{\infty} \frac{d \rho}{(\rho^{\nu}+1)^{\mu}}
\label{eqcont2}
\end{eqnarray}
The integral in the r.h.s. of eq. (\ref{eqcont2}) is namely the
one just computed above for $D=1$, hence:
\begin{equation}
\int^{\infty}_0 \, \frac{d \xi}{(D + \xi^{\nu})^{\mu}}
= 
\frac{1}{\nu} \, B \left( \mu-\frac{1}{\nu}, \frac{1}{\nu} \right) \,
\frac{1}{D^{\mu-1/\nu}}
 \label{eqdcomp}
\end{equation}
\end{itemize}

\section{$b$ and $B$ in terms of $G^{(a)}$ and $V^{(a)}$}\label{detsdetg}

In this appendix, we show how to express the $b$ coefficients
in term of $G^{(a)}$ and $V^{(a)}$. We assume that the dimension of the $\cals$ matrix is $N$ and 
we single out its last line and column to build the associated Gram matrix $G^{(N)}$.
We introduce the $b_i$ \cite{Binoth:2005ff} which are solutions of the equation:
\begin{equation}
  \cals . b = e
  \label{eqdefbcoeff}
\end{equation}
with
\begin{align}
  b &= \left[
  \begin{array}{c}
    b_1 \\
    \vdots \\
    b_N
  \end{array}
\right]
  \label{eqdefvecb} \\
  e &= \left[
  \begin{array}{c}
    1 \\
    \vdots \\
    1
  \end{array}
\right]
  \label{eqdefvece}
\end{align}

\noindent
The $\cals$ matrix can be related to the block matrix $\widehat{\cals}^{(N)}$ 
whose blocks are proportional to the Gram matrix $G^{(N)}$, to the vector $V^{(N)}$ and to the scalar $\cals_{NN}$ defined in sec.~\ref{sectthreepoint} and \ref{sectfourpoint}:
  \begin{align}
    \widehat{\cals}^{(N)}
      &=
      \left[
       \begin{array}{ccc}
         {} - G^{(N)} & | & V^{(N)} \\
        --     & + & - \\
        V^{(N) \, T}  & | & \cals_{NN} \\
      \end{array}
      \right]
      \label{e7}
  \end{align}
The subtraction of column and line $N$ leading from ${\cal S}$ to  
$\widehat{\cal S}^{(N)}$
corresponds to the following matricial product:
\begin{align}
  \widehat{\cal S}^{(N)} &=  L_{-N} \cdot {\cal S} \cdot C_{-N}
  \label{dec1}
\end{align}
with
\begin{equation}
L_{-N} 
=
\left[
 \begin{array}{ccr}
 \vdots            & | & -1 \\
 \cdots \; \mathds{1}_{N-1} \; \cdots & | & \vdots \\
 \vdots            & | & -1 \\
 -----             & + & -- \\
 0 \; \cdots \; 0        & | &  1
 \end{array}
\right] 
\, , \quad
C_{-N} 
= 
L_{-N}^{T} 
\label{dec2}
\end{equation}
Left multiplication of ${\cal S}$ by $L_{-N}$ subtracts line $N$ of 
${\cal S}$ from each of the lines 1 to $N-1$ of ${\cal S}$. 
Likewise right multiplication of ${\cal S}$ by $C_{-N}$ 
subtracts column $N$ of ${\cal S}$ from each of the columns 1 to $N-1$. 
The matrix $L_{-N}$ is invertible ($\det \left( L_{-N} \right) = 1$) and  $L_{-N}^{-1}$ is
the matrix $L_{+N}$ obtained from $L_{-N}$ by replacing all ``$-1$" by ``$+1$" 
in the last column, so that left multiplication by $L_{+N}$ adds line $N$ to 
each of the lines 1 to $N-1$ - and likewise with $C_{+N}=C_{-N}^{-1}$ vs. 
$C_{-N}$.

\vspace{0.3cm}

\noindent
From eq. (\ref{dec1}), it is clear that the two matrices $\cals$ and $\widehat{\cals}^{(N)}$ have the same determinant.
So $\dets$ can be computed by expanding with respect to 
the last line of eq. (\ref{e7}), this yields:
\begin{align}
\dets 
&=
{\cal S}_{NN} \, \det \left( - \, G^{(N)} \right) 
+
\sum_{j=1}^{N-1} (-1)^{N+j}  \, V^{(N)}_{j} \, \overline{G}^{(N)}_{j}
\label{e8}
\end{align}
The quantity
\begin{align}
\overline{G}^{(N)}_{j}
&=
\left|
 \begin{array}{ccc}
  - \, G^{(N)}_{\mbox{\footnotesize column $j$ suppressed}} & \vdots &  V^{(N)} 
 \end{array}
\right|
\nonumber
\end{align}
is in its turn expanded w.r.t. to its last column $V^{(N)}$:
\begin{align}
\overline{G}^{(N)}_{j}
&=
(-1)^{i-1} \, 
\sum_{i=1}^{N-1} \mbox{Cof} \left[ G^{(N)} \right] _{ij} \, V^{(N)}_{i} 
\label{e9}
\end{align}
where $\mbox{Cof} \left[ G^{(N)} \right]$ is the matrix of cofactors of 
$G^{(N)}$.
Substituting (\ref{e9}) into (\ref{e8}) we get:
\begin{align}
\dets
&=
(-1)^{N-1} 
\left\{ 
 {\cal S}_{NN} \, \detg 
 +
 V^{(N) \, T} \cdot \mbox{Cof} \left[ G^{(N)} \right] \cdot V^{(N)}
\right\}
\label{e10-0}
\end{align}
or, in other words:
\begin{align}
  \cals_{NN} + V^{(N) \, T} \cdot \left( G^{(N)} \right)^{-1} \cdot V^{(N)} &=  (-1)^{N-1} \, \frac{\dets}{\detg}
  \label{e10-01}
\end{align}
This is to compare with eq.~(\ref{eqapp4}) reminding that the constant term $C$ of the polynomial $D$ 
is precisely $\cals_{NN}$ (cf.\ eq.~(\ref{eqVJA3}) or eq.~(\ref{eqVJA4})), so we conclude that:
\begin{align}
  \Delta_{\tn} &= (-1)^{N-1} \, \frac{\dets}{\detg}
  \label{eqe10-02}
\end{align}

\vspace{0.3cm}

\noindent
The eq. (\ref{eqdefbcoeff}) can be reformulated in 
terms of $\widehat{\cal S}^{(N)}$. Inverting eq. (\ref{dec1}), we get:
\begin{align}
  \widehat{\cal S}^{(N)} \cdot \widehat{b}
  &= \widehat{e}
  \label{reform1}
\end{align}
with
\begin{align}
\widehat{b} 
\equiv C_{+N} \cdot b =
\left[ 
 \begin{array}{c}
 \beta^{(N)} \\
 -- \\
 B
 \end{array}
\right] 
&\quad \text{and} \quad
\widehat{e} 
\equiv L_{-N} \cdot e =
\left[ 
 \begin{array}{c}
 0 \\
 \vdots \\
 0 \\
 --\\
 1
 \end{array}
\right]
\label{reform2}
\end{align}
\noindent
where $\beta^{(N)}$ is the $N-1$ vector:
\begin{align}
  \beta^{(N)} &= \left[
  \begin{array}{c}
    b_1 \\
    \vdots \\
    b_{N-1}
  \end{array}
\right]
  \label{eqdefbetan}
\end{align}
and
\begin{equation}
  B = \sum_{i=1}^{N} b_i
  \label{eqdefbibb}
\end{equation}

\vspace{0.3cm}

\noindent
We will discuss the solutions of eq. (\ref{reform1}) in the following cases: 1) the determinants of the matrices $\cals$ and $G^{(N)}$ do not vanish, 
2) the determinant of the matrix $\cals$ is different from 0 but the determinant of the Gram matrix $G^{(N)}$ vanishes and 3) both determinants of the $\cals$ and $G^{(N)}$ matrices vanish.

\subsection{Case $\dets \ne 0$ and $\detg \ne 0$}\label{nevervanish}

Lets us parametrise $(\widehat{\cal S}^{(N)})^{-1}$ as
\begin{eqnarray}
(\widehat{\cal S}^{(N)})^{-1}
& = &
\left[
 \begin{array}{ccc}
  Q & | & W \\
  --     & + & - \\
  W^{T}  & | & U \\
 \end{array}
\right]
\label{invshat}
\end{eqnarray}
the solution of eq.~(\ref{reform1}) then reads:
\begin{align}
  \beta^{(N)} &= W 
  \label{eqsoleqshbhehj0} \\
  B &= U
  \label{eqsoleqshbhehj1}
\end{align}
In order to compute the different parameters, equation~(\ref{invshat})
is substituted in the equation 
$\widehat{\cal S}^{(N)} 
\cdot (\widehat{\cal S}^{(N)})^{-1} = \mathds{1}_{N}$. We get
the following linear system:
\begin{eqnarray}
- \, G^{(N)} \cdot Q \, + \, V^{(N)} \cdot W^{T} 
& = & 
\mathds{1}_{N-1}
\label{inv-11}\\
- \, G^{(N)} \cdot W \, + \, V^{(N)} \, U
& = & 
0
\label{inv-21}\\
 Q \cdot V^{(N)} \, + \, {\cal S}_{NN} \, W
& = & 
0
\label{inv-12}\\
V^{(N) \, T} \cdot W \, + \, {\cal S}_{NN} \, U
& = &
1
\label{inv-22}
\end{eqnarray}
{\em Assuming that $G^{(N)}$ is invertible}, eq. (\ref{inv-21}) is solved into 
\begin{eqnarray}
W & = & U \, (G^{(N)})^{-1} \cdot V^{(N)} 
\label{sol21}
\end{eqnarray}
Substituting (\ref{sol21}) into eq. (\ref{inv-22}) we get:
\begin{eqnarray}
1 & = & 
U \, 
\left[ 
 V^{(N) \, T} \cdot (G^{(N)})^{-1} \cdot V^{(N)} \, + \, {\cal S}_{NN}
\right]
\label{sol22}
\end{eqnarray}
i.e.
\begin{equation}
  U = \frac{1}{\Delta_{\tn}} = (-1)^{N-1} \, \frac{\detg}{\dets}
  \label{sol220}
\end{equation}
At this level, the solution of eq.~(\ref{reform1}) is fully determined. Nevertheless, it is interesting to continue and compute the remaining parameter $Q$. For that,
let us substitute (\ref{sol21}) and (\ref{sol220}) into eq. 
(\ref{inv-11}) which becomes:
\begin{eqnarray}
- \, G^{(N)} \cdot Q 
& = & 
\mathds{1}_{N-1} \, - \, 
U \, V^{(N)} \cdot V^{(N) \, T} \cdot (G^{(N)})^{-1} 
\end{eqnarray}
i.e.
\begin{eqnarray}
Q 
& = & 
- \, (G^{(N)})^{-1} 
\, + \, U \, 
\left[ (G^{(N)})^{-1} \cdot V^{(N)} \right] \cdot 
\left[ (G^{(N)})^{-1} \cdot V^{(N)} \right] ^{T} 
\label{sol11}
\end{eqnarray}
As a consistency check, we substitute (\ref{sol11}) into the l.h.s of eq. 
(\ref{inv-12}) which becomes: 
\begin{eqnarray}
\mbox{l.h.s. (\ref{inv-12})}
& = &
- \, (G^{(N)})^{-1} \cdot V^{(N)}
\left\{
 1 \, - \, 
 U \, 
 \left[ 
  V^{(N) \, T} \cdot (G^{(N)})^{-1} \cdot V^{(N)} \, + \, {\cal S}_{NN}
 \right]
\right\}
\nonumber\\
& = &
- \, (G^{(N)})^{-1} \cdot V^{(N)} \,
\underbrace{
 \left[ 1 - U \, (-1)^{N-1} \frac{\dets}{\detg} \right] 
}_{0}
\; = \; 0
\label{test}
\end{eqnarray}
i.e. eq. (\ref{inv-12}) is indeed fulfilled.

\vspace{0.3cm}

\noindent
In a nutshell, to solve eq. (\ref{reform1}), we use results (\ref{eqsoleqshbhehj0}), (\ref{eqsoleqshbhehj1}), (\ref{sol21}), (\ref{sol220})
and read:
\begin{align}
\beta^{(N)} 
&= 
B \, (G^{(N)})^{-1} \cdot V^{(N)} \;\;,\;\; 
\label{inv-inv1}\\
B 
&= (-1)^{N-1} \, \frac{\detg}{\dets}
\label{inv-inv2}
\end{align}
Note that the identity (\ref{inv-inv2}) has already been given in ref.~\cite{Binoth:2005ff} and \cite{Binoth:1999sp}.
In term of $\bbar_i = b_i \, \det{\cals}$, the solutions given by eqs. (\ref{inv-inv1}) and (\ref{inv-inv2}) become
\footnote{Instead of subtracting column and line $N$ while solving eq. (\ref{reform1})
we could have 
subtracted any other column and line $a \neq N$. Eq. (\ref{eqsolbbar}) would then 
\begin{align*}
\bbar_{j \neq a} 
= 
(-1)^{N-1} \, 
\detg \, \left( \left[ G^{(a)} \right]^{-1} \cdot V^{(a)} \right)_{j}
&,&
\overline{b}_{a} 
= (-1)^{N-1} \, \detg \, - \sum_{j \neq a} \bbar_{j} 
\end{align*}
The column vector $\bbar$ defined by (\ref{eqsolbbar}) seems to depend on 
the choice of the line and column singled out but actually it does {\em not}. 
When $\det \, {\cal S} \neq 0$, $\bbar$ fulfils ${\cal S} \cdot \bbar = (\det \, {\cal S}) \, e$
whose solution is manifestly $a$-independent
.}:

\begin{align}
\bbar_{j \neq N} 
= 
(-1)^{N-1} \, 
\detg \, \left( \left[ G^{(N)} \right]^{-1} \cdot V^{(N)} \right)_{j}
&,&
\overline{b}_{N} 
= (-1)^{N-1} \, \detg \, - \sum_{j \neq N} \bbar_{j} 
\label{eqsolbbar}
\end{align}
which is indeed the results used in sec.~\ref{sectthreepoint} and \ref{sectfourpoint}.

\vspace{0.3cm}

\noindent
Although eqs.~(\ref{eqe10-02}) and (\ref{eqsolbbar}) provide the main results of this appendix, we will discuss now the peculiar cases where some 
determinants vanish limiting ourselves to cases that do not lead to Landau singularities and whose associated kinematics is relevant to colliders.

\subsection{Case $\dets \ne 0$ and $\detg = 0$}\label{onevanish}

We now assume  $\detg = 0$, therefore we can no longer proceed as in
sec. \ref{nevervanish} to solve eq. (\ref{inv-11}) and (\ref{inv-21}). 
The system $\{$(\ref{inv-21}),(\ref{inv-22})$\}$ is identical with the
resolution of ${\cal S} \cdot b = e$ in terms of $G^{(N)}$ and $V^{(N)}$ in the
case at hand.
When $\dets \neq 0$ and 
$\detg = 0$, $B$ vanishes, cf.\ eq. (\ref{inv-inv2}), from which we infer 
\begin{equation}\label{equ}
U=0
\end{equation}
To solve eq. (\ref{inv-21}) $W$ shall thus belong to $\mbox{Ker} \, G^{(N)}$. 
In the case at hand 
$\mbox{Ker} \, G^{(N)}$ is one-dimensional\footnote{For $N=3$, $4$, $\dim\left( \text{Ker} \, G^{(N)}  \right) > 1$ leads 
to degenerate kinematics cf.\ the discussion at the end of the subsec.~\ref{bothvanish}} and it is the eigenspace of 
$\mbox{Cof}[G^{(N)}]$ associated with the only non vanishing eigenvalue 
$\gamma_{1}$ of the 
latter. Let $u_{1}$ be a normalised vector spanning $\mbox{Ker} \, G^{(N)}$. 
We parametrise 
\begin{equation}\label{eqv}
W = \nu \, u_{1}
\end{equation}
where $\nu$ is determined by solving eq.(\ref{inv-22}) which, using
(\ref{equ}), reads:
\begin{eqnarray}
\nu & = & \left( V^{(N) \, T} \cdot u_{1} \right)^{-1}
\label{inv-22aa}
\end{eqnarray}
We notice that $V^{(N) \, T} \cdot u_{1} \neq 0$ since otherwise
eq.~(\ref{inv-22}) cannot be fulfilled\footnote{
Notice also that since $\mbox{Cof}[G^{(N)}]$ is a real symmetric matrix, $\mbox{Cof}[G^{(N)}] = \gamma_1 \, u_1 \cdot u_1^{T}$ and then
$\dets = (-1)^{N-1} \, \gamma_{1} \, (V^{(N) \, T} \cdot u_{1})^{2}$ which is by assumption different from zero, so
$V^{(N) \, T} \cdot u_{1} \neq 0$.}.
At this level, we hold all the information to solve eq. (\ref{reform1}). Nevertheless, let us go on and also determine the parameter $Q$.

\vspace{0.3cm}

\noindent
For that, we substitute the expression of $W$ into eq. (\ref{inv-11}) which becomes:
\begin{eqnarray}
- \, G^{(N)} \cdot Q 
& = & 
\mathds{1}_{N-1} \, - \,  
\frac{V^{(N)} \cdot u_{1}^{T}}{V^{(N) \, T} \cdot u_{1}}
\label{inv-22aaa}
\end{eqnarray}
We look for the pseudo inverse, more precisely the Moore-Penrose pseudo inverse (MPPI)
\footnote{{\em Any} 
other pseudo inverse could also 
be used yet the MPPI provides extra properties - namely the MPPI $H$ is
symmetric and commutes with $G^{(N)}$, and $G^{(N)} \cdot H$ is symmetric and 
nilpotent i.e. a projector - convenient for further use.} 
\cite{barnett1990} of $G^{(N)}$, i.e.\ $H$ such that:
\begin{align}
  G^{(N)} \cdot H \cdot G^{(N)} &= G^{(N)} \label{eqmppicond1} \\ 
  H \cdot G^{(N)} \cdot H &= H \label{eqmppicond2} \\
  \left( G^{(N)} \cdot H \right)^{\dagger} &= G^{(N)} \cdot H  \label{eqmppicond3} \\
  \left( H \cdot G^{(N)} \right)^{\dagger} &= H \cdot G^{(N)}  \label{eqmppicond4}
\end{align}
In order for (\ref{inv-22aaa}) to admit solutions the following compatibility
condition has to hold:
\[
  \left[ \mathds{1}_{N-1} - G^{(N)} \cdot H \right] \cdot 
\left[ 
  \mathds{1}_{N-1} \, - \,  
 \frac{V^{(N)} \cdot u_{1}^{T}}{V^{(N) \, T} \cdot u_{1}}
\right]
= 0
\]
$[\mathds{1}_{N-1} - G^{(N)} \cdot H]$ is the (orthogonal) projector onto 
$\mbox{Ker} \, G^{(N)}$ i.e. 
$[\mathds{1}_{N-1} - G^{(N)} \cdot H] = u_{1} \cdot  u_{1}^{T}$ thus we have
indeed
\[
u_{1} \cdot u_{1}^{T} \cdot 
\left[ 
  \mathds{1}_{N-1} \, - \,  
 \frac{V^{(N)} \cdot u_{1}^{T}}{V^{(N) \, T} \cdot u_{1}}
\right]
= 
0
\]
and $Q$ is of the form
\begin{eqnarray}
Q 
& = &
- \, H \cdot 
\left[ 
  \mathds{1}_{N-1} \, - \,  
 \frac{V^{(N)} \cdot u_{1}^{T}}{V^{(N) \, T} \cdot u_{1}}
\right] + \, u_{1} \cdot  n^{T}
\label{sola}
\end{eqnarray}
where $n$ is a so far arbitrary (complex) $(N-1)$-vector. 
Eq.~(\ref{sola}) is a matricial counterpart of the usual unknown $(N-1)$-vector 
spanning Ker $G^{(N)}$ in the singular vectorial case (it is ``the general 
solution of the homogeneous equation $G^{(N)} \cdot X = 0$").
We shall further request $Q$ to be symmetric (as $(\widehat{\cals}^{(N)})^{-1}$
has to be so!). We have:
\begin{eqnarray}
Q - Q^{T} 
& = &
\frac{1}{u_{1}^{T} \cdot V^{(N)}} \, 
\left[ H \cdot V^{(N)} \cdot u_{1}^{T} - u_{1} \cdot V^{(N) \, T} \cdot H \right]
+
\left[ u_{1} \cdot n^{T} - n \cdot u_{1}^{T} \right]
\nonumber\\
& = &
 \left[ 
  \frac{1}{u_{1}^{T} \cdot V^{(N)}} \, H \cdot V^{(N)} - n 
 \right] \cdot 
  u_{1}^{T} 
 -
 u_{1} \cdot 
 \left[
  \frac{1}{u_{1}^{T} \cdot V^{(N)}} \, H \cdot V^{(N)} - n 
 \right]^{T} 
\label{sym}
\end{eqnarray}
This vanishes if and only if
\[
\frac{1}{u_{1}^{T} \cdot V^{(N)}} \, H \cdot V^{(N)} - n 
=
\xi \, u_{1}
\]
i.e.
\begin{equation}\label{n}
n 
= 
\frac{1}{u_{1}^{T} \cdot V^{(N)}} \, H \cdot V^{(N)} - \xi \, u_{1}
\end{equation}
where the scalar $\xi$ is fixed by substituting into eq. (\ref{inv-12}). 
We get:
\[
- H \cdot V^{(N)} 
\left[ 1 - \frac{(u_{1}^{T} \cdot V^{(N)})}{(u_{1}^{T} \cdot V^{(N)})} \right]
+ 
\left[ 
  \frac{V^{(N) \, T} \cdot H \cdot V^{(N)}}{(u_{1}^{T} \cdot V^{(N)})}
  + \frac{{\cal S}_{NN}}{(u_{1}^{T} \cdot V^{(N)})} 
  - \xi \, (u_{1}^{T} \cdot V^{(N)})
\right] 
= 0
\]
i.e.
\begin{equation}\label{xi}
\xi 
= 
\frac{1}{(u_{1}^{T} \cdot V^{(N)})^{2}} \,
\left[ {\cal S}_{NN} \, + \, V^{(N) \, T} \cdot H \cdot V^{(N)} \right]
\end{equation}
Collecting the results of eqs. (\ref{sola}), (\ref{n}) and (\ref{xi}), the parameter $Q$ is given by:
\begin{align}
  Q &= - H \cdot \left[ 
  \mathds{1}_{N-1} \, - \,  
\frac{V^{(N)} \cdot u_{1}^{T}}{V^{(N) \, T} \cdot u_{1}} \right]  \notag \\
&\quad {} + \frac{1}{u_1^T \cdot V^{(N)}}\, u_1 \cdot \left\{ 
  V^{(N)} \cdot H - \frac{\left[  \cals_{NN} + V^{(N) \, T} \cdot H \cdot V^{(N)} \right]}{u_1^T \cdot V^{(N)}} \right\}
  \label{eqresultQ0}
\end{align}
which completes the extraction of $(\widehat{\cal S}^{(N)})^{-1}$.
Using eqs. (\ref{eqresultQ0}), we can verify that eq. (\ref{inv-21}) is indeed verified.

\vspace{0.3cm}

\noindent
To sum up, putting together the results of eqs. (\ref{equ})-(\ref{inv-22aa}) and (\ref{eqresultQ0}),
the solution of eq. (\ref{reform1}) is given by:
\begin{eqnarray}
B 
& = & 
0
\label{inv-noninv1}\\
\beta^{(N)} 
& = & 
\frac{1}{(V^{(N) \, T} \cdot u_{1})} \, u_{1}
\label{inv-noninv2}
\end{eqnarray}
with $u_1$ a normalised vector spanning $\text{Ker} \, G^{(N)}$.
Note that the solution given by eqs. (\ref{inv-inv1}) and (\ref{inv-inv2}) smoothly matches the solution given by eqs. (\ref{inv-noninv1}), (\ref{inv-noninv2}) in the limit $\detg \to 0$ 
as shown in section \ref{sectcomment1}, cf. eqs. (\ref{newB}), (\ref{vlimcont2}).

\subsection{Case $\dets = 0$ and $\detg = 0$}\label{bothvanish}

We seek to express the MPPI
$\widehat{\Sigma}$ of $\widehat{\cal S}^{(N)}$ in terms of $V^{(N)}$ and the 
MPPI $H$ of $G^{(N)}$. We parametrise $\widehat{\Sigma}$ as
\begin{eqnarray}
\widehat{\Sigma}
& = &
\left[
 \begin{array}{ccc}
    Q   & | & W \\
   --   & + & - \\
  W^{T} & | & U \\
 \end{array}
\right]
\label{invshatb}
\end{eqnarray}
we thus have:
\begin{eqnarray}
\widehat{\cal S}^{(N)} \cdot \widehat{\Sigma}
& = &
\left[
 \begin{array}{ccc}
- \, G^{(N)} \cdot Q+V^{(N)} \cdot W^{T}&|&- \, G^{(N)} \cdot W+U \, V^{(N)}\\
   ------------                         &+&  ------------ \\
(Q \cdot V^{(N)}+{\cal S}_{NN} \, W)^{T}&|&V^{(N) \, T} \cdot W+{\cal S}_{NN} \, U\\
 \end{array}
\right]
\label{prodssig}
\end{eqnarray}
Let us warn the reader that we will not determine the MPPI $\widehat{\Sigma}$ 
in general. Since we are interested in solving the eq. $S \cdot  b = e$, 
we restrict our\-selves here to the case when this equation admits solutions 
i.e. when the compatibility condition is fulfilled. 
The compatibility condition for (\ref{reform1}) in terms of the MPPI 
is
$[\mathds{1}_{N} - \widehat{\cal S}^{(N)} \cdot \widehat{\Sigma}] 
\cdot \widehat{e} = 0$. This condition explicitly reads:
\begin{eqnarray}
- \, G^{(N)} \cdot W + U \, V^{(N)} & = & 0
\label{ccomp1}\\
V^{(N) \, T} \cdot W + {\cal S}_{NN} \, U & = &  1
\label{ccomp2}
\end{eqnarray}
Furthermore, the Moore-Penrose extra conditions require 
$\widehat{\cal S}^{(N)} \cdot \widehat{\Sigma}$ to be a projector:
it shall be symmetric, implying
\begin{eqnarray}
Q \cdot V^{(N)} + S_{NN} \, W & = & 0
\label{ccomp3}\\
\left[ - \, G^{(N)} \cdot Q + V^{(N)} \cdot W^{T} \right]^{T} 
& = &
- \, G^{(N)} \cdot Q + V^{(N)} \cdot W^{T}
\label{ccomp4}
\end{eqnarray}
together with the nilpotent request:
\begin{eqnarray}
\left[ - \, G^{(N)} \cdot Q + V^{(N)} \cdot W^{T} \right]^{2} 
& = &
- \, G^{(N)} \cdot Q + V^{(N)} \cdot W^{T}
\label{ccomp5}
\end{eqnarray}
Together with MP conditions 
(\ref{ccomp3})-(\ref{ccomp5}), compatibility conditions 
(\ref{ccomp1})-(\ref{ccomp2}) make the extraction of $\widehat{\Sigma}$  
easier.

\vspace{0.3cm}

\noindent
A solution such that $U=0$ would require
$W$ to belong to Ker $G^{(N)}$ in order to fulfil (\ref{ccomp1}) and not 
be orthogonal to $V^{(N)}$ in order to fulfil (\ref{ccomp2}).
Let us assume Ker $G^{(N)}$ to be one-dimensional and spanned by a normalised 
vector $u_{1}$: one would have
$\mbox{Cof}[G^{(N)}] = \gamma_{1} \, u_{1} \cdot  u_{1}^{T}$ with $\gamma_{1}
\neq 0$, thus 
$V^{(N) \, T} \cdot \mbox{Cof}[G^{(N)}] \cdot V^{(N)} = \gamma_{1} 
(u_{1}^{T}\cdot V^{(N)})^{2} \neq 0$, which would be inconsistent with 
 $\dets = 0$ and $\detg = 0$ simultaneously. Consequently:
\begin{itemize}
\item
either $U=0$ and Rank $G^{(N)}$ has to be $\leq N-3$ thus 
$\mbox{Cof}[G^{(N)}] = 0$ identically,
which corresponds to degenerate kinematics; 
\item
or Rank $G^{(N)} = (N-2)$, in which case $U$ cannot be 0.
\end{itemize}
Let us ignore the first alternative and focus on the second one.
Indeed, the former leads to degenerate kinematics for the case $N = 3,4$
not relevant for colliders. Nevertheless, as explained in the introduction, since a two-loop scalar N-point function
can be written as a sum of double integrals whose integrands are some ``generalised" one-loop Feynman integrals,
the kinematics of the latter depends on two parameters. So, one may wonder if such degenerate kinematics is met
when these parameters run over the unit square.
If relevant, this issue will be tackled later on.

\vspace{0.3cm}

\noindent
In order for (\ref{ccomp1}) to admit solutions, the following compatibility
condition is fulfilled: 
\begin{equation}\label{compcondG}
  \left[ \mathds{1}_{N-1} - G^{(N)} \cdot H \right] \cdot V^{(N)} = 0
\;\; \Leftrightarrow V^{(N)} \perp \mbox{Ker} \, G^{(N)}
\end{equation}
$H$ being the MPPI
of $G^{(N)}$. 
Eq. (\ref{ccomp1}) is then solved in the form:
\begin{equation}\label{solvred3}
W = U \, H \cdot V^{(N)} + k, \;\; 
k \in \mbox{Ker} \, G^{(N)} 
\end{equation}
So far $k$ remains arbitrary.
Notwithstanding, $(V^{(N) \, T} \cdot k) = 0$, substituting 
(\ref{solvred3}) into (\ref{ccomp2}) thus yields:
\begin{equation}\label{solvred4}
U \, 
\left[ 
 {\cal S}_{NN} + \left( V^{(N) \, T} \cdot H \cdot V^{(N)} \right)
\right] 
 = 1
\end{equation}
Let us define $\Omega$ as the factor of $U$ in eq.~(\ref{omega}), namely:
\begin{align}
  \Omega &= 
\left[ 
 {\cal S}_{NN} + \left( V^{(N) \, T} \cdot H \cdot V^{(N)} \right)
\right] 
  \label{eqdefomega}
\end{align}
Since $\widehat{\cal S}^{(N)}$ does admit one unique MPPI 
$\widehat{\Sigma}$, whose $U$ has to obey (\ref{solvred4}), the combination $\Omega$ 
involved in (\ref {solvred4}) {\em is necessarily non zero}, and 
(\ref {solvred4}) fixes $U=\Omega^{-1}$. 

\vspace{0.3cm}

\noindent
Let us now determine $Q$ fulfilling (\ref{ccomp3})-(\ref{ccomp5}).
Using(\ref{ccomp1})-(\ref{ccomp3}), the l.h.s. of 
(\ref{ccomp5}) may be expanded as
\begin{eqnarray}
\lefteqn{\left[ - \, G^{(N)} \cdot Q + V^{(N)} \cdot W^{T} \right]^{2}} 
\nonumber\\
& = &
\left( G^{(N)} \cdot Q \right)^{2} - 
G^{(N)} \cdot \left( Q \cdot  V^{(N)} \right) \cdot W^{T}
\nonumber\\
&&
- \; 
V^{(N)} \cdot \left( W^{T} \cdot G^{(N)} \right) \cdot Q + 
V^{(N)} \cdot \left( W^{T} \cdot V^{(N)} \right) \cdot W^{T}
\nonumber\\
& =&
\left( G^{(N)}  \cdot Q \right)^{2} + 
{\cal S}_{NN} \, U \, V^{(N)} \cdot W^{T} - 
U \, V^{(N)} \cdot V^{(N) \, T} \cdot Q +
\left( 1 - {\cal S}_{NN} \, U \right) \, V^{(N)} \cdot W^{T} 
\nonumber
\end{eqnarray}
Thus condition (\ref{ccomp5}) may be replaced by:
\begin{eqnarray}
\left( G^{(N)} \cdot Q \right)^{2} + \left( G^{(N)} \cdot Q \right) - 
U \, V^{(N)} \cdot V^{(N) \, T} \cdot Q
& = &
0
\label{ccomp5b}
\end{eqnarray}
Inspired by the cases where $\widehat{\cal S}^{(N)}$ is invertible,
let us seek $Q$ of the form
\[
Q = - H + y \cdot  y^{T}
\]
with the $(N-1)$-column vector $y$ to be determined. We get:
\begin{eqnarray}
\left( G^{(N)} \cdot Q \right) \;
& = &
- \left( G^{(N)} \cdot H \right) \; +  \; 
G^{(N)} \cdot y \cdot y^{T}
\label{out1}\\
\left( G^{(N)} \cdot Q \right)^{2}
& = &
\;\;\;\,
\left( G^{(N)} \cdot H \right)^{2} - 
\left( G^{(N)} \cdot H \cdot G^{(N)} \right) \cdot y \cdot y^{T} 
\nonumber\\
&&
+ \, \left( G^{(N)} \cdot y \right) \cdot 
\left[ 
 - \, \left( y^{T} \cdot G^{(N)} \right) \cdot H +
 \left( y^{T} \cdot G^{(N)} \cdot y \right) \, y^{T}
\right]
\label{out2}
\end{eqnarray}
where $H$ is the pseudo inverse of $G^{(N)}$, $(G^{(N)} \cdot H)$ is nilpotent (cf. eqs.~(\ref{eqmppicond1}) and (\ref{eqmppicond2})) and 
``(\ref{out2})+(\ref{out1})" simplifies: 
\begin{equation}\label{out1p2}
\left( G^{(N)} \cdot Q \right)^{2} + \left( G^{(N)} \cdot Q \right)
= 
\left( G^{(N)} \cdot y \right) \cdot 
\left[ 
 - \, \left( y^{T} \cdot G^{(N)} \right) \cdot H +
 \left( y^{T} \cdot G^{(N)} \cdot y \right) \, y^{T}
\right]
\end{equation}
and (\ref{ccomp5b}) reads:
\begin{eqnarray}
\lefteqn{
\left( G^{(N)} \cdot y \right) \cdot 
\left[ 
 - \, \left( y^{T} \cdot G^{(N)} \right) \cdot H +
 \left( y^{T} \cdot G^{(N)} \cdot y \right) \, y^{T}
\right] 
}
\nonumber\\
&&
- \; U \, V^{(N)} \cdot 
\left[ 
 - \, V^{(N) \, T} \cdot H + \left( V^{(N) \, T} \cdot y \right) \cdot y^{T} 
\right]
\, = \,
0
\label{ccomp5c}
\end{eqnarray}
At this point we recall that $V^{(N)}$ fulfils the compatibility condition 
(\ref{compcondG}) and we make the following Ansatz:
\begin{equation}\label{ansatzv}
y = \zeta \, H \cdot V^{(N)}
\end{equation}
with $\zeta$ to be determined.
Let us note that actually, since $Q$ quadratically depends on $y$, only $\zeta^{2}$
not $\zeta$ {\em per se} is involved in the expression of $Q$.
The l.h.s. of eq.(\ref{ccomp5c}), with the help of eq.~(\ref{compcondG}), becomes:
\begin{eqnarray}
  \lefteqn{\text{l.h.s.} \; (\ref{ccomp5c})}
\nonumber\\
& = &
\zeta^{2} 
\left( G^{(N)} \cdot H \cdot V^{(N)} \right) \, 
\nonumber\\
&& \mbox{} \cdot
\left[ 
 - \, \left( V^{(N) \, T} \cdot H \cdot G^{(N)} \right) \cdot H 
 +
 \zeta^{2} \, 
 \left( V^{(N) \, T} \cdot H \cdot G^{(N)} \cdot H \cdot V^{(N)} \right) \cdot 
 V^{(N) \, T} \cdot H
\right] 
\nonumber\\
&&
- \; U \, V^{(N)} \cdot 
\left[ 
 - \, V^{(N) \, T} \cdot H + 
 \zeta^{2} \left( V^{(N) \, T} \cdot H \cdot V^{(N)} \right) \, 
 V^{(N) \, T} \cdot H
\right]
\nonumber\\
& = &
\left( \zeta^{2} - U \right) \,
\left[ - 1 + \zeta^{2} \left( V^{(N) \, T} \cdot H \cdot V^{(N)} \right) \right]
\, V^{(N)} \cdot V^{(N) \, T} \cdot H
\label{ccomp5d}
\end{eqnarray}
which vanishes for
\begin{equation}\label{zeta2}
\zeta^{2}= U
\end{equation}
We shall now enforce the symmetry property (\ref{ccomp4}):
\begin{eqnarray}
\lefteqn{
- \, G^{(N)} \cdot Q + V^{(N)} \cdot W^{T}
}
\nonumber\\
& = &
- \, G^{(N)} \cdot 
\left[ 
 - \, H + \, 
 U \, \left( H \cdot V^{(N)} \right) \cdot \left( H \cdot V^{(N)} \right)^{T}
\right]
+ \,  U \, V^{(N)} \cdot V^{(N) \, T} \cdot H + V^{(N)} \cdot k^{T}
\nonumber\\
& = &
- \, G^{(N)} \cdot H + V^{(N)} \cdot k^{T}
\label{testsym}
\end{eqnarray}
where the last equality is obtained using the compatibility condition 
(\ref{compcondG}).
Due to the fact that the MPPI $H$ of $G^{(N)}$ is symmetric and 
does commute with $G^{(N)}$,
the r.h.s. of (\ref{testsym}) is thus symmetric if and only if $k = 0$ - which now 
completely fixes $W$.

\vspace{0.3cm}

\noindent
Last, we shall check whether (\ref{ccomp3}) is fulfilled.
\begin{eqnarray}
\lefteqn{Q \cdot V^{(N)} + {\cal S}_{NN} \, W}
\nonumber\\ 
& = &
\left[ 
 - \, H +  
 U \, \left( H \cdot V^{(N)} \right) \cdot \left( H \cdot V^{(N)} \right)^{T}
\right] \cdot V^{(N)} + {\cal S}_{NN} \, U \, H \cdot V^{(N)}
\nonumber\\
& = & 
\left[ 
 -1 + U \, 
 \left( {\cal S}_{NN} + V^{(N) \, T} \cdot H \cdot V^{(N)} \right)
\right]
\, H \cdot V^{(N)}
\end{eqnarray}
which indeed vanishes by virtue of (\ref{solvred4}). q.e.d.

\vspace{0.3cm}

\noindent
We have to distinguish between the two cases.

\subsubsection{$\dets = 0$ and $[\mathds{1}_{N} - \widehat{\cal S}^{(N)} \cdot \widehat{\Sigma}]\cdot \widehat{e} = 0$}

If $\dets = 0$, yet with 
$[\mathds{1}_{N} - \widehat{\cal S}^{(N)} \cdot \widehat{\Sigma}] 
\cdot \widehat{e} = 0$ then $\detg = 0$. 
Indeed, were $\detg$ non zero, the solution of eq. 
(\ref{ccomp1}) would be given as in eq. (\ref{inv-inv1}) by 
$\beta^{(N)} = U \, (G^{(N)})^{-1} \cdot V^{(N)}$, whose substitution 
into eq. (\ref{ccomp2}) would again lead to 
$U \, [V^{(N) \, T} \cdot (G^{(N)})^{-1} \cdot V^{(N)} + {\cal S}_{NN}] 
= U \, (-1)^{N-1} \, \dets/\detg = 1$. However, as the 
coefficient of $U$ would vanish, no $U$ could satisfy the latter 
equation, and thus eqs. (\ref{ccomp1}), (\ref{ccomp2}) would have no
solution; this is likewise for (\ref{reform1}), which would contradict the 
compatibility condition.   
Let us restrict ourselves to the case ``dim Ker $G^{(N)} = 1$" (cf. 
the discussion in this subsection). We can then use the results of 
this section to express $\widehat{\Sigma} \cdot \widehat{e}$:
\begin{eqnarray}
U
& = &
\Omega^{-1} \; = \; \left[ {\cal S}_{NN} + 
\left( V^{(N) \, T} \cdot H \cdot V^{(N)} \right) \right]^{-1}
\label{noninv-noninv1}\\
W
& = &
\Omega^{-1} \; H \cdot V^{(N)}
\label{noninv-noninv2}
\end{eqnarray}
The solutions of eq.~(\ref{reform1}) are given by:
\begin{align}
  \widehat{b} &= \widehat{\Sigma} \cdot \widehat{e} + \left[ \mathds{1}_{N} - \widehat{\cal S}^{(N)} \cdot \widehat{\Sigma} \right] \cdot n
  \label{eqsolsbeqe0}
\end{align}
where $n$ is an arbitrary $N$ vector. In terms of the block components of the $\widehat{\Sigma}$ matrix, these solutions become:
\begin{align}
  \beta^{(N)} &= \Omega^{-1} \, H \cdot V^{(N)} + \left( \mathds{1}_{N-1} - G^{(N)} \cdot H \right) \cdot \widetilde{n} \label{eqsolsbeqe1} \\
  B &= \Omega^{-1}
  \label{eqsolsbeqe2}
\end{align}
where $\widetilde{n}$ is a $N-1$ vector formed by the $N-1$ first components of $n$.
The fact that $\detg$ has to vanish can be also seen as 
a direct consequence of the compatibility condition required for 
eq. (\ref{eqdefbcoeff}) equivalent to eq. (\ref{reform1}):
\[
  \left[ \mathds{1}_{N}-{\cal S} \cdot \Sigma \right] \, \cdot e = 0
\]
with $\Sigma$ the MPPI of $\cals$.
If ${\cal S}$ is {\em real}, as can be shown e.g. by diagonalisation of 
${\cal S}$ (and $\Sigma$) this compatibility condition is equivalent to
$e \perp \mbox{Ker} \, {\cal S}$ (where ``$\perp$" is understood as the Euclidean 
scalar product $u \cdot v = \sum_{i=1}^{N} u_{i} v_{i}$): this means that 
{\em all} $u \in \mbox{Ker} \, {\cal S}$
fulfil the condition $e^{T} \cdot u = 0$. This condition has been shown in
appendix E of ref.~\cite{Guillet:2013mta} to imply the existence of an eigenvector\footnote{Such a vector 
can be formed by taking the $N-1$ first components
of any non vanishing vector in Ker ${\cal S}$.} for $G^{(N)}$ with 
corresponding vanishing eigenvalue , hence $\detg = 0$. 

\vspace{0.3cm}

\noindent
This can be extended to a {\em complex} ${\cal S}$ as 
shown in what follows. Albeit not real ${\cal S}$ is symmetric thus admits
a Takagi decomposition \cite{autonne,takagi,siegel,hua,schur,benedetti,Horn:2012} of the form
${\cal S} = F \cdot {\cal S}_{D} \cdot F^{T}$, where ${\cal S}_{D} = $
diag$(s_{j}, j=1,\cdots,N)$ is a real non negative diagonal matrix and $F$ 
is a $N \times N$ unitary matrix. Let us note 
$f_{k}$ the column vectors of $F$ so that the Takagi decomposition of 
${\cal S}$  reads:
\[
{\cal S} = \sum_{j \in J} s_{j} \, f_{j} \cdot f_{j}^{T} 
\]
where $J$ is the set of index values corresponding to the non zero
elements of ${\cal S}_{D}$. Let us also define the set of index values $K$ corresponding to 
vanishing elements of ${\cal S}_{D}$.  
The $f_{j}, j \in K$  are mutually orthogonal, normalised eigenvectors of 
${\cal S}$ associated with 
eigenvalue zero i.e. form an orthonormal basis of Ker ${\cal S}$. A Takagi 
decomposition of the MPPI $\Sigma$ of ${\cal S}$ is provided by:
\[
\Sigma = \sum_{l \in J} s_{l}^{-1} \, f_{l}^{*} \cdot f_{l}^{\dagger}
\]
so that
\[
  \left[ \mathds{1}_{N} \, - \, {\cal S} \cdot \Sigma \right]
=
\sum_{l \in K} f_{l} \cdot f_{l}^{\dagger}
\]
i.e. $[ \mathds{1}_{N} \, - \, {\cal S} \cdot \Sigma]$ is the (orthogonal) 
projector onto Ker ${\cal S}$ and the consistency condition 
$\left[ \mathds{1}_{N} - {\cal S} \cdot \Sigma \right] \cdot e =0$ again means
i.e. $e \perp \mbox{Ker} \, {\cal S}$ (where``$\perp$" now  refers to the 
{\em Hermitian} product 
$u^{\dagger} \cdot v = \sum_{i=1}^{N} u_{i}^{*} v_{i}$).

\vspace{0.3cm}

\noindent
Whether ${\cal S}$ is real or complex, ``$e \perp \mbox{Ker} \, {\cal S}$" 
implies that ``$\det \, {\cal S}=0$" does {\em not} correspond to any 
kinematical singularity. Indeed such a singularity is characterised by a point 
satisfying the Landau conditions \cite{Itzykson:1980rh,eden2002analytic}, which may occur
either on the boundary of the domain of integration (end-point singularity 
as for an infrared or collinear singularity), or 
inside the domain of integration and where a pinching of the integration
hypercontour occurs (pinching singularity as e.g. for a threshold singularity).
To such a point corresponds an eigenvector $x \in$ Ker ${\cal S}$ belonging to 
the first hyperquadrant defined by $\{ \sum_{j=1}^{N} x_{j} > 0, \; \;
x_{i} \geq 0 \;\; \mbox{for all} \; i=1,\cdots,N \}$. 
In contrast,
in the present case, ``$e \perp \mbox{Ker} \, {\cal S}$" prevents any element 
of $\mbox{Ker} \, {\cal S}$ from belonging to this hyperquadrant.
Let us notice that the converse is not true: there are situations where 
$e$ may not be ``$\perp \mbox{Ker} \, {\cal S}$" yet no vector of 
Ker ${\cal S}$ may belong to the first hyperquadrant  either. 
For instance, the fake threshold at
$s=(m_{1}-m_{2})^{2}$ in the one-loop two-point function with distinct internal 
masses $m_{1}^{2} \neq m_{2}^{2}$: all non vanishing two component-vectors of 
$\mbox{Ker} \, {\cal S}$ are found to have components with opposite signs, which
means that the point in the integrand of the one-loop function fulfilling the 
Landau conditions lies outside the integration domain.

\subsubsection{$\dets = 0$ and $[\mathds{1}_{N} - \widehat{\cal S}^{(N)} \cdot \widehat{\Sigma}]\cdot \widehat{e} \ne 0$}

If $\dets = 0$ but the condition 
  $[\mathds{1}_{N}- \widehat{\cal S} \cdot \widehat{\Sigma}] \cdot \widehat{e} = 0$
is not fulfilled, eq. (\ref{reform1}) has no solution. 
This situation occurs 
in cases of practical relevance such the one leading to Landau singularities. 
In deriving the representation of $I^4_{N}$ using the Stokes-type
identity, the coefficients $\bbar$, 
defined when 
$\detg \neq 0$ by
\begin{eqnarray}
\overline{b}_{j \neq a} 
= 
(-1)^{N-1} \, 
\detg \, \left( \left[ G^{(a)} \right]^{-1} \cdot V^{(a)} \right)_{j}
&,&
\overline{b}_{a} 
= (-1)^{N-1} \, \detg \, - \sum_{j \neq a} \overline{b}_{j} 
\label{bbar}
\end{eqnarray}
naturally emerge. When 
$\dets = 0$ and 
$[\mathds{1}_{N}- \widehat{\cal S} \cdot \widehat{\Sigma}] \cdot \widehat{e} \ne 0$,
there is no $b$ fulfilling ${\cal S} \cdot b = e$ yet $\overline{b}$ defined 
by (\ref{bbar}) still exists and each of its components can be seen as the 
limits of the previous case eqs.~(\ref{eqsolbbar}) when $\dets \to 0$, and $\overline{b}$ 
fulfils ${\cal S} \cdot \overline{b} = 0$.

\subsection{Continuity between solutions given by eqs. (\ref{inv-inv1}), (\ref{inv-inv2}) and eqs. (\ref{inv-noninv1}), (\ref{inv-noninv2}) }\label{sectcomment1}

In spite of the {\em discontinuity in the algebraic
treatment} 
to compute the expression of $(\widehat{\cal S}^{(N)})^{-1}$ when 
$\detg = 0$ vs. $\neq 0$, we expect the {\em expression} of 
$(\widehat{\cal S}^{(N)})^{-1}$ itself to enjoy continuity in 
$\detg$. Let us investigate this issue. 

\vspace{0.3cm}

\noindent
The expression $B = (-1)^{N-1} \detg/\dets$ vanishes
continuously in $\detg$.

\vspace{0.3cm}

\noindent
Whatever $\detg$ is, we recall that the real symmetric 
matrices $G^{(N)}$ and $\mbox{Cof}[G^{(N)}]$, which fulfil
$G^{(N)} \cdot \mbox{Cof}[G^{(N)}] = 
\mbox{Cof}[G^{(N)}] \cdot G^{(N)} = 
\detg \, \mathds{1}_{N-1}$,
commute and can be diagonalised using a common Euclidean orthonormal eigenbasis
$\{u^{\prime}_{j}\}_{j = 1, \cdots, N-1}$:
\begin{eqnarray}
G^{(N)} 
& = &
\sum_{j=1}^{N-1} g_{j}^{\prime} \;\; 
u_{j}^{\prime} \cdot u_{j}^{\prime \, T} 
\nonumber\\
\mbox{Cof}[G^{(N)}] 
& = &
\sum_{j=1}^{N-1} \gamma_{j}^{\, \prime} \;\; 
u_{j}^{\prime} \cdot u_{j}^{\prime \, T} 
\nonumber\\
\mathds{1}_{N-1}
& = &
\sum_{j=1}^{N-1} u_{j}^{\prime} \cdot u_{j}^{\prime \, T} 
\nonumber
\end{eqnarray}
where $g_{j}^{\prime} \, \gamma_{j}^{\, \prime} = \detg$ for all
$j=1,\cdots, N-1$.
The label ``$_{j=1}$" labels the (unique) 
eigendirection corresponding to the eigenvalue
that makes $\detg$ vanish. The ``prime notation" stands for
the generic case $\detg~\neq~0$, with corresponding quantities computed
in subsec. \ref{onevanish},
whereas unprimed quantities are the respective values when $\detg$ 
vanishes. We introduce
\begin{equation}\label{Hprime}
H^{\prime} = 
\sum_{j=2}^{N-1} \left( g^{\prime}_{j} \right)^{-1} \, 
u_{j}^{\prime} \cdot u_{j}^{\prime \, T} 
\end{equation}
and rewrite
\begin{eqnarray}
\sum_{k=2}^{N-1} \gamma_{k}^{\, \prime} 
u_{k}^{\prime} \cdot u_{k}^{\prime \, T}
& = &
g_{1}^{\prime} \, \gamma_{1}^{\, \prime} \,
\sum_{k=2}^{N-1} \left(g_{k}^{\prime} \right)^{-1} \, 
u_{k}^{\prime} \cdot u_{k}^{\prime \, T} 
\nonumber\\
& = &
g_{1}^{\prime} \, \gamma_{1}^{\, \prime} \, H^{\prime}
\label{tool}
\end{eqnarray}
so that
\begin{equation}\label{tool2}
\mbox{Cof}[G^{(N)}] 
= 
\gamma_{1}^{\prime} \, 
\left[ 
 u_{1}^{\prime} \cdot u_{1}^{\prime \, T} + 
 g_{1}^{\prime} \, H^{\prime}
\right]
\end{equation}
We can rewrite $\dets$
\begin{eqnarray}
\dets 
& = & 
(-1)^{N-1} \, \gamma_{1}^{\, \prime} 
\left\{
 g_{1}^{\prime} \, 
\left[ 
 {\cal S}_{NN} + \left( V^{(N) \, T} \cdot H^{\prime} \cdot V^{(N)} \right)
\right]
+ \left( u_{1}^{\prime \, T} \cdot V^{(N)} \right)^{2}
\right\}
\label{detsnew}
\end{eqnarray}
It is convenient to introduce
\begin{equation}\label{omega}
\Omega^{\prime}
=
\left[ 
 {\cal S}_{NN} + \left( V^{(N) \, T} \cdot H^{\prime} \cdot V^{(N)} \right)
\right]
\end{equation}
The expression $B = (-1)^{N-1} \detg/\dets$ 
also takes the form:
\begin{eqnarray}
(-1)^{N-1} \, \frac{\detg}{\dets}
& = &
\frac{g_{1}^{\prime}}
{g_{1}^{\prime} \, \Omega^{\prime} + 
 \left( u_{1}^{\prime \, T} \cdot V^{(N)} \right)^{2}}
\label{newB}
\end{eqnarray}
The r.h.s. of eq. (\ref{newB}) vanishes continuously with $g_{1}^{\prime}$
while keeping $(u_{1}^{\prime \, T} \cdot V^{(N)}) \neq 0$. 

\vspace{0.3cm}

\noindent
When $\detg \neq 0$ the expression of 
$W^{\prime} = B \, (G^{(N)})^{-1} \cdot V^{(N)}$, rewritten
\begin{eqnarray}
W^{\prime} 
& = & 
\frac{1}
{g_{1}^{\prime}  \, \Omega^{\prime} + 
 (u_{1}^{\prime \, T} \cdot V^{(N)})^{2}}
\left\{
 u_{1}^{\prime} \; (u_{1}^{\prime \, T} \cdot V^{(N)}) 
+ g_{1}^{\prime} \, H^{\prime} \cdot V^{(N)} 
\right\}
\label{vlimcont1}
\end{eqnarray}
has a smooth limit when $g_{1}^{\prime} \to 0$
while $(u_{1}^{\prime \, T} \cdot V^{(N)})$ is kept $\neq 0$ i.e.
$\detg\to 0$ while $\dets \neq 0$:
\begin{equation}\label{vlimcont2}
W^{\prime} \to \frac{1}{(u_{1}^{T} \cdot V^{(N)})} \, u_{1}
\end{equation}
namely the expression for $W$ found in eqs. (\ref{eqv}), 
(\ref{inv-22aa}). 

\vspace{0.3cm}

\noindent
Let us now focus on $Q$. With $G^{(N)}$ invertible we found (cf.\ eq.~(\ref{sol11}))
\[
Q^{\prime} = 
- \, (G^{(N)})^{-1} + 
B \, (G^{(N)})^{-1} \cdot V^{(N)} \cdot V^{(N) \, T} \cdot (G^{(N)})^{-1}
\]
As just seen, $B \, V^{(N) \, T} \cdot (G^{(N)})^{-1} \to 
(u_{1}^{T} \cdot V^{(N)})^{-1} \, u_{1}^{T}$ but in each term giving $Q^{\prime}$ 
there is an extra $(G^{(N)})^{-1}$ which becomes wild onto the $u_{1}$ 
direction, thus $Q^{\prime}$ written in this form is an indeterminate $\infty - \infty$. 
We rewrite $Q^{\prime}$ splitting the various contributions on
$u_{1}^{\prime}$ vs. the $u_{j \geq 2}^{\prime}$:
\begin{eqnarray}
- \, Q^{\prime} 
& = &
\left\{ 
  \left( g^{\prime}_{1} \right)^{-1} \, u_{1}^{\prime} \cdot u_{1}^{\prime \, T}
 +
 \sum_{j=2}^{N-1} 
 \left( g^{\prime}_{j} \right)^{-1} \, u_{j}^{\prime} \cdot u_{j}^{\prime \, T}
\right\}
\nonumber\\
&&
- \, \frac{(-1)^{N-1}}{\dets} \, 
\left\{
  \left( g^{\prime}_{1} \right)^{-1} \,
 u_{1}^{\prime}  \cdot  u_{1}^{\prime \, T} \cdot V^{(N)}
 +
 \sum_{j=2}^{N-1} \left( g^{\prime}_{j} \right)^{-1} \, 
 u_{j}^{\prime} \cdot u_{j}^{\prime \, T} \cdot V^{(N)}
\right\} 
\nonumber\\
&& 
\;\;\;\;\;\;\;\;\;\;\;\;\;\;\;\;\;\;\;
\mbox{} \cdot \left\{
 \gamma_{1}^{\, \prime} \, 
  V^{(N) \, T} \cdot u_{1}^{\prime} \cdot  u_{1}^{\prime \, T}
 +
 \sum_{k=2}^{N-1} \gamma_{k}^{\, \prime} \,
 V^{(N) \, T} \cdot u_{k}^{\prime} \cdot  u_{k}^{\prime \, T}
\right\}
\label{contQ1}
\end{eqnarray}
Using eqs. (\ref{Hprime}) and  (\ref{tool}), eq. (\ref{contQ1}) becomes:
\begin{eqnarray}
- \, Q^{\prime} 
& = &
\left\{ 
  \left( g_{1}^{\prime} \right)^{-1} \, u_{1}^{\prime} \cdot u_{1}^{\prime \, T}
 +
H^{\prime}
\right\}
\nonumber\\
&&
- \; \frac{(-1)^{N-1} \, \gamma_{1}^{\, \prime}}{\dets} \, 
\left\{
  \left( g_{1}^{\prime} \right)^{-1} \, u_{1}^{\prime} \; 
 \left( u_{1}^{\prime \, T} \cdot V^{(N)} \right)
 +
 H^{\prime} \cdot V^{(N)}
\right\} 
\nonumber\\
&& 
\;\;\;\;\;\;\;\;\;\;\;\;\;\;
\mbox{} \cdot \left\{
 \left( V^{(N) \, T} \cdot u_{1}^{\prime} \right) \; u_{1}^{\prime \, T}
 +
 g_{1}^{\prime} \, V^{(N) \, T} \cdot H^{\prime}
\right\}
\label{contQ2}
\end{eqnarray}
which can be recast into
\begin{eqnarray}
- \, Q^{\prime} 
& = &
H^{\prime} +
\left( g_{1}^{\prime} \right)^{-1} \, u_{1}^{\prime} \cdot u_{1}^{\prime \, T}
\left\{ 
 1 \; -  \; \frac{(-1)^{N-1}}{\dets} \, \gamma_{1}^{\, \prime} \, 
 \left( V^{(N) \, T} \cdot u_{1}^{\prime} \right)^{2} \,
\right\} 
\nonumber\\
&&
- \; \frac{(-1)^{N-1}}{\dets} \, \gamma_{1}^{\, \prime} \, 
\left( u_{1}^{\prime \, T} \cdot V^{(N)} \right)
\left\{
 u_{1}^{\prime} \cdot V^{(N) \, T} \cdot H^{\prime}
 +
 H^{\prime} \cdot V^{(N)} \cdot u_{1}^{\prime \, T}
\right\} 
\nonumber\\
&& 
- \; \frac{(-1)^{N-1}}{\dets} \, 
\gamma_{1}^{\, \prime} \, g_{1}^{\prime} \,
H^{\prime} \cdot V^{(N)} \cdot V^{(N) \, T} \cdot H^{\prime}
\label{contQ3}
\end{eqnarray}
Using (\ref{detsnew}) we rewrite the first bracket in eq. (\ref{contQ3}) as
\begin{eqnarray}
\lefteqn{
 g_{1}^{\prime} \, 
 \left\{ 
  1 \; -  \; \frac{(-1)^{N-1}}{\dets} \, \gamma_{1}^{\, \prime} \, 
  \left( V^{(N) \, T} \cdot u_{1}^{\prime} \right)^{2}
 \right\} 
}
\nonumber\\
& = &
\frac{(-1)^{N-1}}{\dets} \, \gamma_{1}^{\, \prime} \, \Omega^{\prime}
\label{contQ4}
\end{eqnarray}
in which the indeterminate $\infty - \infty$ has been cancelled leaving 
a net contribution with a finite limit when $g_{1}^{\prime}$ will 
be sent to $0$. Finally using (\ref{detsnew}) 
$Q^{\prime} $ takes the following form:
\begin{eqnarray}
- \, Q^{\prime} 
& = &
H^{\prime} + 
\frac{\Omega^{\prime}}
{g_{1}^{\prime} \, \Omega^{\prime} + 
\left( u_{1}^{\prime \, T} \cdot V^{(N)} \right)^{2}} \, 
u_{1}^{\prime} \cdot u_{1}^{\prime \, T}
\nonumber\\
&&
\;\;\;\;
- \;  \frac{\left( u_{1}^{\prime \, T} \cdot V^{(N)} \right)}
{g_{1}^{\prime} \, \Omega^{\prime} + 
\left( u_{1}^{\prime \, T} \cdot V^{(N)} \right)^{2}} \, 
\left[
 u_{1}^{\prime} \cdot V^{(N) \, T} \cdot H^{\prime}
 +
 H^{\prime} \cdot V^{(N)} \cdot u_{1}^{\prime \, T}
\right] 
\nonumber\\
&& 
\;\;\;\;
- \; \frac{g_{1}^{\prime}}
{g_{1}^{\prime} \, \Omega^{\prime} + 
\left( u_{1}^{\prime \, T} \cdot V^{(N)} \right)^{2}} \, 
H^{\prime} \cdot V^{(N)} \cdot V^{(N) \, T} \cdot H^{\prime}
\label{contQ5}
\end{eqnarray}
In the limit $g_{1}^{\prime} \to 0$ while 
$(u_{1}^{\prime \, T} \cdot V^{(N)})$ kept $\neq 0$, 
the last term of (\ref{contQ5}) $\to 0$ and
\begin{eqnarray}
Q^{\prime} 
& \to & 
H + \;
\frac{\Omega}{\left( u_{1}^{T} \cdot V^{(N)} \right)^{2}} \;
u_{1} \cdot u_{1}^{T}
\nonumber\\
&& 
\;\;\;\;\;
\; - \; 
\frac{1}{\left( u_{1}^{T} \cdot V^{(N)} \right)} \,
\left[ 
 u_{1} \cdot \left( V^{(N) \, T} \cdot H \right) 
 + \left( H \cdot V^{(N)} \right) \cdot u_{1}^{T} 
\right] 
\label{limdetg0}
\end{eqnarray}
i.e. namely the form found in eq.~(\ref{eqresultQ0}),  q.e.d.!

\vspace{0.3cm}

\noindent
This study has yielded interesting byproducts: eq. (\ref{tool2}) relates 
in some sense the matrix $\mbox{Cof}[G^{(N)}]$ (which provides the inverse
of $G^{(N)}$ as long as $\detg \neq 0$), and the MPPI $H$ of 
$G^{(N)}$ when $\detg =0$, 
leading to eqs. (\ref{detsnew}) and (\ref{newB}). Eq. (\ref{contQ5}) will also help understand what make the
case ``$\detg = 0, \dets \neq 0$" vs. the combined limit
``$\detg = 0, \dets = 0$ simultaneously" depart from each
other, as it interpolates between these two limit cases in some sense.

\vspace{0.3cm}

\noindent
Let us compare $\widehat{\Sigma}$ obtained in subsec.\ \ref{bothvanish} with 
$(\widehat{\cal S}^{(N)})^{-1}$ extracted in subsec.\ \ref{nevervanish}. 
If one formally imposes the compatibility condition 
$(u_{1}^{\prime \, T} \cdot V^{(N)})=0$ first
while keeping $g_{1}^{\prime} \neq 0$, then sends $g_{1}^{\prime} \to 0$,
the ingredients $W$ and $Q$ building $\widehat{\Sigma}$ are simply
obtained from those building $(\widehat{\cal S}^{(N)})^{-1}$ by replacing 
$(G^{(N)})^{-1}$ by $H^{\prime}$; $U$ is obtained likewise by reconsidering 
eq. (\ref{newB}): 
the dependence in $g_{1}^{\prime}$ that causes the
indeterminate 0/0 in $B$ cancels and the r.h.s. of (\ref{newB}) is just
$\Omega^{\prime \, -1} \to \Omega^{-1}$. 
With the {\em prior} formal imposition of compatibility condition 
(\ref{compcondG}), the limit
$g_{1}^{\prime} \to 0$ is the {\em combined} limit 
$\dets \to 0$, $\detg \to 0$ keeping their ratio 
$\Omega$ fixed cf. (\ref{newB}). 
This remark helps us understand how the case addressed in this subsec. and the
one covered in the previous subsec. of this appendix depart from each other. 
The latter instead was the limit $g_{1}^{\prime} \to 0$ while 
$(u_{1}^{\prime \, T} \cdot V^{(N)})$ was kept $\neq 0$ so that 
$\dets$ was kept non zero. Keeping 
$(u_{1}^{\prime \, T} \cdot V^{(N)}) \neq 0$ and sending $g_{1}^{\prime} \to 0$ 
(as done in the previous subsec.) then sending 
$(u_{1}^{\prime \, T} \cdot V^{(N)}) \to 0$ vs. imposing 
$(u_{1}^{\prime \, T} \cdot V^{(N)}) = 0$ first then sending $g_{1}^{\prime} \to 0$
(as done in the present subsec.) faces a mismatch between the orderings 
of limits. This is best seen considering the various scalar factors that 
weight the tensor structures decomposing $Q^{\prime}$ in eq. (\ref{contQ5}).
Let us consider for instance the last term in eq. (\ref{contQ5}): 
\[
\mbox{last term of} \,(\ref{contQ5}) 
= 
\frac{g_{1}^{\prime}}
{g_{1}^{\prime} \, \Omega^{\prime} + 
\left( u_{1}^{\prime \, T} \cdot V^{(N)} \right)^{2}} \, 
H^{\prime} \cdot V^{(N)} \cdot V^{(N) \, T} \cdot H^{\prime}
\]
taking $g_{1}^{\prime} \to 0$ while keeping 
$(u_{1}^{T} \cdot V^{(N)}) \neq 0$ makes this term drop from 
the expression of $Q^{\prime}$. In contrast, taking first 
$(u_{1}^{\prime \, T} \cdot V^{(N)}) = 0$ makes $g_{1}^{\prime}$ cancel 
from the scalar factor which then has the finite limit $\Omega^{-1}$ when 
$g_{1}^{\prime} \to 0$, hence the contribution 
$\propto H \cdot V^{(N)} \cdot V^{(N) \, T} \cdot H/ \Omega$ to $Q$ in the
``combined limit" case.
Conversely, let us consider the third term in eq. (\ref{contQ5}) given by:
\[
\mbox{third term of} \,(\ref{contQ5}) 
= 
\frac{\left( u_{1}^{\prime \, T} \cdot V^{(N)} \right)}
{g_{1}^{\prime} \, \Omega^{\prime} + 
\left( u_{1}^{\prime \, T} \cdot V^{(N)} \right)^{2}} \, 
\left[
 u_{1}^{\prime} \cdot V^{(N) \, T} \cdot H^{\prime}
 +
 H^{\prime} \cdot V^{(N)} \cdot u_{1}^{\prime \, T}
\right]
\]
When taking $g_{1}^{\prime} \to 0$ while keeping 
$(u_{1}^{T} \cdot V^{(N)}) \neq 0$ this term yields the finite contribution
$\propto [ u_{1} \cdot V^{(N) \, T} \cdot H + H\cdot V^{(N)} \cdot u_{1}^{T}] /
(u_{1}^{\prime \, T} \cdot V^{(N)})$ to $Q^{\prime}$. In contrast, taking first 
$(u_{1}^{\prime \, T} \cdot V^{(N)}) = 0$ makes this term drop from 
the expression of $Q^{\prime}$ in the ``combined limit" case.

%\section{Two basic integrals \label{appendJ}} 
\section{Integrals occurring in the computation of $L_3^4$ and $L_4^4$ \label{appendJ}} 

In what follows $A$ and $B$ are assumed dimensionless and complex valued, the
signs of their real parts are unknown, and the signs of their imaginary parts 
may or may not be the same either. 

\subsection{First kind}

The computation of $I_{3}^{4}$ involves the computation in closed
form of the following kind of integral:
\begin{equation}
K = \int^{\infty}_0 \frac{d \xi}{(\xi+A) \, (\xi+B)}
\label{eqdefk1}
\end{equation} 
After partial fraction decomposition the r.h.s. of eq. (\ref{eqdefk1}) becomes:
\begin{equation}
K = \frac{1}{B-A} \, 
\int^{\infty}_0 d \xi \, \left[ \frac{1}{\xi+A} - \frac{1}{\xi+B} \right]
\label{eqdefk2}
\end{equation}
The r.h.s. of eq. (\ref{eqdefk1}) converges at infinity although each term 
separately grows logarithmically. Let us introduce a cut off $\Lambda$ and 
write eq. (\ref{eqdefk2}) as:
\begin{equation}
K = \lim_{\Lambda \rightarrow +\infty} \widetilde{K}(\Lambda), \;\;\;\;\; 
\widetilde{K}(\Lambda) 
\; = \; 
\frac{1}{B-A} \, 
\int^{\Lambda}_0 d \xi \, \left[ \frac{1}{\xi+A} - \frac{1}{\xi+B} \right]
 \label{eqdefk3}
\end{equation}
The $\xi$ integration yields:
\begin{align}
\widetilde{K}(\Lambda) &= 
\frac{1}{B-A} \, 
\left[ 
 \ln \left( \frac{1+A/\Lambda}{A} \right) 
 - 
 \ln \left( \frac{1+B/\Lambda}{B} \right) 
\right] 
\notag 
\end{align}
so that 
\begin{equation}
K 
= 
\frac{1}{B-A} \, \left[ \ln \left( B \right) - \ln \left( A \right) \right]
\label{eqdefk4}
\end{equation}
regardless of the signs of $\Im(A)$ vs. $\Im(B)$.

\subsection{Second kind}

The integral
\begin{equation}
  J(\nu) 
  = 
  \int^{+\infty}_0  
   \frac{d \xi}{\left(\xi^{\nu}+A \right) \, \sqrt{\xi^{\nu}+B}}
  \label{eqdefj1}
\end{equation}
arises in some intermediate step 
in the computation of three- and four-point functions, for real 
 masses. When no internal masses are vanishing it arises for 
$\nu = 2$. The integral need not be
computed in closed form and shall instead be 
recast in an alternative, handier form cleared from any radical. 

\vspace{0.3cm}

\noindent
In the case of real masses, $\Im(A)$ and $\Im(B)$ are always of the same sign. Whenever it is the case, 
the use of the celebrated Feynman ``trick" is justified and leads to:
\begin{equation}\label{feynmantrick}
\frac{1}{\left(\xi^{2}+A \right) \, \sqrt{\xi^{2}+B}}
=
\frac{1}{2} \, 
\int^1_0 
\frac{d x \, x^{-\frac{1}{2}}}{(\xi^{2}+(1-x) \, A + x \, B)^{3/2}}
\end{equation}
$J(2)$ is readily rewritten as:
\begin{equation}
J(2) 
= 
\frac{1}{2} \; 
\int^{+\infty}_0 d \xi \, 
\int^1_0 
\frac{d x \, x^{-\frac{1}{2}}}{(\xi^{2}+(1-x) \, A + x \, B)^{3/2}}
\label{eqdeffuncj1}
\end{equation}
Then the $\xi$ integration is performed first, using eq. (\ref{eqFOND1}) of
appendix B. Performing the change of variable $z = \sqrt{x}$ in the result 
obtained yields:
\begin{equation}
J(2) = \int^1_0 \frac{dz}{(1-z^2) \, A + z^2 \, B}
\label{eqdeffuncjp2}
\end{equation}

\section{Basic integrals in terms of dilogarithms and
logarithms: $K$-type integrals}\label{appF}

The computations of the various $N$-point functions in closed form can be
reduced to the calculation of integrals of simple types. The $K$-type is 
of the form
\[
K 
=  
\int^a_b du \, 
\frac{\ln ( A \, u^2 + B) - \mbox{``subtracted term"}}{u^2 - u_0^2}
\]
where $u_0^2 \neq - B/A$. 
In the real mass case, $A$ is real and 
the non vanishing 
imaginary part of the complex quantity $B$ is infinitesimal. 
The integration contour runs along the positive real axis, 
more precisely, this case involves the segment between $[0,1]$.

\vspace{0.3cm}

\noindent
In the real mass case, $u_0^2$ may primitively appear in 
calculations either as a real parameter, or slightly shifted away from the real 
axis by an infinitesimal contour prescription $\Im(u_0^2) = i \, s \lambda$ with
$s = \pm$. 
In the calculation of three-point functions presented in this article $K$-type 
integrals naturally arise with real parameters $u_0^2$ always in conjunction 
with a ``subtracted term" equal to $\ln ( A \, u_0^2 + B)$, so that the poles at 
$u = \pm u_0$ are actually fake. Alternatively, the calculation of four-point 
functions involves $K$-type integrals with parameters $u_0^2$ slightly shifted 
away from the real axis, which makes the integrals well-defined even with a 
vanishing ``subtracted term". In the latter case however, framing the 
calculation with a vanishing ``subtracted term" may result in an annoying 
mismatch between the infinitesimal imaginary part of $B$ shifting the 
integration contour from the cut of $\ln(A \, u^2 + B)$ in the numerator, 
and the contour prescription for the pole $1/(u^2- u_0^2)$. 
This mismatch may lead to possible apparent ambiguities 
on the signs of infinitesimal imaginary parts of some arguments in $\dilog$ 
functions, such as $\dilog[(u_0 \pm 1)/(u_0 - \baru)]$ with 
$\baru = \sqrt{-B/A}$ when the signs of the infinitesimal $\Im(u_0)$ and  
$\Im(\baru)$ happen to be the same while $\Re[(u_0 \pm 1)/(u_0 - \baru)] > 1$. 
Sordid fiddling would then be required to solve these ambiguities. 
This annoyance can be avoided by framing the calculation so as to 
involve $K$-type integrals with a ``subtracted term" equal to
$\ln(A \, \Re(u_0^{2}) + B)$ instead of 0, which effectively allows to 
ignore the infinitesimal imaginary part in $u_0^2$, as will be shown below.

\vspace{0.3cm}

\noindent
We hereby compute the above type of integral.
This appendix often makes use of the identity
\begin{align}
\ln(z) &= \ln(-z) + i \, \pi \, S(z) \;\; , \;\; S(z) \equiv \sign(\Im(z))
\label{eqdeflnzlnmz}
\end{align}

\noindent
Let us parametrise $u_0^2$ by $\Re(u_0^2) = u^{2 \, R}_0$ and
$\Im(u_0^2) = s \lambda$ with $s = \pm$ and note $u_0 \equiv \sqrt{u^{2}_0}$. 
We define:
\begin{align}
K^R_{0,1}(A,B,u_0^2) 
&= \int^1_0 du \, 
\frac{\ln(A \, u^2 + B) - \ln(A \, u^{2 \, R}_0 + B)}{u^2 - u_0^{2}} 
\label{eqdefkrab1b}
\end{align}
Using partial fraction decomposition, eq.\ (\ref{eqdefkrab1b}) can be written:
\begin{align}
K^R_{0,1}(A,B,u_0^2) &= \frac{1}{2 \, u_0} \, \int^1_0 du \, 
\left( \frac{1}{u-u_0} - \frac{1}{u+u_0} \right) \, 
\left( \ln(A \, u^2 + B) - \ln(A \, u^{R \, 2}_0 + B) \right)
  \label{eqdefkrab2b}
\end{align}
\noindent
Introducing $\baru^2 \equiv -B/A$ and $S_u \equiv S(-\baru^2)$, the logarithmic 
term $\ln (A \, u^2+B)$ can be split as\footnote{This comes from the splitting of $\ln (ab)$ with $a$ real 
not necessarily $>0$ and $b$ is complex non real, for which \cite{tHooft:1978jhc}
\[
\ln(ab) 
= 
\ln(a - i \, \lambda \, S_b) + \ln(b), \;\; 
S_b = \sign \left( \Im(b) \right)
\]
}
\begin{align}
\ln (A \, u^2+B) 
&= \ln(A-i \, \lambda \, S_u ) + \ln(u - \baru) + \ln(u+\baru)
\label{eqarrange1}
\end{align}
It is convenient to introduce $\tuz$ defined by:
\[
  \tuz \equiv
  \left\{
  \begin{array}{ccc}
    i \, s \, \sqrt{-u^{2 \, R}_0} & \text{if} & u^{2 \, R}_0 < 0 \\
    & & \\
    \sqrt{u^{2 \, R}_0} & \text{if} & u^{2 \, R}_0 > 0 \\
  \end{array}
  \right.
\]
The subtracted term can in its turn be split as
\begin{align}
\ln(A \, u^{2 \, R}_0 + B) 
&= \ln(A-i \, \lambda \, S_u ) + \ln(\tuz - \baru) + \ln(\tuz+\baru) + 
\eta(\tuz-\baru,\tuz+\baru)
\label{eqarrange2a}
\end{align}
where the function $\eta(z_1,z_2)$ is given by:
\begin{align}
  \eta(z_1,z_2) &= \ln(z_1 \, z_2) - \ln(z_1) - \ln(z_2) \notag \\
  &= \left\{
  \begin{array}{cl}
    -2 \, i \, \pi & \text{if} \; \Im(z_1) \geq 0, \; \Im(z_2) \geq 0 \; \text{and} \; \Im(z_1 \, z_2) < 0 \\
    2 \, i \, \pi & \text{if} \; \Im(z_1) < 0, \; \Im(z_2) < 0 \; \text{and} \; \Im(z_1 \, z_2) \geq 0 \\
    -2 \, i \, \pi & \text{if} \; \Im(z_1) = 0, \; \Im(z_2) = 0, \; \Re(z_1) < 0 \; \text{and} \;  \Re(z_2) < 0 \\
    0 & \text{otherwise}
  \end{array}
  \right.
  \label{eqdefeta01}
\end{align}
Incorporating identities (\ref{eqarrange1}) and 
(\ref{eqarrange2a}) into eq. (\ref{eqdefkrab2b}), we get:
\begin{align}
&K^R_{0,1}(A,B,u_0^2)
\notag\\
&= \frac{1}{2 \, u_0} \, 
\left[ 
 \bR(u_0,\baru,\tuz) + \bR(u_0,-\baru,\tuz) - 
 \bR(-u_0,\baru,-\tuz) - \bR(-u_0,-\baru,-\tuz) 
 \vphantom{\frac{u_0+1}{u_0}} 
\right. 
\notag \\
&\quad {} \quad {} \quad {} \;\;
\left.
 - \eta(\tuz-\baru,\tuz+\baru) \, \ln \left( \frac{u_0-1}{u_0} \right)
 + \eta(-\tuz-\baru,-\tuz+\baru) \, \ln \left( \frac{u_0+1}{u_0} \right) 
\right] 
\label{eqdefkrab3b}
\end{align}
where we have defined:
\begin{equation}
  \bR(u_0,\baru,\tuz) = \int^1_0 du \, \frac{\ln(u-\baru) - \ln(\tuz-\baru)}{u-u_0}
\label{eqdeffctr1}
\end{equation}
Using eq. (\ref{eqdeflnzlnmz}),  
$[\bR(u_0,\baru,\tuz) - \bR(-u_0,-\baru,-\tuz)]$ can be further recast into
\begin{align}
&\bR(u_0,\baru,\tuz) - \bR(-u_0,-\baru,-\tuz)
\notag \\
& \quad{} = \int^1_{-1} du \, \frac{\ln(u-\baru) - \ln(\tuz-\baru)}{u-u_0} 
 + i \, \pi \, \left[ S(\baru) + S(\tuz-\baru) \right] \, 
 \ln \left( \frac{u_0}{1+u_0} \right)
\label{eqdefdifr2}
\end{align}
For convenience, let us define 
\begin{equation}
\bR^{\prime}(u_0,\baru,\tuz) 
= 
\int^1_{-1} du \, \frac{\ln(u-\baru) - \ln(\tuz-\baru)}{u-u_0}
\label{eqdeffctr1bis}
\end{equation}
A comment is in order here. 
One shall be cautious that in eqs. (\ref{eqdeffctr1}) and 
(\ref{eqdeffctr1bis}), the denominators of the integrands
(pole terms) involve $u_0$ {\em whereas} $\tuz$ is involved in the numerators 
(logarithmic subtracted terms). Notwithstanding, \\
{\bf i)} when $\Re(u_0^2) < 0$, the pole in eqs. (\ref{eqdeffctr1}) and 
(\ref{eqdeffctr1bis}) is well away the integration contour, the 
contour prescription of the pole is irrelevant and $u_0$ can be
harmlessly traded for $\tuz$; \\
{\bf ii)} when $\Re(u_0^2) > 0$, $\tuz$ is real and $u_0 = \tuz + i \, s \, \lambda$, 
in which case the contour prescription of the pole proves also 
to be ineffective. Indeed, let us rewrite the r.h.s. of eq. 
(\ref{eqdeffctr1bis}) explicitly as:
\[
\int^1_{-1} du \, 
\frac{\ln(u-\baru) - \ln(\tuz-\baru)}{u - \tuz - i \, s \, \lambda}
\]
the numerator vanishes $\propto (u - \tuz)$ at $u=\tuz$ so that the 
pole at $u=\tuz$, whether $\tuz$ is on the integration contour or not, 
is actually fake. Only the prescription coming from the infinitesimal
$\Im(\baru)$ in $\ln(u-\baru)$ matters, which determines on which
side of the cut the integration contour runs. 
Thus $\bR^{\prime}(u_0,\baru,\tuz)$ is equivalently\footnote{
The limit $\lambda \to 0$ of the integrand pointwise in $u$ provides an 
``integrable hat in the variable $u$" which controls the limit $\lambda \to 0$ 
of the integral defining $\bR^{\prime}(u_0,\baru,\tuz)$, towards the value 
$R^{\prime}(\baru,\tuz)$, cf. Lebesgue's theorem of dominated convergence.} 
given by:
\begin{equation}
  \bR^{\prime}(u_0,\baru,\tuz) = \int^1_{-1} du \, 
\frac{\ln(u-\baru) - \ln(\tuz-\baru)}{u-\tuz} \equiv R^{\prime}(\baru,\tuz)
\label{eqdeffctr10}
\end{equation}
This is so because the subtracted term involves $\tuz$, not $u_0$. 
If instead the subtracted term were 
$\ln(u_0-\baru)$ a potential conflict might occur between the signs of 
$\Im(u_0)$ vs. $\Im(\baru)$. This might result into a non vanishing residue 
$[\ln(\tuz-\baru) - \ln(u_0-\baru)]$ whenever $\tuz-\Re(\baru) <0$ and 
sign$(\Im(u_0-\baru)) =$ sign$(\Im(\baru))$. 
No such 
subtlety is faced with complex masses as will be met in the second article, for which poles 
and cuts are generically well away from the integration contour.

\vspace{0.3cm}

\noindent
In conclusion, eq. (\ref{eqdeffctr1bis}) can be safely traded for 
eq. (\ref{eqdeffctr10}) in all cases. The integral $R^{\prime}(y,z)$ can be 
computed following the main line of the 
computation in appendix B of ref \cite{tHooft:1978jhc}, more generally for two complex 
numbers $y$ and $z$ ($z$ can be real or complex, whereas $\Im(y)$ shall be 
kept non vanishing) as this will also be relevant in the complex mass case.
We first make the change of variable $x=u-y$:
\begin{align}
R^{\prime}(y,z)  
&= \int^{1-y}_{-1-y} dx \, \frac{1}{x+y-z} \, \left( \ln(x) - \ln(z-y) \right)
\label{eqdefrprime1}
\end{align}
Next we deform the straight contour $[-1-y,1-y]$ of integration in eq. 
(\ref{eqdefrprime1}) into the dihedron $[-1-y,0] \cup [0,1-y]$ in the complex
$x$-plane. This procedure is legitimate since 
1) the pole at $x=z-y$ is fake thus we do not have to care about the location 
of $(z-y)$ w.r.t. the triangle $\{-1-y,0,1-y\}$, and 
2) none of the paths $[0, \pm 1 - y]$ crosses the cut of $\ln(x)$. 
We split the integral as 
\[
\int^{1-y}_{-1-y} = \int^{1-y}_{0} - \int^{-1-y}_{0}
\] 
and $R^{\prime}(y,z) $ becomes:
\begin{align}
R^{\prime}(y,z)  
&= 
\int^{1-y}_{0} dx \, \frac{1}{x+y-z} \, \left( \ln(x) - \ln(z-y) \right) \notag \\
&\quad {} - 
\int^{-1-y}_{0} dx \, \frac{1}{x+y-z} \, \left( \ln(x) - \ln(z-y) \right)
\label{eqdefrprime2}
\end{align}
We then make the changes of variable $x=(1-y) \, t$ in the first integral 
and $x=(-1-y) \, t$ in the second one, so that $R^{\prime}(y,z) $ now reads:
\begin{align}
R^{\prime}(y,z)  
&= \int^{1}_{0} dt \, 
\left\{ 
 \frac{d}{dt} \, \ln \left( 1 + t \, \frac{1-y}{y-z} \right) 
\right\} \, 
\left( \ln( (1-y) \, t) - \ln(z-y) \right) 
\notag \\
&\quad {} - 
\int^{1}_{0} dt \, 
\left\{ 
 \frac{d}{dt} \, \ln \left( 1 - t \, \frac{1+y}{y-z} \right) 
\right\} \, 
\left( \ln( -(1+y) \, t) - \ln(z-y) \right)
\label{eqdefrprime3}
\end{align}
The two integrals of eq. (\ref{eqdefrprime3}) are performed by
integration by parts. Using the identity
$\dilog(w) = -\dilog(1-w) + \pi^2/6 - \ln(w) \, \ln(1-w)$ 
we finally get:
\begin{align}
R^{\prime}(y,z) 
&= 
\dilog \left( \frac{z+1}{z-y} \right) - 
\dilog \left( \frac{z-1}{z-y} \right) 
\notag \\
&\quad {} 
+ 
\eta \left( -1-y,\frac{1}{z-y} \right) \, \ln \left(\frac{z+1}{z-y} \right) 
- 
\eta \left( 1-y,\frac{1}{z-y} \right) \, \ln \left(\frac{z-1}{z-y} \right)
\label{eqdefrprime4}
\end{align}
The quantity $[\bR(u_0,-\baru,\tuz)-\bR(-u_0,\baru,-\tuz))]$ can be computed 
along the same lines, and we end up with:
\begin{flalign}
\bR \left( u_0,\baru,\tuz \right) + &\bR \left( u_0,-\baru,\tuz \right) 
=  \bR \left( -u_0,\baru,-\tuz \right) + \bR\left( -u_0,-\baru,-\tuz \right) 
 \notag \\
&\quad {} \quad {}\quad {}\quad {} 
+ \calf \left( \tuz,\baru \right) +  i \, \pi \, 
\left[ S\left( \tuz-\baru \right) + S \left( \tuz+\baru \right) \right] \, 
\ln \left( \frac{u_0}{1+u_0} \right) 
  \label{eqdefrprime5}
\end{flalign}
where
\begin{align}
\calf(x,y)  
&= 
\dilog \left( \frac{x+1}{x-y} \right) - 
\dilog \left( \frac{x-1}{x-y} \right) + 
\dilog \left( \frac{x+1}{x+y} \right) - 
\dilog \left( \frac{x-1}{x+y} \right) 
\notag \\
&
\quad {} 
- \eta \left( 1-y,\frac{1}{x-y} \right) \, 
\ln \left(\frac{x-1}{x-y} \right) + 
\eta \left( -1-y,\frac{1}{x-y} \right) \, 
\ln \left(\frac{x+1}{x-y} \right) 
\notag \\
&
\quad {}
- 
\eta \left( 1+y,\frac{1}{x+y} \right) \, 
\ln \left(\frac{x-1}{x+y} \right) + 
\eta \left( -1+y,\frac{1}{x+y} \right) \, 
\ln \left(\frac{x+1}{x+y} \right) 
%\notag \\
\label{eqdefcalf2a}
\end{align}
Putting everything together in eq. (\ref{eqdefkrab3b}) things can be further 
rearranged using eq. (\ref{eqdeflnzlnmz}), by noting that, for any two complex
numbers $a$ and $b$:
\begin{equation}
  \eta(a,b)-\eta(-a,-b) = - i \, \pi \, \left[ S(a) + S(b) \right]
  \label{eqdefdiffeta}
\end{equation}
We finally get\footnote{$u_0$ is harmlessly 
replaced by $\tuz$ in the prefactor $1/(2 u_0)$ in the expression of 
$K^R_{0,1}(A,B,u_0^2)$.} for $K^R_{0,1}(A,B,u_0^2)$:
\begin{align}
  K^R_{0,1}(A,B,u_0^2) &= \frac{1}{2 \, \tuz} \, \left[ \calf(\tuz,\baru) - \eta(\tuz-\baru,\tuz+\baru) \, \ln \left( \frac{u_0-1}{u_0+1} \right) \right]
  \label{eqdefkrab3bbis}
\end{align}
In particular when $\tuz$ is real all the $\eta$ functions appearing explicitly or through $\calf(\tuz,\baru)$ in eq.~(\ref{eqdefkrab3bbis}) vanish.

\section{Prescription for the imaginary part of $\dets$: real masses i.e. infinitesimal $\Im$ part}\label{ImofdetS}

\noindent
In this appendix, we discuss the sign of the imaginary part
of $\dets$ for the real mass case.
$I_{N}^{4}$ involves the quadratic form
\begin{equation}
{\cal Q}(z) 
= - z^{T}.{\cal S}.z -  i \lambda, \;\;\;\; z^{T} = [z_{1}, \cdots, z_{N}]
\label{e1}
\end{equation}
with $\lambda$ real positive\footnote{$\lambda$ being infinitesimal or
finite is irrelevant here}.
The explicit calculation of $I_{N}^{4}$ involves the square root of 
$\dets$ which carefully keeps track of the $i \lambda$ prescription in 
${\cal Q}(z)$.
    
\vspace{0.3cm}

\noindent
Since the $z_{j}$ sum to 1, ${\cal Q}(z)$ can be rewritten:
\begin{eqnarray}
{\cal Q}(z) 
& = &
-  z^{T}.{\cal S}.z -  i \lambda \, 
\left( \sum_{i=1}^{N}z_{i} \right) \left( \sum_{j=1}^{N}z_{j} \right)
 = - z^{T}.\left[ {\cal S} + i \lambda E \right] .z
\label{e2a}
\\
E_{ij} & = & 1 \;\; \mbox{for all} \; i,j = 1, \cdots, N
\label{e2b}
\end{eqnarray}
Thus, the sought-after prescription is provided by:
\begin{equation}
\dets : \, = \det \left[ {\cal S} + i \lambda E \right]
\label{e3}
\end{equation}
${\cal S}$ is implicitly assumed to be invertible in what follows. 

\vspace{0.3cm}

\noindent
Let us write the r.h.s. of eq. (\ref{e3}) as: 
\begin{eqnarray}
\det \left[ {\cal S} + i \lambda E \right]
& = & 
\left( i \lambda \right)^{N}
\dets \,
\det 
 \left[ 
  {\cal S}^{-1} \cdot E 
  \, + \, 
  \left( \frac{1}{i \lambda} \right)  \mathds{1}_{N} 
 \right]
\label{e4}
\end{eqnarray}
The 2$^{nd}$ determinant in the r.h.s.\ of eq.~(\ref{e4}) is the value of the 
characteristic polynomial of the matrix $-K$ with $K = {\cal S}^{-1} \cdot E$ evaluated at 
$x = 1/(i \lambda)$. 
The matrix $K$ is given by
\begin{eqnarray}
K
& = &
\left[
 \begin{array}{ccccc}
 b_{1}  & \cdots & b_{1} & \cdots &  b_{1}\\
 \vdots &        & \vdots&        & \vdots\\
 b_{j}  & \cdots & b_{j} & \cdots &  b_{j} \\
 \vdots &        & \vdots&        & \vdots\\
 b_{N}  & \cdots & b_{N} & \cdots &  b_{N}
 \end{array}
\right]
\label{e5}\\
b_{j} 
& = & 
\sum_{k=1}^{N} \left( {\cal S}^{-1} \right) _{jk}
\label{e5c}\\
\mbox{tr} [ K ] 
& = & 
\sum_{j=1}^{N} b_{j} = B = (-1)^{N-1} \, \frac{\detg}{\dets}
\label{e5b}
\end{eqnarray}
(for eq. (\ref{e5b}) cf.\ appendix~\ref{detsdetg})
i.e.\ all lines of ${\cal S}^{-1} \cdot E$ are proportional to each other, 
thus ${\cal S}^{-1} \cdot E$ has rank 1: all its eigenvalues but one are zero.
Consequently, the coefficients of all but the first two monomials of highest
and next-to-highest degrees in $x$ of  
$\det \left[  K  \, + \,  x \, \mathds{1}_{N}  \right]$ vanish and
\begin{eqnarray}
  \det \left[  K  \, + \,  x \, \mathds{1}_{N}  \right]
& = &
x^{N} + x^{N-1} \, \mbox{tr} \left[ K \right]
\nonumber\\
& = &
x^{N} + x^{N-1} \, B
\label{e6}
\end{eqnarray}
One thus has:
\begin{eqnarray}
\det \left[ {\cal S} + i \lambda E \right]
& = & 
\dets + i \lambda  \, B \, \dets
\nonumber\\
& = &
\dets + i \lambda \, (-1)^{N-1} \detg 
\label{e7-0}
\end{eqnarray}
i.e. the prescription for the sign of ``the $\Im$ part of $\dets$" 
is ``$+ \lambda \, \mbox{sign}\left( (-1)^{N-1} \detg \right)$'' 
(as long as $\detg \neq 0$). 
Let us also notice that the proof holds whether $\lambda$ is infinitesimal or
finite.
We can remark that the prescription coming naturaly for all $\Delta_{\tn}$ in sec.\ \ref{sectthreepoint} and \ref{sectfourpoint}, namely a $+ i \, \lambda$, 
after applying the Stokes-type identity is the one obtained by taking into account the prescription given by eq.~(\ref{e3}) 
(supplemented by eq. (\ref{e7-0})) into eq.~(\ref{eqe10-02}).

\bibliographystyle{unsrt}
\bibliography{../biblio,../publi}

\end{fmffile}

\end{document}